\documentclass[journal]{IEEEtran}
\usepackage{amsmath,amsfonts}
\usepackage{algorithmic}
\usepackage{hyperref}
\usepackage{tabularx}
\usepackage{booktabs}

\usepackage[linesnumbered, ruled, noend]{algorithm2e}
\usepackage{soul} % 导入 soul 包
\usepackage{color, xcolor} % 颜色包，color 必须导入，xcolor 建议导入

\usepackage{array}

\usepackage{textcomp}
\usepackage{stfloats}
\usepackage{url}
\usepackage{verbatim}
\usepackage{graphicx}
\usepackage[caption=false,font=footnotesize,labelfont=sf,textfont=sf]{subfig}
\usepackage{multirow}
\usepackage[figuresright]{rotating}

\begin{document}

\title{A Correlated Data-Driven Collaborative Beamforming Approach for Energy-efficient IoT Data Transmission}

% \title{A Query-Driven Communication Scheme for Wireless Sensor Networks via Routing Protocols and Collaborative Beamforming}

\author{Yangning~Li,
    Hui Kang,
    Jiahui~Li,
    Geng Sun,~\IEEEmembership{Senior Member,~IEEE,}
    Zemin Sun,
    Jiacheng Wang, \\
    Changyuan Zhao,
    Dusit Niyato,~\IEEEmembership{Fellow,~IEEE}
    
    \thanks{This work is supported in part by the National Natural Science Foundation of China (62172186, 62272194, 62471200), in part by the Science and Technology Development Plan Project of Jilin Province (20240302079GX), in part by the National Research Foundation, Singapore, in part by Infocomm Media Development Authority under its Future Communications Research \& Development Programme (FCP-NTU-RG-2022-010 and FCP-ASTAR-TG-2022-003), in part by Singapore Ministry of Education (MOE) Tier 1 (RG87/22 and RG24/24), in part by the NTU Centre for Computational Technologies in Finance (NTU-CCTF), in part by the RIE2025 Industry Alignment Fund - Industry Collaboration Projects (IAF-ICP) (Award I2301E0026), administered by A*STAR, in part by Alibaba Group and NTU Singapore through Alibaba-NTU Global e-Sustainability CorpLab (ANGEL), in part by the Postdoctoral Fellowship Program of China Postdoctoral Science Foundation (GZC20240592), in part by the China Postdoctoral Science Foundation General Fund (2024M761123), and in part by the Scientific Research Project of Jilin Provincial Department of Education (JJKH20250117KJ). \textit{(Corresponding authors: Jiahui Li and Geng Sun.)}}

    \thanks{Yangning Li, Hui Kang, Jiahui Li, and Zemin Sun are with the College of Computer Science and Technology, Jilin University, Changchun 130012, China, and also with Key Laboratory of Symbolic Computation and Knowledge Engineering of Ministry of Education, Jilin University, Changchun 130012, China (e-mails: ynli23@jlu.edu.cn; kanghui@jlu.edu.cn, lijiahui@jlu.edu.cn, sunzemin@jlu.edu.cn).}

    \thanks{Geng Sun is with the College of Computer Science and Technology, Jilin University, Changchun 130012, China, and with Key Laboratory of Symbolic Computation and Knowledge Engineering of Ministry of Education, Jilin University, Changchun 130012, China; he is also affiliated with the College of Computing and Data Science, Nanyang Technological University, Singapore 639798 (e-mail: sungeng@jlu.edu.cn).}

    \thanks{Jiacheng Wang, Changyuan Zhao and Dusit Niyato are with the College of Computing and Data Science, Nanyang Technological University, Singapore (e-mails: jiacheng.wang@ntu.edu.sg; zhao0441@e.ntu.edu.sg; dniyato@ntu.edu.sg). }
        
    }

\maketitle

\begin{abstract}
An expansion of Internet of Things (IoTs) has led to significant challenges in wireless data harvesting, dissemination, and energy management due to the massive volumes of data generated by IoT devices. These challenges are exacerbated by data redundancy arising from spatial and temporal correlations. To address these issues, this paper proposes a novel data-driven collaborative beamforming (CB)-based communication framework for IoT networks. Specifically, the framework integrates CB with an overlap-based multi-hop routing protocol (OMRP) to enhance data transmission efficiency while mitigating energy consumption and addressing hot spot issues in remotely deployed IoT networks. 
% Following this, we formulate an efficient data transmission optimization problem, which considers both the energy-efficient routing of data and the optimal selection of CB nodes for uplink communication. 
Based on the data aggregation to a specific node by OMRP, we formulate a node selection problem for the CB stage, with the objective of optimizing uplink transmission energy consumption. Given the complexity of the problem, we introduce a softmax-based proximal policy optimization with long-short-term memory (SoftPPO-LSTM) algorithm to intelligently select CB nodes for improving transmission efficiency. 
Simulation results show that the proposed OMRP improves network lifetime by 17\% compared to benchmark routing protocols, while the SoftPPO-LSTM method for CB node selection achieves an 8.3\% increase in throughput over benchmark algorithms. 
% \color{black}
The results also reveal that the combined OMRP with the SoftPPO-LSTM method effectively mitigates hot spot problems and offers superior performance compared to traditional strategies.
\end{abstract}


\section{Introduction}

\IEEEPARstart{T}{he} Internet of Things (IoT) has been widely adopted in various monitoring and control applications, such as environmental monitoring~\cite{yang2023environment}, energy sensing~\cite{qin2024cdt}, and industrial automation~\cite{yang2024effective}. However, as the scale of IoT networks expands, these IoT devices generate massive data volumes, leading to challenges in wireless data harvesting and dissemination as well as energy management~\cite{verma2017survey}. In particular, there is significant data redundancy among the massive data packets due to spatial and temporal correlations, especially in homogeneous IoT networks~\cite{jan2021marginal, begum2023data}. Thus, it is important to manage these vast amounts of data in IoT networks to keep energy efficient and eliminate data redundancy. 

\par Generally, the existing deployment approaches usually place the base station (BS) or sink centrally within the IoT network, and employ the routing protocols to ensure orderly data routing from extensive networks to the BS. However, this scheme may be unsuitable in the remotely deployed IoT networks which are far from the BS or sink. In such cases, the IoT nodes that are closer to the BS often bear a heavier load of relay tasks. Thus, using traditional routing protocols results in excessive energy consumption for the last hop due to geographic distance and data accumulation, which exacerbates hot spot issues~\cite{rawat2023survey}. Moreover, the excessive long-distance links also increase the energy consumption of the IoT devices for data transmission. 

\par In this case, collaborative beamforming (CB) can enhance transmission gain for long-distance links between IoT devices and a remote BS. Specifically, multiple IoT devices can form a virtual antenna array (VAA) and synchronize their transmissions, thereby achieving constructive interference at the location of the remote BS. In this case, an ideal CB setup with $N$ collaborating nodes can achieve an $N^2$ fold increase in power received at the destination. Conversely, for a given threshold of received power, the required transmit power can be decreased by a factor of $1/N^2$~\cite{ochiai2005collaborative}. Thus, CB can compensate for the insufficient transmission gains of IoT devices. Meanwhile, the sufficient transmission gain allows the selection of collaborating nodes to be independent of their geographic distribution, which can significantly mitigate the hot spot issue.

\par However, existing research on CB in IoT networks has not fully integrated the characteristics of the IoT networks. Specifically, most studies focus on optimizing beam patterns to improve transmission performance (\textit{e.g.},~\cite{sun2020improving, smida2023dual, jung2019secrecy, wang2021uplink}), while some also address the network lifetime through multi-slot CB optimization (\textit{e.g.},~\cite{bao2019stochastic, sun2019energy, liu2023energy, hasan2023power}). Nevertheless, few studies consider the data characteristics and routing challenges within IoT networks. For instance, in some homogeneous CB-based IoT systems, there is data redundancy among devices, which should ideally be eliminated before uploading the data to the BS. Besides, expecting each node to use CB after data is generated to communicate with a remote BS is impractical due to the potential for collisions and interference. Thus, managing the efficient and organized upload of data is crucial. Furthermore, the optimization of CB should account for the long-term data and energy management of the IoT network, which has received little attention. 
\par Accordingly, we propose a novel, data-driven CB-based approach for data harvesting and dissemination. This approach reduces energy consumption in data transmission by controlling the routing of data sharing among IoT nodes and selecting CB nodes for uplink communication over a timeline. We summarize the key novelty and main contributions of this paper as follows.

% \color{black}
\begin{itemize}

\item \textit{CB-based Data-driven Long-term Data Harvesting and Dissemination System for IoT}: We introduce CB into a data-driven system for long-term data harvesting and dissemination in IoT networks. In this system, the IoT network aggregates data through routing, eliminates redundancy via data fusion during routing, and then forms a VAA to upload the data to the remote BS. This process is repeated until the IoT nodes deplete their energy. The system is designed from a data-driven perspective, considering the data generation, fusion, and output to the BS. To the best of our knowledge, this is the first exploration of CB-based IoT optimization from a data-driven perspective.

\item \textit{Energy-efficient Routing Protocol Based on Data Redundancy Estimation}: We propose an overlap-based multi-hop routing protocol (OMRP) as the routing strategy for the system. Specifically, OMRP adjusts cluster head selection, fusion order, and relay strategy based on data redundancy estimation, thereby optimizing the data aggregation process of the network. As such, a single execution of OMRP aggregates the data from the entire IoT network to a designated node, thereby enhancing energy efficiency and reducing redundancy.

\item \textit{Deep Reinforcement Learning (DRL)-based Node Selection Method Adapted to OMRP}: Proximal Policy Optimization with Long Short-Term Memory (SoftPPO-LSTM) method is used to optimize node selection in the CB stage. After the OMRP routing protocol is executed, the SoftPPO-LSTM method intelligently selects CB nodes to form a VAA. Following the synchronization of data and strategies among CB nodes, the network data is uploaded to the remote BS using CB, ensuring efficient and coordinated transmission.
\item \textit{Simulation and Performance Evaluation}: 
Simulation results show that the proposed OMRP improves network lifetime by 17\% compared to benchmark routing protocols, while the SoftPPO-LSTM method for CB node selection achieves an 8.3\% increase in throughput over benchmark algorithms. Furthermore, the combined OMRP with the SoftPPO-LSTM method effectively mitigates the hot spot problem, offering superior performance compared to traditional strategies, such as those based on distance or energy.

\end{itemize}

\par The rest of this article is organized as follows. Section \ref{sec:related_works} reviews the previous work on routing protocols and CB. Section \ref{sec:sys_model} introduces the system models. Section \ref{sec:problem_fomu} presents the problem description and analysis. Section \ref{sec:proposed_omrp} presents the proposed OMRP. Section \ref{sec:proposed_eppo} presents the proposed SoftPPO-LSTM node selection method. Section \ref{sec:simulation_analysis} gives the simulations of our proposed methods. Section~\ref{sec:discussion} presents the discussion and we conclude this paper in Section \ref{sec:conclusion}.

\section{Related Works} 
\label{sec:related_works}
% 总起段，我们的目标是：数据驱动的、节能的、远距离通信方案
% 然而这些因素通常被单独考虑
\par In this work, we aim to establish a data-driven, long-term, and energy-efficient communication scheme between IoT and the remote BS through the use of CB and routing protocols. However, these approaches and factors are often considered separately in the literature. To illustrate the novelty of our work, we briefly review some key existing studies.

% （分层）路由
% eg: Xu \emph{et al.}~\cite{xu2021robust}
\subsection{Hierarchical Routing in IoT}

% 1、IoT中的（分层）路由作用：复杂网络的数据传输
% 2、分层路由的一些策略：
%      1）分簇：簇头选择 LEACH和R-LEACH
%      2）多跳：减小传输距离从而节能 D2CRP和CMSTR
%      3）感知维持和消除冗余：
%           wang2012coverage、tao2013flow、song2016coverage
\par In IoT, hierarchical routing is an effective network management strategy for handling data transmission issues in complex networks consisting of a large number of nodes. There are several strategies to enhance the efficiency and performance of hierarchical routing protocols. 
% 1）分簇
Firstly, some studies improved the performance of IoT by considering cluster strategy. Heinzelman \emph{et al.}~\cite{heinzelman2000application} first introduced the concept of clustering and cluster head (CH) selection, the proposed LEACH protocol randomly selected CHs in a round-robin approach and balanced the network energy consumption through IoT nodes taking turns being CHs. Behera \emph{et al.}~\cite{behera2019residual} proposed the R-LEACH protocol, which considered the energy level of nodes and the number of CHs in the CH selection process, and reduced broadcast energy consumption by using the headset mechanism.
% 2）再谈谈多跳策略
Secondly, some studies considered multi-hop transmission, which reduces the transmission power of a single node by reducing the communication distance for each node, thereby reducing energy consumption and extending the service life of the node. Chen \emph{et al.}~\cite{chen2022d2crp} introduced D2CRP, leveraging two-hop neighbor information for cluster formation to enhance energy efficiency. Lin \emph{et al.}~\cite{lin2023cmstr} proposed CMSTR, a multi-chain routing scheme, which studied the shortest Hamiltonian path problem to establish intra-cluster routing and adopted the multi-hop mode to establish inter-cluster routing.
% 3）感知维持和消除冗余
Thirdly, research on data within IoT hierarchical routing often centered on perception maintenance and eliminating redundancy. Wang \emph{et al.}~\cite{wang2012coverage} studied the energy-redundantly covered metric on CH selection for the cluster-based routing protocols, and attempted to simulate the most efficient area coverage tessellation, network coverage lifetime to improve network coverage lifespan. Tao \emph{et al.}~\cite{tao2013flow} and Song \emph{et al.}~\cite{song2016coverage} studied the coverage overlap factor to improve both power efficiency and coverage preservation. However, existing works may not fully consider routing protocols from the perspective of data redundancy, which can lead to unnecessary energy consumption and reduced network efficiency.

% Add
% 我们要实现的是 数据具有空间相关性的 query-driven模型
\subsection{Data-Driven and Data Fusion in IoT}

% 1、解释 IoT中的 Data-Driven: methodology
% 2、相关举例：query-driven、event-driven、data redundancy
\par In IoT, data-driven is a core methodology that focuses on leveraging large amounts of data collected from numerous IoT nodes to guide decision-making and optimize operations\cite{Kavitha2021AiII}. For example, Snigdh \emph{et al.}~\cite{snigdh2021energy} proposed a multiple-tree architecture to enhance energy efficiency and extend the lifespan by optimizing query-driven data transmissions. Biswas \emph{et al.}~\cite{biswas2020true} developed an event-driven, fault-tolerant routing algorithm, addressing real event detection and erroneous measurements through distributed event detection and multi-objective routing optimization, enhancing both the efficiency and accuracy of data transmission. Jan \emph{et al.}~\cite{jan2021marginal} proposed a data-driven aggregation mechanism that effectively addresses spatial correlations in resource-constrained networks, reducing data redundancy and enhancing accuracy through a dual-layer processing architecture at the node and cluster head levels.

% 1、解释IoT中的data fusion
% 2、举例：雷达与多光谱图像融合（应用）、犹豫模糊熵算法、模糊算法选择性融合数据
\par Data fusion technology refers to the process of integrating multiple sets of data originating from different sources in order to produce more consistent, accurate, and useful information \cite{begum2023data}. For example, Saeidi \textit{et al.}~\cite{saeidi2014fusion} employed the Dempster-Shafer theory to fuse airborne LiDAR data with multispectral imagery, enhancing land cover feature extraction. Their work demonstrated the efficiency and accuracy of the methodology in integrating multi-sensor data without prior knowledge. Wang \emph{et al.}~\cite{wang2019new} introduced a data fusion algorithm based on hesitant fuzzy entropy, leveraging hesitant fuzzy entropy to enhance decision accuracy by effectively reducing data redundancy and optimizing energy consumption. This approach improved the quality of fused data, ensuring robust and efficient network performance. Yu \emph{et al.}~\cite{yu2024data} developed a clustering-based data fusion algorithm that incorporated fuzzy logic to enhance data accuracy and reduce redundancy. Their method optimized network energy consumption and extended network lifetime by efficiently clustering nodes using a butterfly-optimized fuzzy algorithm and selectively fusing data through a fuzzy logic controller.

% 我们要实现的是 数据具有空间相关性的 query-driven模型
% 详细的定义在III叙述

\par In our study, we consider spatial correlation in a homogeneous IoT network, where geographically closer nodes tend to collect more redundant data. Data fusion technology is used in the routing process to eliminate redundancy. In this case, a data-driven perspective can be applied to the data collection and harvesting process of the IoT network. A detailed description of the modeling process is provided in Section~\ref{sec:sys_model}.

% CB在IoT中的使用
\subsection{CB in IoT for Energy and Performance}

% 1、波束图优化
% 2、节能优化
% 现有的一些工作考虑了CB在WSN中的使用，bala bala... 但还没有同时考虑路由和CB的
\par Some existing work considers the use of CB in IoT. For example, 
Haro \emph{et al.}~\cite{haro2013energy} designed a virtual beamformer via convex optimization, offering both centralized and distributed solutions to extend network lifetime and meet the quality of service requirements by using a random energy consumption model.
Sun \emph{et al.}~\cite{sun2019energy} mitigated the maximum sidelobe level by employing hybrid discrete and continuous optimization strategies, alongside both centralized and distributed methodologies, thereby improving transmission range and energy efficiency. 
Furthermore, Sun \emph{et al.}~\cite{sun2020improving} developed a multi-objective optimization framework for mobile networks, effectively optimizing the maximum side-lobe level, transmission power, and motion energy consumption for distributed CB, achieving enhanced transmission distance and energy efficiency with minimized motion energy expenditure. Bao \emph{et al.}~\cite{bao2019stochastic} introduced a software-defined energy harvesting architecture for CB communications, leveraging a reinforcement learning algorithm to optimize beamforming performance while ensuring long-term operation and energy efficiency. However, existing methods primarily focus on the short-term optimization of data transmission through CB, without considering the routing, data fusion, and network management processes prior to transmission. This oversight may lead to interference and conflicts with other nodes during the CB process, especially in large IoT networks with a limited number of base stations.

\subsection{Comparison with Existing Studies}

\par In traditional hierarchical routing methods such as LEACH and PEGASIS, the focus is mainly on balancing energy consumption through cluster head selection and multi-hop communication. However, these approaches do not address data redundancy caused by spatial or temporal correlations. Similarly, CB studies optimize beam patterns without considering network-level factors such as routing efficiency and long-term energy consumption. Moreover, studies for IoT networks often treat routing and CB as independent processes, missing the opportunity to jointly optimize the entire communication process.

\par Different from the existing works, our framework uniquely combines routing protocols and CB techniques, thereby enabling a holistic optimization of the communication process from data generation to uplink transmission. Additionally, our approach dynamically adjusts strategies based on data redundancy estimation. Furthermore, we address the entire sequence of events, from the initial data generation by IoT nodes to the final upload to the remote BS, thus allowing our DRL-based method to achieve long-term network optimization from a data-driven perspective.

% Section
% System Model
%
\section{System Model}
\label{sec:sys_model}

% 总起段，本章结构

% \par In this section, we first present the overview of the proposed data-driven IoT data dissemination system. Then, we present the system models, including ... . Note that a comprehensive list of the main notations used in this paper is shown in Table~\ref{tab:notation}.

\par In this section, we first present the overview of the considered data-driven IoT data harvesting and dissemination system. Then, we present the system models, including the CB-based communication model, energy consumption model, and data correlation and fusion model. Note that a comprehensive list of the main mathematical notations used in this paper is shown in Table~\ref{tab:notation}.

% 要体现：
% 1、Overlap range
% 2、CB

% \begin{figure}
%     \centering
%     \begin{minipage}{0.99\linewidth}
%         \centering
%         \includegraphics[width=\linewidth]{images/0825_sys_a.pdf}
%         % \usepackage[font=normalsize,labelfont=sf,textfont=sf]{subcaption}
%         % \captionsetup{justification=centering}
%         % \caption{}
%         \label{fig:CB_sys_fig:sub1}
%     \end{minipage}
%     \begin{minipage}{0.99\linewidth}
%         \centering
%         \includegraphics[width=\linewidth]{images/0825_sys_b.pdf}
%         % \captionsetup{justification=centering}
%         % \caption{}
%         \label{fig:CB_sys_fig:sub2}
%     \end{minipage}
%     \caption{The CB-based data harvesting and dissemination system overview. (a) The routing process is activated by a query from the BS. (b) Sketch map and geometrical configuration of CB process after aggregating the data to the sink node.}
%     \label{fig:CB_sys_fig}
% \end{figure}

\begin{figure}
    \centering
        \centering
        \includegraphics[width=\linewidth]{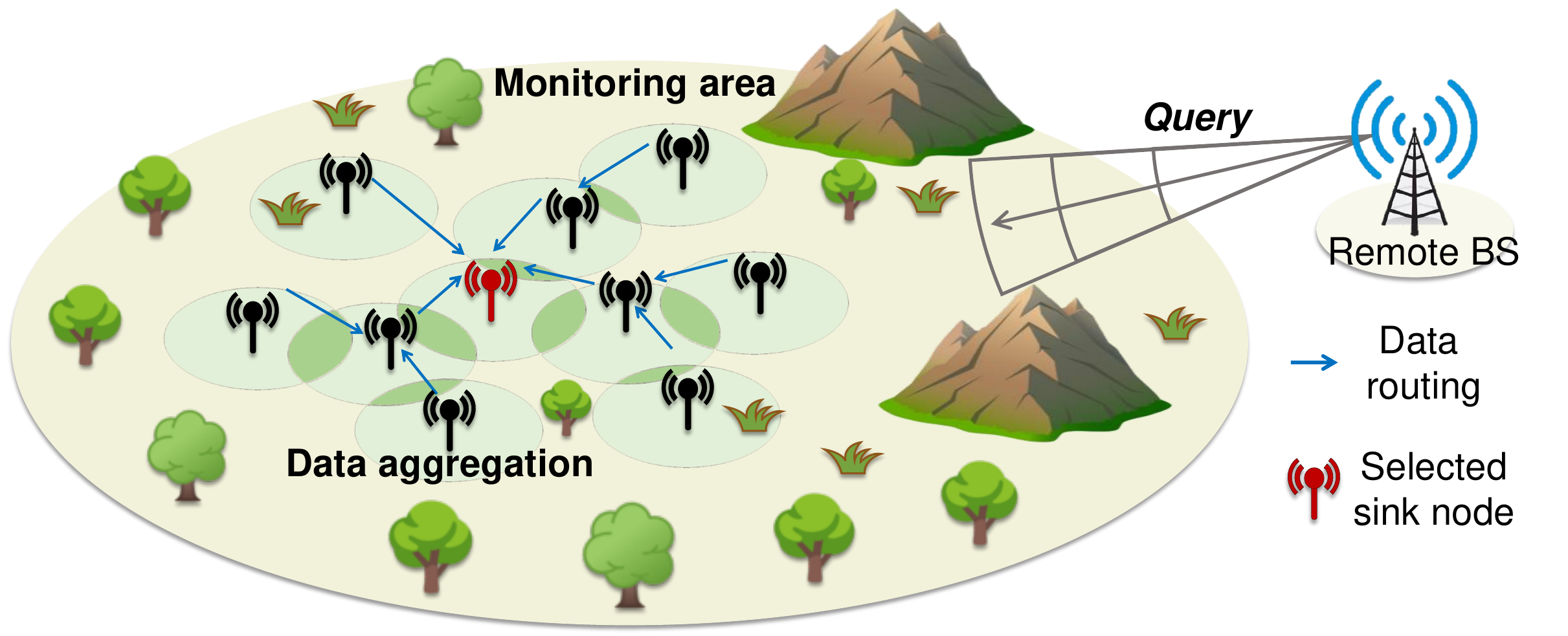}
    \subfloat(a)
        \label{fig:CB_sys_fig:sub1}
        \centering
        \includegraphics[width=\linewidth]{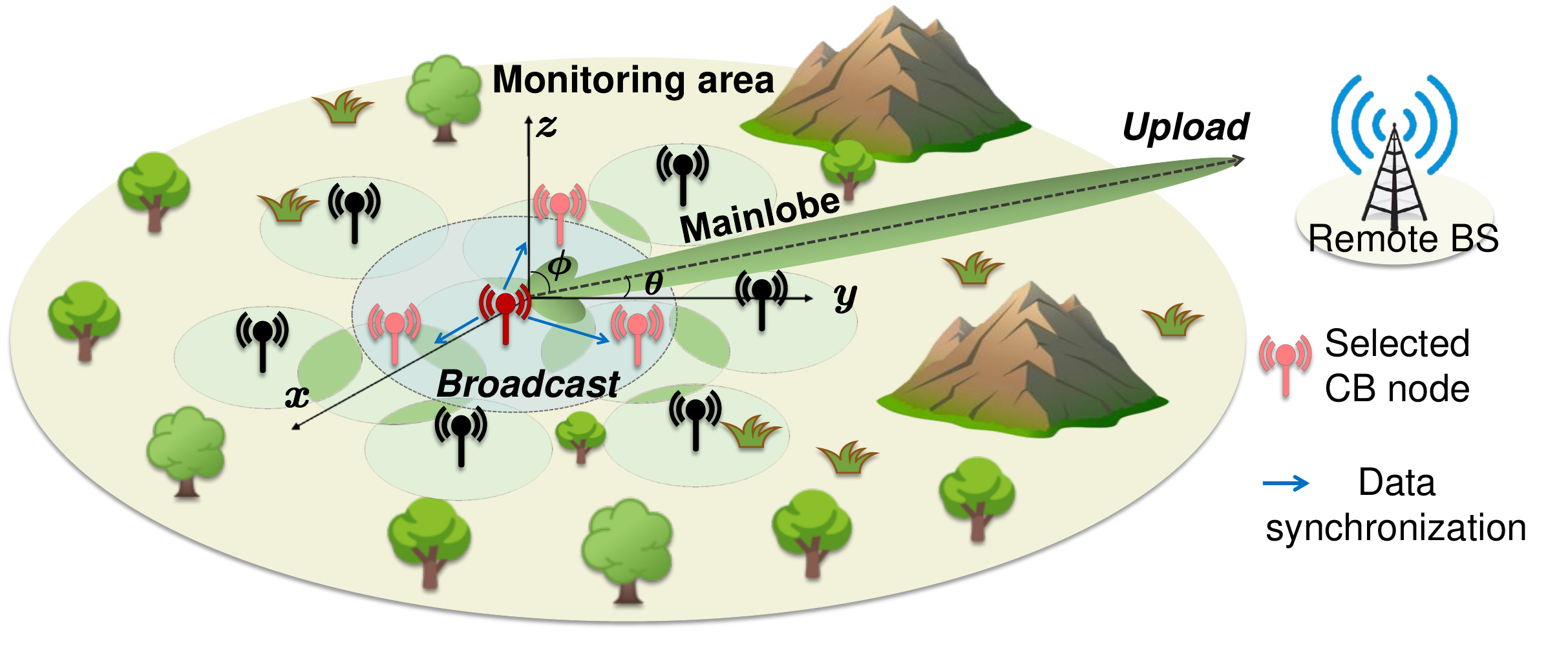}
    \subfloat(b)
        \label{fig:CB_sys_fig:sub2}
    \caption{The CB-based data harvesting and dissemination system overview. (a) The routing process is activated by a query from the BS. (b) Sketch map and geometrical configuration of CB process after aggregating the data to the sink node.}
    \label{fig:CB_sys_fig}
\end{figure}

% \sethlcolor{pink}\hl{[0719 TO ADD New Notations]}

\subsection{System Overview}

% 第一段交代清楚场景 + 宏观的运行机制
\par As is shown in Fig.~\ref{fig:CB_sys_fig}, a CB-based data harvesting and dissemination system is considered to serve a large homogeneous IoT cluster, in which the monitor area contains massive IoT nodes. The assumptions of the system are as follows:
\begin{itemize}
   
   \item The IoT nodes are static and energy-limited. Once deployed, there is no external energy supply. 

   \item Each node supports basic data processing and fusion capabilities, and is equipped with a single omnidirectional antenna operating at a specific frequency for CB.

   \item The network handles homogeneous data, which is common in environmental monitoring scenarios that involve images, temperature maps, or point-cloud data~\cite{fei2025deep, john2023evaluation, adade2024advanced}.

   \item Each IoT node has its own location and monitoring area. The precise locations of the IoT nodes can be obtained through positioning systems or innate settings, and there may be overlapping areas among nodes.
   
   \item There are spatial correlations among the IoT nodes with overlapped monitoring areas. Data redundancy can be eliminated by IoT nodes through data fusion~\cite{hua2008optimal}.
\end{itemize}

\par Based on these assumptions, we denote the set of IoT nodes as $\mathcal{N} = \{1, 2, \cdots, N_{IoT}\}$, and the monitoring area of node $i$ with a radius of $r$ is $A_i$. The neighbor list of node $i$ is the node whose monitoring areas overlap with node $i$, and their set is defined as $\mathcal{F}_i = \{j | d_{i,j} < 2r, j \in \mathcal{N}\}$, where $d_{i,j}$ is the Euclidean distance of node $i$ and node $j$. Due to the insufficient transmit power of the IoT nodes and complex terrestrial network environments, it is difficult for a single node to transmit data to a remote BS directly. Therefore, these IoT nodes will gather the data packets through routing policy and communicate with the remote BS by performing CB.

% 第二段把系统怎么运行交代清楚
\par As such, in a certain mission round, the data sensed by the IoT nodes needs to be aggregated and transmitted to the remote BS for the purposes of data backup and perception maintenance. Due to the long-range transmission distance between the IoT cluster and the BS, the minimum receive power threshold at the remote BS is difficult to meet by the transmit power of a single IoT node. Alternately, multiple IoT nodes are expected to form a VAA and perform CB for uplink data transmission. In this work, we define the sink nodes for intro-cluster data aggregation. Accordingly, all IoT nodes need to first use a certain hierarchical routing protocol to transmit data to the sink node, and then the sink node selects candidate nodes and initiates CB to transmit the data to the remote BS.

% 这里是运行机制的细节
\par
\begin{figure*}
    \centering
    \includegraphics[width=0.99\textwidth]{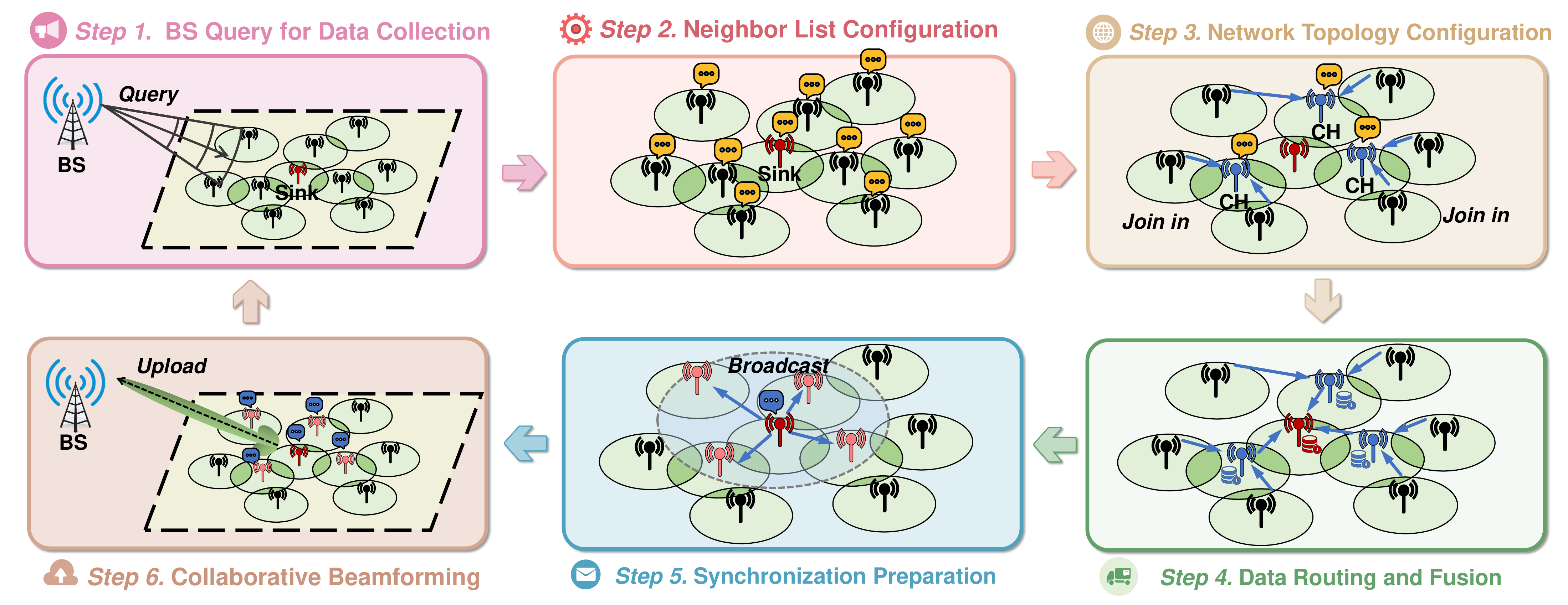}
    \caption{Network model and operating mechanisms. Each round of data harvesting consists of six steps. Routing and fusion are configured and performed in steps 3 and 4, while synchronization and CB are performed in steps 5 and 6.}
    \label{system_steps}
\end{figure*}

 Fig.~\ref{system_steps} shows the designed operation mechanism for the considered CB-based data harvesting and dissemination system. Based on the above analysis, we divide a mission round into six steps, which are listed as follows:

\begin{itemize}

\item \textit{Step 1}: The remote BS broadcasts a message to the network to request the collection of monitor data. The message specifies the sink node of the network for this round. 

\item \textit{Step 2}: After receiving the message, each IoT node re-broadcasts it in order to search for neighbors.

\item \textit{Step 3}: The IoT nodes execute a specific hierarchical routing protocol to determine the network topology and time division multiple access (TDMA) schedule. 

\item \textit{Step 4}: The IoT nodes perform data routing and fusion according to the established network topology and TDMA schedule. Note that data fusion is performed at the receiver, and the fused data of the network is finally collected at the sink node.

\item \textit{Step 5}: The sink node executes the node selection strategy based on the collected network information, and the selected nodes are regarded as beamforming nodes. The sink node broadcasts data and strategies to the beamforming nodes for CB preparation. 

\item \textit{Step 6}: The beamforming nodes perform CB and send the fused data to the remote BS.

\end{itemize}

\par In such operation mechanisms, the sink node retries the node selection strategy if the beamforming nodes fail during \textit{Step 5}. Likewise, the BS reselects a new sink node if CB data is not received for an extended period. Additionally, the BS can identify fault nodes from the received data and broadcast fault information to the network for updates.

% 第三段介绍一下坐标系啥的
\par Without loss of generality, we consider a 3D Cartesian coordinate system, and the positions of the antenna of $i$th IoT node and the remote BS are represented as $(x^{IoT}_{i}, y^{IoT}_{i}, h^{IoT})$ and $(x^{BS}, y^{BS}, h^{BS})$, respectively. Subsequently, we detail the key models with respect to data transmission and fusion including the CB-based communication model, energy model, and data correlation and fusion model.
% \sethlcolor{pink}\hl{[TODO 0708]}

\begin{table}[htbp]
\centering
\caption{{Notations of the Main Conceptions}}
\resizebox{\linewidth}{!}{
\begin{tabular}{c|cl}
\toprule[1.2pt]
& \textbf{Symbols} & \textbf{Definition} \\ 
\midrule[0.9pt]
\multirow{5}{*}{\rotatebox[origin=c]{90}{\textbf{System Model} \quad \quad  \quad \quad \quad \quad \quad \quad \quad}}    

& $A_i$       & The monitoring area of IoT node $i$  \\
& $h^{IoT}$, $h^{BS}$    & Altitudes of the IoT nodes and the remote BS\\ 
& $\mathcal{N}$  & The set of the IoT nodes \\
% & $h^{BS}$    & Altitudes of the remote BS \\ 
% & $M_i$       & The fusion sequence of IoT node $i$  \\
% & ${m}_{ij}$   & The $j$th node in the fusion sequence for node $i$ \\
& $N_{IoT}$, $N_{CB}$         & The total number of IoT nodes and that of nodes performing CB \\
& $d_{i, j}$         & The Euclidean distance between nodes $i$ and $j$  \\
& $\mathcal{F}_i$ & The neighbor list of node $i$ \\
& $\mathrm{AF}(\phi, \theta, I)$ & Array factor \\
& $G_r$, $G_t$ & Receiving gain at the BS and transmitting gain of CB \\
& $P_r$, $P_t$ &   Received power at the BS and transmitting power of CB \\
& $M_i$, ${m}_{ij}$ & The fusion sequence of node $i$ and its $j$th node \\
% & $N_{CB}$    & The number of IoT nodes perform CB \\
& $q_0$       & Energy dissipation-data fusion        \\ 
& $r$         & The monitoring radius of IoT nodes  \\
& $E_{elec}$  & Energy dissipation-electronic circuit        \\ 
& $\varepsilon_{fs}$, $\varepsilon_{mp}$    & Energy dissipation of free space and mutipath amplifiers   \\ 
% & $\varepsilon_{mp}$        & Energy dissipation-mutipath amplifier       \\ 
% $R_c$         & Data spatial correlation coefficient of the circle sensing range  \\ 
& $C^{j}_{i}$ & Data packet size after the fusion with the $j$th node in $M_i$ \\
& $\alpha_{i, j}$    & The fusion rate between two autonomous nodes $i$ and $j$  \\
& $\beta_{i, j}$     & Distance factor for node $i$ selecting node $j$ as a relay node \\
& $E_0$       & Initial energy of each IoT node\\ 
& $\rho_{i}$  & Ratio of the overlapping area between node $i$ and its neighbors \\
& $\mathcal{G}_t $      & The topology of the network in the $t$th round \\
& $\mathcal{N}_t$, $\mathcal{E}_t$  & The set of alive nodes and directed links in the $t$th round \\  
& $T_i$       & The survival time of node $i$ \\
& $T_{(i)}$       & The $i$th element in the sorted survival times of the nodes \\
% &   & The network state set of directed links in the $t$th round \\

\midrule[1.2pt]  
\multirow{6}{*}{\rotatebox[origin=c]{90}{\textbf{Algorithm} \quad \quad \quad \quad }}
& $P$ &  The lower bound ratio of CH nodes to all nodes in OMRP \\
& $K$ &  The amplification factor in OMRP \\
& $\mathcal{S}, \boldsymbol{s}$ & State space and state vector of environment \\
& $\mathcal{A}, \boldsymbol{a}$ & Action space and action vector of agent \\
& $\mathcal{R}, \boldsymbol{r}$ & Reward space and reward \\
& $\mathcal{P}$ & State transition probability of environment \\
& $\gamma$ & Discount factor \\
% & ${d}_{i,s}(t)$  & The distance between node $i$ and the sink in the $t$th round \\
& ${e}_{i}(t)$ & The residual energy of node $i$ in the $t$th round \\
% & $\mathbb{E}_{\pi_{\theta}}$  & The expected reward with respect to the policy 
% $\pi_{\theta}$\\
% & $L^{c l i p} ( \theta)$ & The clipped objective function of PPO algorithm \\
& $v_{i}(t)$ & The score of node $i$ as a beamforming node in the $t$th round \\
& $\zeta_{1}$, $\zeta_{2}$ & The weights assigned to the throughput and the
energy cost \\
& ${\theta_{Q_{new}}}$, ${\theta_{Q_{old}}}$  & The parameters of the new and old actor networks \\
& ${\theta_c}$, ${\theta_f}$  &     The parameters of the critic network and the feature network\\
& $lr$ & Learning rate of the training process\\
% &    & The parameters of the feature network \\ 
\bottomrule[1.2pt]
\end{tabular}}
\label{tab:notation}
\end{table}

% \par In the designed operation mechanism, three models need careful consideration: the CB-based Communication Model, the Energy Model in the Transmission Process of IoTs, and the Data Differentiation and Fusion Model of IoTs. These will be described in detail below.

\subsection{CB-based Communication Model}

% \par To facilitate communication between IoT nodes and the remote BS, CB is employed to achieve high gain and ensure the receiving power threshold of the remote BS is effectively met. Considering the long distance between the sensing area and the remote BS, two-ray multipath fading model is used as follows: 

% \begin{equation}
% \label{beam_two}
% P_{r}=\frac{P_{t} G_{t} G_{r} h_{t}^{2} h_{r}^{2}}{d^{4}},
% \end{equation}

% \noindent where $P_r$, $P_t$ ,$G_r$, $G_t$ represent receiving power, total transmitting power, receiving gain of BS and transmitting gain of CB, respectively. $h_t$, $h_r$ are the altitudes of the nodes and BS, and $d$ is the distance between BS and the collaborating nodes center.

% \sethlcolor{pink}\hl{[TODO]} Fig.~\ref{fig_sys_model}(\subref{sys.sub.2}) 

\par In the CB-based communication model, the positions, transmission power, and initial phase of beamforming nodes determine the performance of communication. Specifically, we let $(\phi,\theta)$ denote any direction centered on the VAA, where $\phi \in [0, \pi]$ and $\theta \in [-\pi, \pi]$ are the elevation and azimuth angles, respectively. According to the electromagnetic wave superposition principle, the array factor (AF) of $N_{CB}$ nodes can be approximated as follows \cite{jayaprakasam2015psogsa}:
\begin{equation}
\label{AF}
\mathrm{AF}(\phi, \theta, I)=\sum_{k=1}^{N_{\mathrm{CB}}} I_{k} e^{j \Psi_{k}} e^{j \frac{2 \pi}{\lambda} d_{k}(\phi, \theta)},
\end{equation}

\noindent where $\lambda$ is is the wavelength, $I_k$ and $\Psi_k$ are the excitation current weight and initial phase of the $k$th node, respectively. Furthermore, $\Psi_k$ is defined as follows:
\begin{equation}
\label{Psi}
\Psi_{k}=-\frac{2 \pi}{\lambda}d_{k}(\phi, \theta),
\end{equation}

\noindent where $d_{k}(\phi, \theta)$ is the Euclidean metric between the $k$th IoT node and the target BS, and it is expressed as follows:
\begin{equation}
\label{dk}
 d_{k}(\phi, \theta) = \sqrt{L^{2}+r_{k}^{2}-2r_{k}L\sin(\theta)\cos(\phi-\delta_{k})},
\end{equation}
where $r_k$ and $\delta_{k}$ collectively define the polar coordinates of the $k$th node.

\par Considering the long distance between the sensing area and the remote BS, the two-ray multipath fading model is used, and the received power at the remote BS is given as follows: 
\begin{equation}
\label{beam_two}
P_{r}=\frac{P_{t} G_{t} G_{r} h_{t}^{2} h_{r}^{2}}{d^{4}},
\end{equation}

\noindent where $P_t$, $G_r$, and $G_t$ represent total transmitting power, receiving gain of BS, and transmitting gain of CB, respectively. Moreover, $h_t$ and $h_r$ are the heights of the IoT nodes and BS, and $d$ is the distance between BS and the center of collaborating nodes.

\par Therefore, the transmission rate from beamforming nodes to BS is as follows:
\begin{equation}
\label{beam_r}
R_{BS}=B\log_{2}\left(1+\frac{P_r}{\sigma^2}\right),
\end{equation}

\noindent where $B$ is the bandwidth, and $\sigma^2$ is the noise power.

\par Based on the above analyses, more beamforming nodes and higher excitation current weight lead to higher received power and transmission rate. However, this will result in more energy consumption during the transmission process.

\subsection{Energy Model in the Transmission Process of IoTs}

\par The energy consumption during the transmission process mainly consists of two parts, which are short-distance communication between nodes and long-distance CB communication between the nodes and the BS. Specifically, the typical first-order radio model~\cite{heinzelman2002application} is used for inter-node communication. Short-distance communication between the nodes considers the free space model and long-distance transmission follows the multipath fading model. The energy required to send and receive $b$-bit packets over a distance $d$ is given as follows:
% \begin{equation}
% \label{radio_trans}
% E_{T}(b,d)=\left\{\begin{array}{l l}{{b\cdot E_{e l e c}+b\cdot\varepsilon_{f s}\cdot d^{2},\quad d<d_{0}}}\\ {{b\cdot E_{e l e c}+b\cdot\varepsilon_{a m p}\cdot d^{4},\quad d\geq d_{0},}}\end{array}\right.
% \end{equation}
\begin{equation}
\label{eq:radio_trans}
E_{T}(b,d) = 
\begin{cases}
b \cdot E_{{elec}} + b \cdot \varepsilon_{{fs}} \cdot d^2, &  d < d_0 \\
b \cdot E_{{elec}} + b \cdot \varepsilon_{{amp}} \cdot d^4, &  d \geq d_0
\end{cases},
\end{equation}
\begin{equation}
\label{eq:radio_rec}
E_{R}(b)=b\cdot E_{elec},
\end{equation}

\noindent where $E_{elec}$ denotes the per bit energy dissipation that is used for running the transmitter and receiver electronic circuits. Moreover, $\varepsilon_{fs}$ and $\varepsilon_{amp}$ represent the energy of amplifier for free-space fading and multipath fading, respectively. Finally, $d_0$ is the distance threshold to determine the two fading models. 

\par Besides, the total energy consumption of the beamforming nodes during the CB process is as follows:
\begin{equation}
\label{eq:beam_energy}
E^{}_{CB}=\sum_{k=1}^{N_{\mathrm{CB}}} I_k^2P_0 \cdot \frac{C}{R_{BS}},
\end{equation}
\noindent where $P_0$ is the maximum transmission power with an excitation current of 1, and $C$ is the size of the data packet after data fusion at the sink node and is also the final packet size to be sent in the CB process.

\par In general, a larger data volume in transmission leads to more energy consumption. Therefore, it is crucial to eliminate redundancy and fuse data efficiently.

\subsection{Data Correlation and Fusion Model of IoTs}
% 定义 correlation model 并给出最一般的fusion rate, 理由:距离越近数据冗余应该越多 【TODO: 补参考文献】
\par
\begin{figure}[ht]
    \centering
    \includegraphics[width=0.90\linewidth]{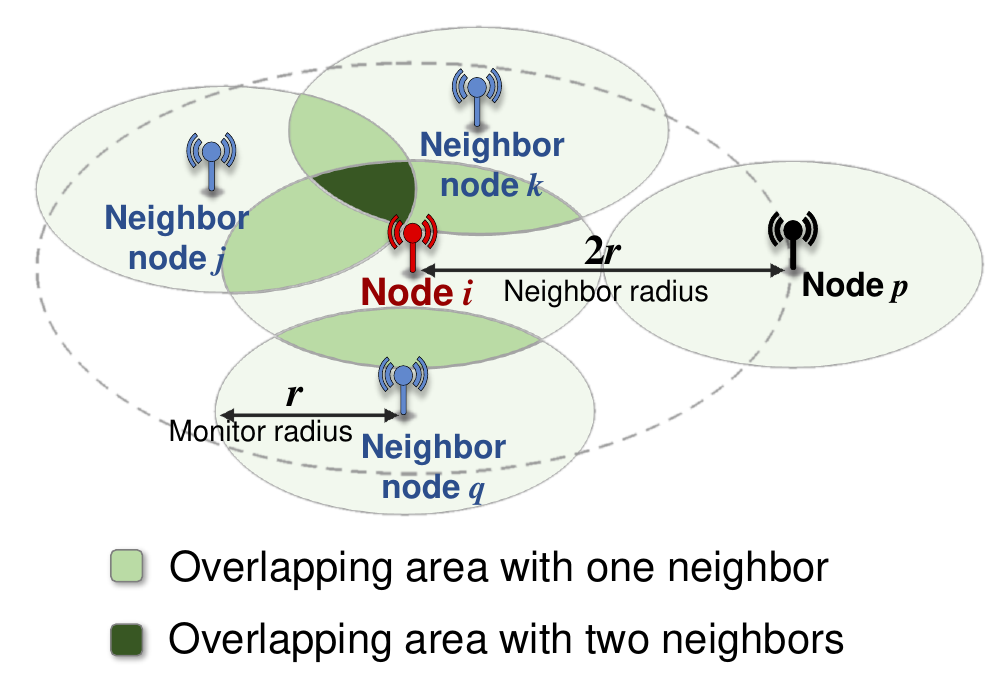}
    \caption{Overlapping area of node $i$ with its neighbors $j$, $k$ and $q$. Note that each node has a monitor radius of $r$.}
    \label{sys_overlaping_fig}
\end{figure}

 In this model, data collected by individual nodes in each round exhibit spatial correlation, and these nodes can eliminate the redundancy through data fusion \cite{hua2008optimal}. Generally, a proximity of two independent nodes leads to a bigger overlapping monitoring area and tends to increase the redundancy of the data they collect. Without loss of generality, we use spatial monitoring areas to define the correlation model. Specifically, the data redundancy rate between two nodes is zero if there is no overlapping monitoring area. On the contrary, as illustrated in Fig.~\ref{sys_overlaping_fig}, overlapping areas within their sensing ranges become evident. The magnitude of the overlapping area serves as an indicator of the data redundancy of these nodes. In this case, the fusion rate $\alpha_{i, j}$ between two autonomous nodes $i$ and $j$ is defined as follows:
\begin{equation}
\label{fusion_rate}
\alpha_{i, j} =  \frac{1}{A_{i}} | A_{i} \cap A_{j} |.
\end{equation}

% 灰色线条改0920

% 进一步论述：数据融合过程可以进一步执行
% 本质：数据融合过程中消除的数据冗余本质上来自于空间感知数据的重叠。
\par Moreover, the fused data packet can be further fused with another packet, the fusion rate is determined by the spatial distribution represented by the data packets. For example, let $M_{i}=\left\langle m_{i 0}, m_{i 1}, \ldots, m_{i\left|M_{i}\right|}\right\rangle$ denote the fusion sequence of node $i$, where ${m}_{ij}$ denotes the $j$th node in the fusion sequence for node 
$i$, and then the fusion rate between node $i$ and node $m_{ij}$ is as follows:
\begin{equation}
\label{fusion_rate_multi}
\alpha_{i,j} = \frac{1}{A_{m_{ij}}} \bigcup_{k = 0}^{j - 1} A_{m_{ik}} \cap A_{m_{ij}},
\end{equation}

\noindent where $A_{m_{ij}}$ is the monitoring area of node ${m_{ij}}$. Following this, the size of the fused data packet at node $i$ in this process changes as follows:
\begin{equation}
\label{fusion_rate_multi}
C^{j}_{i} = C^{j - 1}_{i}+(1- \alpha_{i,j})|A_{j}|,
\end{equation}

\noindent where $C^{j}_{i}$ represents the size of the data packet after data fusion between node $i$ and the $j$th node in $M_i$. Note that the data redundancy eliminated in the data fusion process essentially arises from the overlap of spatial sensing data. In this scenario, regardless of the routing methods used, the size of the data packets collected by the IoT nodes, which have the same spatial distribution, remains constant after efficient data fusion. The fused data will have a smaller data size, thereby saving energy during the next hop transmission process.

\par In the fusion process, downstream nodes (\textit{i.e.}, the nodes that receive the data packets) can scan two packets separately, eliminate redundancy, and generate the fused packet. In this process, the fusion energy cost is calculated as follows: 
\begin{equation}
\label{fusion_energy}
E_{fusion} = q_0(C_1+C_2),
\end{equation}
\noindent where $q_0$ is the standard fusion cost~\cite{luo2006adaptive}, and $C_1$ and $C_2$ are the sizes of two packets to be fused, respectively.

\section{Problem Description and Analysis}
\label{sec:problem_fomu}

\par In this section, we provide a comprehensive description and analysis of the problem, focusing on the challenges associated with maximizing the network lifetime and total throughput to the remote BS. We describe the roles of nodes, formulate the network topology and node selection mathematically, and introduce our motivation for using routing and beamforming methods to jointly optimize network performance.

\subsection{Problem Description}

% 主要目标
% 描述每轮节点的职责（都在干啥）
\par The main goal of the considered system is to maximize the network transmission efficiency to the remote BS. This efficiency is determined by two key factors, namely network lifetime and throughput. In each round of tasks, the system needs to assign different working roles to nodes based on their residual energy and locations within the network as efficiently as possible.
% , thereby maintaining network operation, achieving optimal energy utilization, and extending the network lifetime and throughput. 
% 节点的职责描述
\par Nodes in the considered system can perform multiple roles. On the one hand, nodes can function as both senders and receivers of data. Upon receiving data from upstream nodes (\textit{i.e.}, the nodes that send the data packets), these nodes might either forward it directly as relay nodes or send it to the downstream nodes after fusing the data. These varying roles shape the topology of the network and routing process, and influence the energy consumption in data fusion and transmission. On the other hand, nodes may be required to act as beamforming nodes during the CB phase to enhance signal gain and energy efficiency, or just to remain idle to save energy. The assignment of beamforming roles impacts both the quality of communication and the residual energy of the nodes.

\subsection{Problem Analysis}

% 因此需要考虑时序，引出基于t定义的数学符号
% 使用 0，1，2编码三种状态：0-空边，1-转发，2-融合后转发 
\par To address these varying roles and their impacts effectively, it is essential to make informed decisions about network topology and node selection at different stages of the network lifetime, as decisions may need to adapt over time to ensure optimal performance. 

\par Mathematically, let a directed graph $\mathcal{G}_t = (\mathcal{N}_t, \mathcal{E}_t)$ denote the topology of the network in the $t$th round, where $\mathcal{N}_t$ is the set of alive nodes, and $ \mathcal{E}_t =\{x_{ji} | x_{ji} \in \{0, 1, 2\}, j \in \mathcal{N}_t, i\in \mathcal{N}_t\}$ is the state set of directed links. In particular, the state variable $x_{ji}$ indicates whether to select node $j$ to send data to node $i$ and perform data fusion, and $x_{ji}$ is defined as follows:
\begin{equation}
\label{eq:x_ji}
x_{ji} = \begin{cases} 
0, & j \text{ has no directed links to } i \\
1, & j \text{forwards data to } i, i \text{ forwards it}  \\
2, & j \in M_i, i \text{ fuses the data and forwards it}
\end{cases},
\end{equation}

\noindent where the difference between state $x_{ji} = 2$ and 
state $x_{ji} = 1$ is whether the node $i$ fuses the data from the upstream node $j$ or not. 

% 总结了一下归一化能耗，引出后面 f1-网络生存时间 f2-网络总吞吐量
% 6个部分：路由传输、路由接收、数据融合、CB、数据同步时的接收、数据同步时的broadcast(当选sink)
\par In summary, Eqs.~\eqref{eq:radio_trans}, \eqref{eq:radio_rec}, \eqref{eq:beam_energy}, \eqref{fusion_energy} and \eqref{eq:x_ji} provide the energy consumption of the IoT nodes in the considered system. Specifically, the network topology and role assignment jointly affect the transmission, reception, and fusion energy consumption in the routing phase. In addition, the final packet size and the selection of beamforming nodes affect the transmission and reception energy consumption for data synchronization (See Step 5 of Fig.~\ref{system_steps}). Finally, the excitation current weight of the beamforming node and the final packet size determine the beamforming energy consumption in the CB process. Considering the six energy consumption components mentioned above, the normalized total power consumption of node $i$, denoted as $W_i$, is calculated as follows:
\begin{equation}
\label{eq:W_i}
\begin{aligned}
W_i = \frac{1}{E_0} & \Bigg[ E_{T}(C^{|M_i|}_{i}, d_{i,h}) + \sum_{j \in M_i} E_{R}(C^{|M_j|}_{j}) \\
& + \sum_{j \in M_i} q_0(C^{j - 1}_i + C^{|M_j|}_j) + I_i^2 P_0 \cdot \frac{C^{|M_{s}|}_{s}}{R_{BS}} \\
& + \overline{E}_{R}(C^{|M_{s}|}_{s}) + \overline{E}_{T}(C^{|M_s|}_{i = s}, d_{i, m}) \Bigg],
\end{aligned}
\end{equation}
\noindent where $s$, $h$ and $\smash{C^{|M_{s}|}_{s}}$ denote the sink node, the next hop of node $i$ and the final packet size, respectively. Additionally, $E_0$, $\smash{ \overline{E}_{R}(C^{|M_{s}|}_{s})}$ and $\smash{ \overline{E}_{T}(C^{|M_s|}_{i = s}, d_{i, m})}$ are the amount of initial energy, the average received and broadcast energy consumption when node $i$ acts as a beamforming node or a sink node, respectively.

\par Based on the above analysis, this problem involves long-term network topology management and role assignment, making it difficult to solve with a single method. In addition, existing studies usually consider routing and CB strategies of IoT networks separately. This motivates the joint design of two methods for solving the routing and collaborative beamforming problems.

\par Next, we introduce the routing and data fusion method based on OMRP, which manages network topology and aggregates fused data at the sink node. Then, we present the CB method based on SoftPPO-LSTM, which selects beamforming nodes based on node distribution and residual energy. The two methods jointly optimize the network lifetime and total throughput to the remote BS.

% 第四章：OMRP协议
% 总分结构 -> 介绍协议三个阶段
% \section{The Proposed OMRP for Optimizing $\mathbb{E}$}
\section{The Proposed Overlap-based Multi-hop Routing Protocol for Routing and Data Fusion}
\label{sec:proposed_omrp}

% 0727 new
% \par In this section, we optimize the decision variable $\mathbb{E}$ of the problem~\eqref{eq:formulation} based on the location and residual energy of the nodes. Since the sink node and the node energy change with each round, the decision variable $\mathbb{E}$ needs to change over time accordingly. As such, the decision variable $\mathbb{E}$ can be optimized by proposing a hierarchical routing protocol. Therefore, we seek to propose a protocol, namely, OMRP, to optimize $\mathbb{E}$. 

\par In this section, we introduce the proposed OMRP routing method, which is based on the location and residual energy of nodes. The OMRP method is performed during Steps 3 and 4 in Fig.~\ref{system_steps}. During these steps, OMRP organizes the network into clusters according to the distribution of CHs. These CHs are responsible for gathering and fusing data packets from the cluster nodes before transmitting the data to the sink node via direct or multi-hop communication. The protocol leverages spatial data correlation to direct CH selection, data fusion sequence, and relay schemes within the IoT network. Similar to other classic protocols such as the LEACH protocol and its latest variants~\cite{yang2023environment, heinzelman2000application, chen2022d2crp}, OMRP is designed with three stages, which are the set-up stage, the formation stage, and the data routing stage, and we detail them as follows.

\color{black}

\subsection{Set-up Stage}

% 0727 new
\par In the set-up stage, the proposed OMRP aims to update the neighbor list $\mathcal{F}$ and overlapping degree $\rho$ of each node, thereby determining the CHs of the network. To this end, each node is configured with an initial neighbor list $\mathcal{F}_i$ when the network is first deployed. Then, if a node approaches energy depletion, this node will broadcast the \emph{SLP\_notify} message to its neighbors, prompting them to update their neighbor lists. Based on the updated neighbor lists, the overlapping degree of these nodes can be determined. Let $\rho_i$ represent the ratio of the overlapping area with its neighbors to the total sensing area of node $i$. This ratio is given by
\begin{equation}
\label{eq:rho_i}
\rho_{i}=\frac{1} {A_{i}} \bigcup_{j \in \mathcal{F}_i} A_{i} \cap A_{j}, 
\end{equation}

\noindent where $A_i$ and $\mathcal{F}_i$ are the sensing area and neighbor list of node $i$, and $A_{i} \cap A_{j}$ denotes the area where node $i$ overlaps with its neighbor $j$. Fig.~\ref{sys_overlaping_fig} illustrates an example of the overlapping area between a node and its three neighbors. Evidently, we have $0 \le \rho_i \le 1$, and when $\rho_i = 1$, it means that the sensing area of node $i$ is completely covered by its neighbors.

% 论述 overlapping degree 的数据层次含义: 
% 1、与 redundancy 正相关
% 2、更可能位于网络中心 -> 做CH更有优势

\par Furthermore, a node with a higher overlapping degree indicates that its sensing area is more extensively covered by neighboring nodes, resulting in increased redundancy in the sensed data during the fusion process. In other words, nodes with a larger overlapping degree add fewer unique data elements to the final packet set after fusion, due to the higher redundancy in their sensed data. Additionally, since a higher overlapping degree suggests the presence of more neighboring nodes, it is likely located at the center of a cluster, making it a more suitable candidate to be selected as a CH.

% \par By considering the overlapping degree $\rho$, the threshold function of LEACH is improved as follows:

\par Therefore, we introduce the overlapping degree $\rho$ in the CH election process. Similar to the CH election mechanism in the LEACH protocol, we design a threshold function to determine the probability of a node being selected as a CH. This threshold function is defined as follows:
\begin{equation}
\label{eq:thresholdfunction}
T(i)=\left\{\begin{array}{l l}
{{\displaystyle\frac{P}{1-P\times(r\mathrm{~mod}\frac{1}{P})}K^{\rho_{i}} }}
&{{\mathrm{~}\ i\in G}}\\ 
{{\displaystyle0,}}
&{{\text{otherwise}}}
\end{array}\right.,
\end{equation}

\noindent where $P$ and $K$ are two constants representing the lower bound ratio of CH nodes to all nodes and an amplification factor with a minimum value of 1, respectively. Moreover, $r$ is the number of rounds since the last election of the cluster head, and the elements in the set $G$ are the nodes that have not successfully campaigned for CH nodes in the latest $r\mod(1/P)$ round. 

\par As such, in the CH election process, each node generates a random number between 0 and 1, determining the CH election result by comparing it with $T(i)$. Specifically, if the random number is less than the threshold value $T(i)$, the node is automatically elected as a CH and notifies other nodes by broadcasting the $CH\_notify$ message within the network. Note that $1 \le K^{\rho_{i}} \le K$ is satisfied in the Eq.~\eqref{eq:thresholdfunction}, where $K^{\rho_{i}}$ serves as the amplification factor to increase the likelihood of nodes with larger overlapping degrees being elected as CHs. Based on the above analysis, this approach increases the possibility of nodes with geographical advantages being elected as CHs, thereby improving the network energy efficiency. 
% Notably, this threshold function behaves equivalently to that of the LEACH protocol when all nodes have equal overlapping degrees.

% 本阶段确定：拓扑和TDMA表
\subsection{Network Formation Stage}

% 总述：这个 Stage 干什么 - 簇内外的 拓扑+TDMA
% 分述：先说簇内 ：CH 和 CM各自干什么
%   1、member 加入 cluster
%   2、CH根据 overlapping degrees 确定TDMA排序 -> 更节能
\par This stage primarily establishes the communication topology and timing schedule within clusters and between clusters. 

% 簇内的通信拓扑和时序安排
\par On the one hand, the communication topology and timing schedule within the cluster are determined by the CHs. Specifically, the CHs broadcast $CH\_notify$ messages within the network, following which the member nodes assess the signal strength of the broadcast and determine which cluster to join. Then, each member node transmits $JOIN\_IN$ message to its CH, following which the CH will extract the overlapping degrees from the received $JOIN\_IN$ messages and sort them in descending order. Finally, the CHs calculate and broadcast the TDMA schedule according to this order, allowing member nodes to transmit data in separate time slots to avoid data collision. In this scenario, node data packets with a higher overlapping degree undergo an earlier fusion, leading to more energy savings in the subsequent fusion process due to the reduced fused packet size.

\begin{figure}
    \centering
    \includegraphics[width=0.99\linewidth]{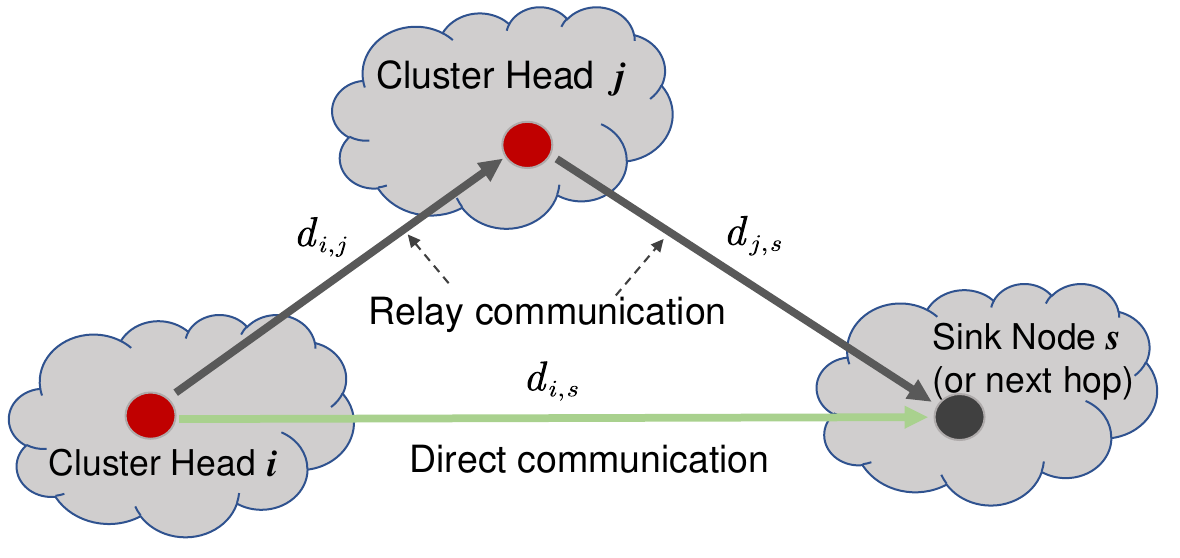}
    \caption{Direct or relay communication method in inter-cluster routing.}
    \label{relay}
\end{figure}

%  再说簇间 - single-hop or multi-hop
%  簇间的通信拓扑和时序安排
\par On the other hand, the communication topology and timing schedule between clusters are determined by the communication scheme between CHs and the sink node. Typically, CHs may forward data packets to the sink node via single-hop or multi-hop methods. Although relay communication can significantly reduce transmission energy consumption by reducing the distance, incompletely fused data will cause additional receiving energy consumption according to Eq.~\eqref{eq:radio_rec}. Consequently, the selection of the next hop for the CHs in OMRP is based on the specific network cluster distribution.

\par Accordingly, we will clarify the principles and mechanisms of next-hop selection in OMRP by analyzing a typical scenario shown in Fig.~\ref{relay}. Specifically, CH $i$ has two methods to communicate with the sink node (or next hop), either treating CH $j$ as a relay node or direct communication. Assume that the distance between all nodes is less than the threshold $d_0$ and the transmission energy dissipation follows the free-space fading model. According to Eqs.~\eqref{eq:radio_trans}, \eqref{eq:radio_rec} and \eqref{fusion_energy}, the energy consumption of the relay scheme is given by
\begin{equation}
    \label{relay-sis}
    \underbrace{2bE_{elec}}_{j, s\ recieve}+\underbrace{ 2bE_{elec}+b\varepsilon _{fs}\left\{ d_{i,j}^{2}+d_{j,s}^{2} \right\}}_{i-j-s\ trans} +bE_{fusion},
\end{equation}

\noindent where $d_{i,j}$ is the distance between node $i$ and node $j$. While the energy consumption for direct communication is given by
\begin{equation}
    \label{relay-dir}
    \underbrace{bE_{elec}}_{s\ recieve}+ \underbrace{bE_{elec}+b\varepsilon _{fs}d_{i,s}^{2}}_{i\ trans}+bE_{fusion}.
\end{equation}

\noindent As can be seen, the energy difference between the two methods is defined as $\varDelta$, which is given by
\begin{equation}
    \label{delta}
    \varDelta =2bE_{elec}+b\varepsilon _{fs}\left\{ d_{i,j}^{2}+d_{j,s}^{2}-d_{i,s}^{2} \right\}.
\end{equation}

\noindent This difference determines which method is more energy-efficient. To facilitate the choice of an optimal relay node, we define the distance factor for node $i$ selecting node $j$ as a relay node as $\beta_{i,j} = d_{i,s}^{2}-\left\{ d_{i,j}^{2}+d_{j,s}^{2} \right\}$, and the candidate relay node is determined by $\beta_{max} = \text{max} \{ \beta_{i,1}, \beta_{i,2}, ..., \beta_{i,j} \}$. According to Eq.~\eqref{delta}, the relay scheme remains more energy-efficient than the direct communication scheme when $\beta_{max} > \frac{2E_{elec}}{\varepsilon_{fs}}$. In this case, the proposed OMRP utilizes the energy-saving relay scheme. The detailed derivation process is shown in Appendix A. We summarize the designed set-up and network formation stages in Algorithm \ref{alg:omrp1}.

% 0826 感觉太罗嗦了，直接删掉似乎也行
% 0731 暂时加个 Appendix 吧【当distance 大于 d_0, 一样适用 】
% \par Moreover, when the distance between nodes exceeds the threshold $d_0$, the candidate relay node determined by $\beta_{max}$ remains optimal, and the relay scheme remains more energy-efficient when $\beta_{max} > \frac{2E_{elec}}{\varepsilon_{fs}}$. The detailed derivation process is shown in Appendix A. We summarize the designed set-up and network formation stage in Algorithm \ref{alg:omrp1}.

\begin{algorithm}[t]
\caption{ \\ Set-up and Network Formation of OMRP}
\label{alg:omrp1}
\textbf{Initialization:} 
\text{Receive query message from the BS calculate} $\rho_i$ \text{according to Eq.~\eqref{eq:rho_i}. $CH \gets \emptyset, SN \gets \emptyset$. } \\

\textbf{// Set-up Stage}

\If{\text{IoT node} $i$ \text{runs out of energy}}{
    \text{$i$ broadcasts $SLP\_notify$ message;} \\
}
\If{\text{IoT node} $j$ receives $SLP\_notify$ message from neighbor $k$}{
    \text{Recalculate} $\rho_j$ \text{without considering neighbor} $k$ \text{according to }Eq.~\eqref{eq:rho_i}; \\
}
\If{$i$ \textbf{is in} $G$}{
     $T(i) \gets K^{\rho_i}*p/(1-p(r \bmod 1/p))$;
}
\Else{ $T(i) \gets 0$;}
\textbf{// Network Formation Stage} \\
\text{IoT node $i$ listens to $CH\_notify$ message;}\\
\If{$i$'s random number $< T(i)$}{
    \text{$i$ broadcasts $CH\_notify$ message;}\\
    \text{Remove $i$ from $G$;} \\
    \text{Add $i$ to $CH$;} \\
}
\Else{
    \text{Add $i$ to $G$;} \\
    \text{Add $i$ to $SN$;} \\
}
\If{$i \in CH$}{
    \For{each $c_j \in CH$}{
        \text{Add $(\beta_j, c_j)$ to $D_i$;} \\
    }
    \text{Find $\beta_{max}$ in $ D_i$ and get $c_{max}$;}  \\
    \If{$\beta_{max} > {2E_{elec}}/{\varepsilon_{fs}}$}{
        $i$ sends a $RELAY$ message to $c_{max}$; \\
        Form relay agreement between $i$ and $c_{max}$;
    }
}
\ElseIf{$i \in SN$}{
    $i$ joins the closest cluster;
}
Start data transmission; \\
$r \gets r + 1$;
\label{alg1}
\end{algorithm}

\subsection{Data Routing Stage}

\par In this stage, each node will act as a sender or receiver at a specific time according to the topology and timing schedule determined in the network formation stage, and eventually all the data packets in the network will be collected and forwarded to the sink node. On the one hand, during the allocated time slot for each member node in the cluster, the sensed data is transmitted to its CH while only the transmitting node remains active, and other member nodes in the cluster turn off their transceivers to save energy. Upon receiving a new data packet, the CH performs data fusion and stores the fused data. On the other hand, after collecting all the packets of their cluster or if the time exceeds the timing schedule, the CHs employ carrier sense multiple access (CSMA) for the subsequent hop transmission and forward the packets to the sink node. Finally, the sink node fuses the received packets and generates the final packets of this round. We summarize the data routing stage in Algorithm \ref{alg:omrp2}.

% 
% \subsection{Summary of OMRP in the System}
% \par The proposed OMRP method involves three main stages, which are summarized as follows:
% \begin{itemize}

%     \item \textit{Set-up Stage:} Nodes update their neighbor lists and overlapping degrees, and CHs are elected based on overlapping degree and energy level. This stage ensures an optimized initial network configuration for routing and data fusion.
    
%     \item \textit{Network Formation Stage:} Communication topologies within clusters and between clusters are established. TDMA schedules are generated to avoid data collision, and inter-cluster relay schemes are determined based on energy-saving criteria.
    
%     \item \textit{Data Routing Stage:} Data packets are transmitted from nodes to their respective CHs, where data fusion occurs. The fused data is forwarded to the sink node via direct or relay communication, optimizing energy consumption and packet size.
%     This structured approach significantly enhances energy efficiency and reduces redundancy in IoT networks.

% \end{itemize}

\color{black}
\par After executing OMRP during the third and fourth steps of the system (See Fig.~\ref{system_steps}), the data packets of the network are gathered and fused at the sink node. The next section will introduce the proposed SoftPPO-LSTM-based CB method, which is designed for data transmission to the remote BS.

\begin{algorithm} [t]
\caption{ \\ Intracluster and Intercluster Routing of OMRP}
\label{alg:omrp2}
% \begin{algorithmic}
    \For{$c_j \in CH$}{
        \text{Listen} $JOIN\_IN\text{ messages}$ \text{from $SN$;} \\
        % \STATE \text{receive} $RELAY\_MSG\text{s}$ \text{from cluster head nodes}
         Get $\mathcal{R} =\{\rho_1,\rho_2,...,\rho_m\}$ from $JOIN\_IN\text{ messages}$; \\
         \text{Sort} $\mathcal{R}$ \text{in descending order;} \\
         Broadcast timing schedule according $\mathcal{R}$;
    }

    \If{$i \in SN$}{
        Transmit data to $i$'s CH by schedule; 
    }
    \If{$i \in CH$}{
        \While{$time <$ timing schedule}{
            $i$ listens and receives data; \\
            \If{$i$ receives data from cluster member}{
                 Fuse data;
            }
        }
        Transmit data to next hop using CSMA;
    }
% \end{algorithmic}
\end{algorithm}

% 
% \section{Design of CB Policy }
% \section{The Proposed SoftPPO-LSTM CB Method for Optimizing $\mathbb{I}$}
\section{The Proposed SoftPPO-LSTM for CB}

% // new 

% \begin{table}[tb]
% \centering
% \caption{Notations of the Main Conceptions in the DRL Algorithm}
% \begin{tabular}{l l}
% \toprule
% \textbf{Notation} & \textbf{Definition} \\ \hline
% $A_{\pi_{\theta}}$       & The advantage function \\
% ${a}_t$ & The action of the $t$th round \\
% ${d}_{i,s}(t)$  & The distance between node $i$ and the sink in the $t$th round \\
% ${e}_{i}(t)$ & The residual energy of node $i$ in the $t$th round \\
% $\mathbb{E}_{\pi_{\theta}}$  & The expected reward with respect to the policy $\pi_{\theta}$\\
% $r_t$ & The reward of the $t$th round \\
% ${s}_t$     & The state of the $t$th round \\
% $L^{c l i p} ( \theta)$ & The clipped objective function of PPO algorithm \\
% $v_{i}(t)$ & The score of node $i$ for acting as a beamforming node \\
% $\theta$ & The policy parameters to optimize \\
% ${\theta_{Q_{new}}}$  & The parameters of the actor network \\
% ${\theta_{Q_{old}}}$ &  The parameters of the old actor network \\
% ${\theta_c}$  &     The parameters of the critic network \\
% ${\theta_f}$   & The parameters of the feature network \\
% \bottomrule
% \end{tabular}
% \label{tab:notation_drl}
% \end{table}

\label{sec:proposed_eppo}
% 总结段

% 
\par In this section, we present the proposed SoftPPO-LSTM, which is adapted to the routing method. The SoftPPO-LSTM method is applied at the beginning of Step 5 to generate a node selection scheme, which is then executed during Steps 5 and 6 in Fig.~\ref{system_steps}. Specifically, the sink node collects and aggregates the data, selects enough adjacent nodes as beamforming nodes, synchronizes the data and CB strategy with them, and collaboratively transmits the data to the remote BS using CB. 
% \color{black}
During this stage, the selection and excitation current weights of the nodes need to be optimized. In this paper, we utilize the DRL method to obtain an optimal CB policy. The motivations for using DRL, the node selection problem formulation, the Markov decision process (MDP) formulation, and our proposed SoftPPO-LSTM algorithm are as follows.

\begin{figure*}
    \centering
    \includegraphics[width=0.99\textwidth]{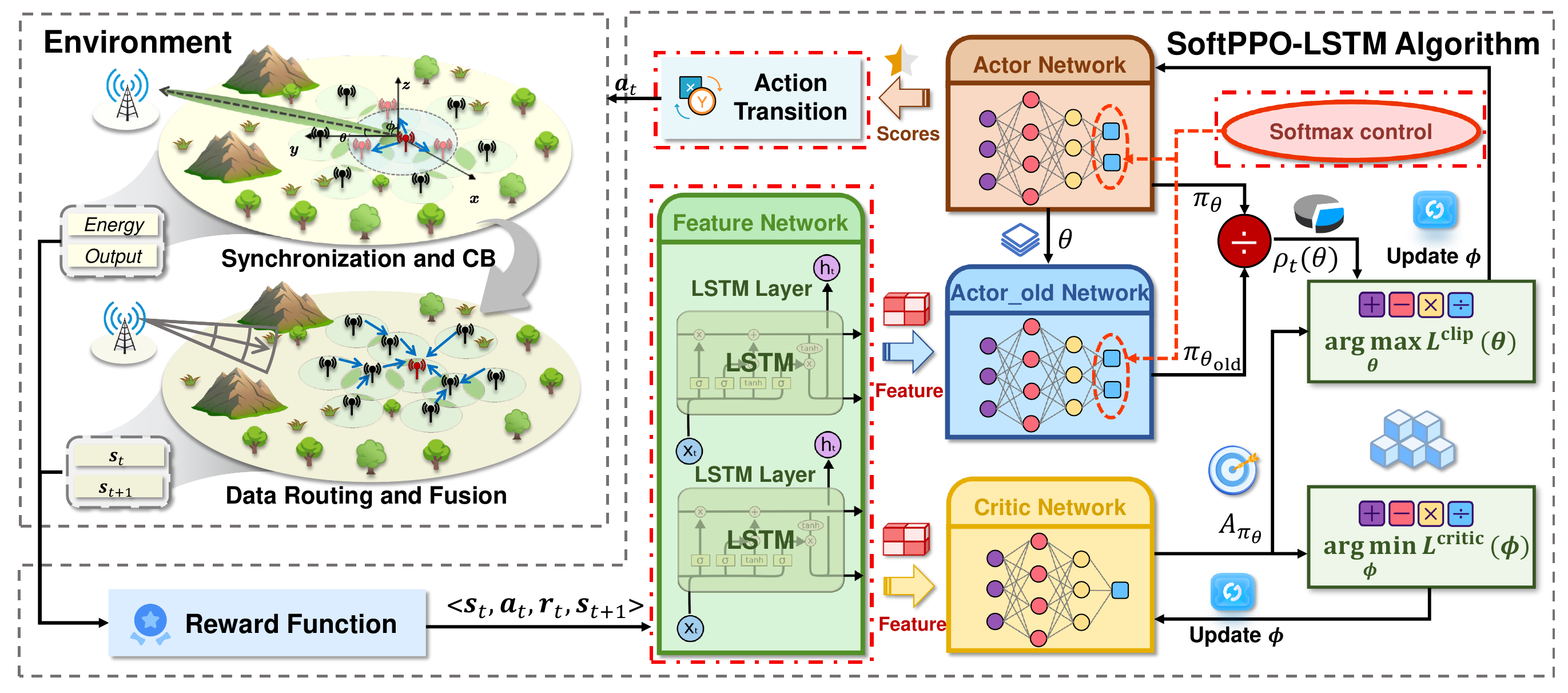}
    \caption{Framework of SoftPPO-LSTM model in the training phrase.}
    \label{network}
\end{figure*}

\color{black}

% \color{black}

\subsection {Motivations for Using DRL}
% 分析问题，CB策略需要考虑因素很多
% 分析因素: distance、energy、location
\par For the CB scheme, there are several factors, including distance, residual energy, and node location, that need to be considered. For example, the nodes far away from the sink node will cause excessive transmission energy consumption in the data synchronization process. Moreover, the nodes with low residual energy participating in CB may accelerate their energy depletion. In addition, the nodes located at the network edge are typically not near the sink node. Note that the sink nodes of different rounds may change, and the residual energy of the nodes may also decrease at different rates. The dynamic nature of the above factors requires an online method to determine the CB strategy. Similarly, since the considered optimization objective is on a long-term time scale of multiple rounds, the CB strategy needs to balance both the long-term and short-term benefits.

% 0909废弃
% \par Considering the aforementioned reasons, it is challenging to design a powerful CB strategy thereby optimizing $\mathbb{I}$. To address these challenges, we leverage a framework to offer a policy, as DRL excels at intuitively learning node strategies for different geographic contexts, thereby overcoming the limitations of traditional methods. To this end, we first transfer the problem of optimizing the decision variable $\mathbb{I}$ as an MDP in the following. 

% 

\par Considering the aforementioned reasons, DRL is a suitable method for CB because it excels at intuitively learning node strategies for different geographic contexts, thereby overcoming the limitations of traditional methods. 
% Therefore, we first formulate the DRL optimization problem in the following subsection.

\subsection {CB Node Selection Problem Formulation}

\par The DRL algorithm optimizes the network lifetime and throughput by configuring beamforming nodes during the CB process. According to Eq.~\eqref{eq:W_i}, the survival time of node $i$ can be expressed as $T_i = 1/W_i$. We sort the survival times in ascending order as $\left\langle T_{(1)}, T_{(2)}, ..., T_{(n)} \right\rangle$, and define network lifetime as the time when the number of nodes that have run out of energy exceeds the proportion $p$ of the total network nodes. The network lifetime should be maximized, which can be expressed as follows:
\begin{equation}
\label{eq:f1}
f_1(\mathbb{I}) = T_{(\lceil p(n - 1) \rceil + 1)},
\end{equation}

\color{black}
% E-网络拓扑 I-激励电流 是决策变量
\noindent where $\mathbb{I} = \{I_{t,k} | 0 < t \le T_{(\lceil p(n - 1) \rceil + 1)}, k \in \mathcal{N}_t\}$ denotes excitation current weight sequence during the network lifetime.

\par Accordingly, let $\overline{C}_{sink}$ be the average fused data packet size at the sink node, the network total throughput to the remote BS can be expressed as follows:
\begin{equation}
\label{eq:f2}
f_2(\mathbb{I}) = \overline{C}_{sink} T_{(\lceil p(n - 1) \rceil + 1)}.
\end{equation}

\par As can be seen, the two objectives are controlled by the same decision variable $\mathbb{I}$. The multi-objective-problem based on the aforementioned two objectives is formulated as follows:
\begin{subequations}
\label{eq:formulation}
\begin{align}
    \underset{\mathbb{I}}{\max} \ & F = ( f_1, f_2 ), \\
    \text{s.t.} \ & 0 \leq I_{t, k} \leq 1, t \le f_1 \ \forall k \in \mathcal{N},
    \label{eq:formulation:subeq1}\\
    & x_{ji} \in \{0, 1, 2\}, \ \forall i,j \in \mathcal{N}, 
    \label{eq:formulation:subeq2}\\
    & 0 \leq \alpha_{i,j} \leq 1, \ \forall i,j \in \mathcal{N},
    \label{eq:formulation:subeq3}\\
    & d_{ij} \leq d_{max}, \ \forall x_{ji} \in \{0, 1, 2\}, i,j \in \mathcal{N},
    \label{eq:formulation:subeq4}\\
    & 0 < p \le 1
    \label{eq:formulation:subeq5}.
\end{align}
\end{subequations}

\par Based on the aforementioned analyses, we transform
the problem of optimizing the decision variable $\mathbb{I}$ into an MDP in the following.

% \color{black}
% MDP构建
\subsection {MDP Formulation}

\par Mathematically, an MDP can be denoted as a tuple $(\mathcal{S},\mathcal{A},\mathcal{P},\mathcal{R},\gamma )$, in which $\mathcal{S}$, $\mathcal{A}$, $\mathcal{P}$, $\mathcal{R}$, and $\gamma$ denote the state set, action set, state transition probability, reward function and discount factor, respectively. Here, we detail the design of state space, action space, and reward in our model.

% 状态空间包含所有节点与Sink node的距离+剩余能量，一共4N维度，其中2N是动态的
\par \textit{1) State Space:} 
% The state space is designed to encompass critical spatial observations affecting system performance. Specifically, 
The state space includes residual energy and distance to the sink node, as these factors directly influence action decisions, further affecting the lifespan of the nodes, and the energy cost of the CB process. Moreover, the location information of the node is included in the state space, as nodes in different locations have varying degrees of location advantage. As such, the state $s_t$ is as follows:
% \begin{equation}
% \label{eq:state}
%     s_t = \{\mathbb{E}_t, \mathbb{D}_t, {\mathbb{X}_t}, \mathbb{Y}_t\},
% \end{equation}
% \begin{equation}
% \label{eq:state1}
%     \boldsymbol{s}_t = \{{e}_{i}(t), {d}_{i,s}(t) | i \in \mathcal{N}\} \cup \{x_{i}, y_{i} | i \in \mathcal{N}\}
% \end{equation}
% \noindent where $\mathbb{E}_t$ and $\mathbb{D}_t$ represent the residual energy vector and the distance vector to the sink node, respectively, which can be obtained or calculated from the upstream packets of the nodes. Moreover, $\mathbb{X}_t$ and $\mathbb{Y}_t$ represent the static two-dimensional coordinates of the node's location, respectively. Therefore, the total dimension of the state space is $4N$, of which the dimension of the dynamically updated state is $2N$.
\begin{equation}
\label{eq:state1}
    \boldsymbol{s}_t = \{{e}_{i}(t), {d}_{i,s}(t) | i \in \mathcal{N}\},
        % \boldsymbol{s}_t = \{{e}_{i}(t), {d}_{i,s}(t), x^{IoT}_{i}, y^{IoT}_{i} | i \in \mathcal{N}\},
\end{equation}

%  分析 state space 的维度，区分了动态部分和静态部分
\noindent where ${e}_{i}(t)$ denotes the residual energy of node $i$, and ${d}_{i,s}(t)$ is the distance between node $i$ and the sink node. These values can be derived or computed from the upstream packets of the nodes. 
% Additionally, $x^{IoT}_{i}$ and $y^{IoT}_{i}$ correspond to the static two-dimensional location coordinates of node $i$. Consequently, the state space has a total dimension of $4N$, with $2N$ being the dynamic component.

% 0730 需要对纯离散的分析有多少种选择
% 动作空间的本质是子集选择，动作空间的含义是对每个节点进行打分
\par \textit{2) Action Space:} After obtaining the state, the sink agent selects beamforming nodes. To ensure the received signal power at the remote BS is above the threshold for stable communication, the total power of beamforming nodes must be sufficient. Assuming all nodes participating in CB use rated power and perform perfect beamforming, only the selection of IoT nodes needs to be optimized. With the discrete action space, each node can be selected or unselected, resulting in an enormous action space size of $2^N$. In this case, DRL methods face challenges in solving the problem, particularly when the number of IoT nodes $N$ is large. Therefore, we convert the subset selection action into an $N$-dimensional continuous scoring vector for each node, substantially simplifying the problem-solving process. As such, the action of the sink agent $a_t$ is as follows:
\begin{equation}
\label{eq:action}
    \boldsymbol{a}_t = \{ v_{i}(t) | i \in \mathcal{N} \},
\end{equation}
\noindent where $0 \le v_{i}(t) \le 1$ is the score of node $i$ for acting as a beamforming node. In this case, the subset selection solution is generated from $\boldsymbol{a}_t$ by sampling from the $N$-dimensional score vector, and the sampling satisfies the constraint \eqref{eq:formulation:subeq1}.

% $a_t = \{{\mathbb{V}}_t^{\mathcal{N}}\}$.

% 奖励函数包含吞吐量和能耗
\par \textit{3) Reward Function:} To extend the network lifetime and increase total throughput, the reward function considers the network energy consumption and the throughput delivered to the remote BS in each round, which is given as follows:
\begin{equation}
\label{eq_reward}
    \boldsymbol{r}_{t}=\zeta_{1}{C}_{t} - \zeta_{2}\sum_{i = 1}^{N}({e}_{i}(t) - {e}_{i}(t + 1)),
\end{equation}

\noindent where ${C}_{t}$ is the throughput of the $t$th round, and ${e}_{i}(t)$ denotes the residual energy of node $i$ in the $t$th round. The parameters $\zeta_{1}$ and $\zeta_{2}$ are the weights assigned to the throughput and the energy cost, respectively.

% 【0730 这里能谈的限制条件只有(17b),因此我直接把(17b)放在Action Space末尾了】
% 描述，解释, MDP也考虑了路由协议，引出SoftPPO-LSTM
\par As can be seen, the MDP captures the essential aspects of the considered system by incorporating a comprehensive state space, a practical action space, and a carefully designed reward function. Note that the OMRP will influence the new state of the next round. This enables the DRL agent, which operates at the sink node, to adapt to the OMRP during training and improve the selection of decision variables $\mathbb{I}$. In the following, we will introduce the \textit{SoftPPO-LSTM Algorithm}, which is tailored to leverage this MDP framework for enhanced network lifetime and throughput.

% \sethlcolor{pink}\hl{[TODO Divide into system model..]}

\subsection{SoftPPO-LSTM Algorithm}

% \sethlcolor{pink}\hl{[0715 Mark]}
\par In this subsection, we adopt PPO as the solving framework and present an enhanced version of the PPO algorithm, namely, SoftPPO-LSTM. SoftPPO-LSTM integrates two key features which are softmax control and LSTM layers to simplify the solution space and facilitate continuous and stable learning over long rounds.
% 简化成了一句话
% 
% 简单介绍PPO框架
\par \textit{1) PPO Algorithm:} PPO is a state-of-the-art reinforcement learning algorithm based on policy. The objective of PPO is to improve the policy parameters for achieving high state values, which is expressed as follows:
\begin{equation}
\label{j_theta}
    \operatorname* {m a x}_{\theta} ~ J ( \theta)=\mathbb{E}_{\pi_{\theta}} [ \sum_{t=1}^{T} \gamma^{t} r_{t} ].
\end{equation}

% \color{black}
% 简单介绍PPO特色：clip函数、超参数以及总结
\par Moreover, PPO enforces a clip on the ratio of the new policy probability to the old policy probability, thereby ensuring that policy updates are within a safe and bounded range. To this end, PPO improves the policy using a surrogate objective that constrains policy updates. Let $A_{\pi_{\theta}}$ be the advantage function, then the constraint is expressed as follows:
\begin{equation}
\label{Lclip}
L^{c l i p} ( \theta)=\mathbb{E}_{t} [ \operatorname* {m i n} ( \rho_{t} ( \theta) A_{\pi_{\theta}} ( s_{t}, a_{t} ), \rho_{t}^{c l i p} ( \theta) A_{\pi_{\theta}} ( s_{t}, a_{t} ) ) ], 
\end{equation}

\noindent where $\theta$ and $\rho_{t}(\theta)$ are the policy parameters and the ratio of new and old policy probabilities, respectively. Moreover, $\rho_{t}^{clip}(\theta) = clip(\rho_{t}(\theta), 1-\epsilon, 1+\epsilon)$ is a clipped value of $\rho_{t}(\theta)$, where $\epsilon$ is a hyperparameter that controls the size of the policy update. As such, PPO combines this surrogate objective with multiple epochs of data to iteratively update the policy while avoiding large policy deviations, resulting in stable learning. 

% 原版PPO不行，较长的episode导致信任问题，env中的启发式路由策略也具备较强随机性，对长时间序列的策略学习造成了一定的挑战。
\par However, in solving the considered MDP, one episode contains hundreds of timesteps, which lead to credit assignment issues~\cite{chen2023balancing}. At the same time, the packet routing in the environment is heuristic, and the long-term adaptability of the agent to the routing strategy is also a challenge. Furthermore, our system contains hundreds of IoT nodes, leading to a vast state and action space within the MDP framework. In particular, it is challenging for the PPO agent to produce continuous action space variables with hundreds of dimensions during hundreds of timesteps. In this case, we improve the PPO algorithm to make it suitable for our MDP in the following.

% 
% 0512 两个改进点拆分叙述了
% softmax: 1、有助于处理大规模离散动作空间问题 2、平滑处理梯度稳定训练过程

\par \textit{2) Softmax Control-based Enhancement:} 
% Due to the complexity of our problem, traditional PPO algorithms face challenges in handling our large-scale transformed action spaces, leading to inefficient exploration and unstable policy updates. 
We introduce softmax control to enhance learning stability for our large-scale discrete action spaces. On the one hand, softmax operations excel in handling classification selections and reducing the volatility of neuron signals, making them well-suited for our optimization problem. The softmax function is defined as follows:
\begin{equation}
\label{eq:softmax}
\mathrm{Softmax}(x) = {\frac{e^{x_{i}}} {\sum_{i=1}^{N} e^{x_{i}}}}.
\end{equation}

\noindent where the input vector $x$ is converted to a probability distribution. In this case, softmax facilitates a nuanced understanding of the relative importance of the original action and guides the training process toward a more systematic exploration of these actions. On the other hand, it ensures stable policy updates by smoothing backpropagation gradients, which is well-suited for our optimization problem with sharp gradients due to the discrete action spaces.

%by smoothing the gradients during the backpropagation phase, softmax prevents drastic changes in policy updates that can lead to volatile training dynamics. This smoothing effect ensures a more gradual and stable convergence of the model and adapts to situations with sharp gradients.
% Consequently, the use of softmax contributes to a more robust and dependable learning trajectory, enhancing the overall performance and reliability of the DRL model.

% 【0801 这段写的有点长，不确定是否要分段】
% LSTM：1、有助于处理长序列决策 2、适应部分可观察的环境，推断隐藏状态，帮助适应routing policy

\par \textit{3) LSTM-based Network Structure:} We introduce LSTM layers into the feature network of our SoftPPO-LSTM algorithm to enhance the temporal processing and inference capabilities. On the one hand, LSTMs excel in guiding training over long episodes with time-series data, making them well-suited for our environments that unfold over extended timesteps. \color{black} Specifically, the LSTM unit consists of a cell, an input gate, an output gate, and a forget gate. The cell remembers values over arbitrary time intervals, while the gates regulate the flow of information into and out of the cell. This structure enables the LSTM to manage temporal dependencies by selectively filtering information at each time step. The operations of an LSTM cell at time step $t$ can be described as follows:
\begin{align}
&f_t = \sigma(W_f \cdot [h_{t-1}, x_t] + b_f), 
\label{eq:lstm:forget}\\
&i_t = \sigma(W_i \cdot [h_{t-1}, x_t] + b_i), 
\label{eq:lstm:input}\\
&\tilde{C}_t = \tanh(W_C \cdot [h_{t-1}, x_t] + b_C), 
\label{eq:lstm:candidate}\\
&C_t = f_t * C_{t-1} + i_t * \tilde{C}_t, 
\label{eq:lstm:cell}\\
&o_t = \sigma(W_o \cdot [h_{t-1}, x_t] + b_o), 
\label{eq:lstm:output}\\
&h_t = o_t * \tanh(C_t) 
\label{eq:lstm:hidden}
\end{align}

\noindent where $\sigma$ denotes the sigmoid function, and $*$ denotes an element-wise multiplication. The variables $f_t$, $i_t$, and $o_t$ are the forget, input, and output gates, respectively. The variables  $\tilde{C}_t$, $C_t$, and $h_t$ are the candidate cell state, the cell state, and the hidden state, respectively. On the other hand, LSTMs are well-suited for environments with partial observability because they can infer hidden states from sequences of observations. In our scenario, the environment first executes a heuristic-based routing policy to determine the current state, which is then provided to the DRL agent as an observation. In this case, the ability of LSTM to infer hidden states from sequences of partial observations fits well with this heuristic routing method, as it enables the DRL model to capture underlying patterns and dynamics and helps the agent better predict future states and make informed decisions. Consequently, this approach not only strengthens the strategic depth of the agent but also improves adaptability and overall performance in dynamically changing and partially observable environments.

% The softmax operation plays a pivotal role in facilitating the continuous representation of large-scale discrete action spaces. By transforming the actor network's output into a probability distribution, softmax facilitates a clear comprehension of action priorities, thereby augmenting the decision-making process's interpretability and enhancing the model's resilience to variations in input through decision interface smoothing. 
% Concurrently, LSTM networks provide a strategic advantage by adeptly managing sequential decision-making challenges and  long-term dependencies. Particularly in environments characterized by partial observability, LSTMs enhance the agent's capacity to infer hidden states and adapt strategies accordingly.This capability enhances the model's adaptability and performance in complex decision-making landscapes.

% 【0731】这里似乎不太好分析时间复杂度
% 主要步骤，对着模型图讲
\par\textit{4) Main Steps of SoftPPO-LSTM algorithm for the CB Process:} As shown in Fig.~\ref{network}, SoftPPO-LSTM employs two neural networks, \textit{i.e.}, an actor neural network which is responsible for policy learning and a critic neural network that estimates the state value function and the advantage function. To ensure stable policy updates, SoftPPO-LSTM maintains an actor old network with identical parameters to calculate the clipping ratio. The current state $s_t$ is fed into the feature network to produce features, which are then input into the actor network and critic network to generate the action $a_t$ and the critic value, respectively. Following the execution of this action, the environment generates the corresponding reward $r_t$ and the next state $s_{t + 1}$. This process forms a fundamental transition within the framework of reinforcement learning.

\par The main steps of the proposed CB policy based on the SoftPPO-LSTM algorithm are shown in Algorithm~\ref{alg:cbpolicy}. Specifically, the algorithm begins by initializing the neural network parameters. Then, SoftPPO-LSTM proceeds to execute episodes of interactions with the environment. Within each episode, the environment and state are reset. For each time slot, the algorithm generates actions, interacts with the environment, and receives the reward and new state observations. Moreover, the algorithm periodically updates the actor and critic neural network parameters. This process repeats until all specified episodes are completed, and the algorithm continuously refines its policy for reinforcement learning tasks. 

% 
 
%% TODO 1118
\par\textit{4) Complexity Analysis of SoftPPO-LSTM:} The computational and space complexity of SoftPPO-LSTM during training and execution phases are analyzed. 

\par The computational complexity of SoftPPO-LSTM during the training phase is $\mathcal{O}((1 + 2MT)(|\boldsymbol{\theta_{Q_{new}}}| + |\boldsymbol{\theta_c}| + |\boldsymbol{\theta_f}|) + (1 + 2MT/d)|\boldsymbol{\theta_{Q_{old}}}| + NMT(V+2))$, which can be summarized as follows:

\begin{itemize}

\item \textit{Network Initialization:} This phase involves the initialization of network parameters. Specifically, the computational complexity is expressed as $\mathcal{O}(|\boldsymbol{\theta_{Q_{new}}}| + |\boldsymbol{\theta_c}| + |\boldsymbol{\theta_f}| + |\boldsymbol{\theta_{Q_{old}}}|)$, where the $| \cdot | $ operation represents the number of parameters in the networks.

\item \textit{Action Transition:} This phase entails transforming actions according to the output scores of the actor network, and its complexity is $\mathcal{O}(NMT)$. Here, $M$ denotes the number of training episodes, $T$ is the number of steps per episode, and $N$ is the number of IoT nodes. 

\item \textit{Reward Calculation and State Transitions:} The computational complexity of reward calculation and state transitions is $\mathcal{O}(NMT(V+1))$, where $V$ represents the complexity of interacting with the environment.

\item \textit{Network Update:} The updating phase is divided into three main parts that are frequent updates of the feature network and critic network, as well as less frequent updates of the actor network. Thus, the complexity of this phase is calculated as $\mathcal{O}(MT(2|\boldsymbol{\theta_{Q_{new}}}|) + MT(2|\boldsymbol{\theta_c}|) + MT(2|\boldsymbol{\theta_f}|) + MT/d(2|\boldsymbol{\theta_{Q_{old}}}|))$, where $d$ is the updating interval of the old actor network.

\end{itemize}

\par Besides, the space complexity of SoftPPO-LSTM during the training phase is $\mathcal{O}(|\boldsymbol{\theta_{Q_{new}}}| + |\boldsymbol{\theta_{Q_{old}}}| + |\boldsymbol{\theta_f}| + |\boldsymbol{\theta_c}| + 2|\boldsymbol{s}| + |\boldsymbol{a}|))$, where $|\boldsymbol{s}|$ and $|\boldsymbol{a}|$ denote the dimensions of the state and action spaces, respectively.

\par During the execution phase, the computational complexity of SoftPPO-LSTM is $\mathcal{O}(MT(|\boldsymbol{\theta_{Q_{old}}}| + |\boldsymbol{\theta_f}|) + NMT)$, which can be contributed by action selection and transition according to the current state using the feature and actor network. Moreover, the space complexity during the execution phase is $\mathcal{O}(|\boldsymbol{\theta_{Q_{old}}}| + |\boldsymbol{\theta_{f}}| + |\boldsymbol{s}|)$ since the feature and actor network parameters need to be stored in memory for action selection.

% \color{black}

% To begin with, the computational complexity of network initialization is $\mathcal{O}(|\boldsymbol{\theta_{Q_{new}}}| + |\boldsymbol{\theta_c}| + |\boldsymbol{\theta_f}| + |\boldsymbol{\theta_{Q_{old}}}|)$, where $| \cdot | $ operation represents the number of parameters in the networks. Besides, the computational complexity of action transformation is $\mathcal{O}(NMT)$, where $M$ denotes the number of training episodes, $T$ is the number of steps per episode, and $N$ is the number of IoT nodes. Moreover, the computational complexity of reward and calculation and state transitions is $\mathcal{O}(NMT(V+1))$, where $V$ represents the complexity of interacting with the environment. 

% Finally, the computational complexity of network updating is $\mathcal{O}(MT|\boldsymbol{\theta_{Q_{new}}}| + MT|\boldsymbol{\theta_c}| + MT|\boldsymbol{\theta_f}| + MT/d|\boldsymbol{\theta_{Q_{old}}}|)$, where $d$ is the updating interval of the old actor network. Therefore, the computational complexity of SoftPPO-LSTM in the training phrase is $\mathcal{O}((MT+1)(|\boldsymbol{\theta_{Q_{new}}}| + |\boldsymbol{\theta_c}| + |\boldsymbol{\theta_f}|) + (1 + MT/d)|\boldsymbol{\theta_{Q_{old}}}| + NMT(V+2))$.

\begin{algorithm}[!t]
\caption{CB Policy based on SoftPPO-LSTM}
% \begin{algorithmic}[1]
\label{alg:cbpolicy}
% \KwIn{Policy network $\pi_\theta(s, a)$, Value network $v_\phi(s)$}
    % Initialize $\theta_0$, $\phi_0$ randomly; \\
    Initialize actor network $\boldsymbol{Q_{new}}$ with parameters $\boldsymbol{\theta_{Q_{new}}}$, and then copy it to $\boldsymbol{Q_{old}}$; \\ 
    Initialize the feature network denoted as $\boldsymbol{\varepsilon_{f}}$ with parameter $\boldsymbol{\theta_f}$, and the critic network denoted as $\boldsymbol{\varepsilon_{c}}$ with $\boldsymbol{\theta_c}$; \\
    $step \gets 0$; \\
    \For{training episode $ = 1$ \textbf{\text{to}} $M$}{
        % Reset the environment and execute Algorithm.\ref{alg:omrp1}, \ref{alg:omrp2} \\
        Reset the environment and execute Algorithms.~\ref{alg:omrp1} and~\ref{alg:omrp2} to get the initial state $\boldsymbol{s_0}$; \\
        % \Repeat        
        % \Until{environment is terminated}
        
        \For{$round\ t = 1$ \textbf{\text{to}} $T$}{
            % Call Algorithm.~\ref{alg:omrp1}, \ref{alg:omrp2} and obtain state $\boldsymbol{s_t}$; \\
            Select action $\boldsymbol{a_t} = \{ v_{i}(t) | i \in \mathcal{N}\}$ \\
            Transform $\boldsymbol{a_t}$ into a node selection set by softmax and action sampling;\\
            Execute CB strategy in the environment and generate the reward $\boldsymbol{r_t}$ by using Eq.~\eqref{eq_reward}; \\
            Execute routing Algorithm.~\ref{alg:omrp1} and~\ref{alg:omrp2} to generate the the next state $\boldsymbol{s_{t+1}}$; \\
            % Update state $s_t = s_{t+1}$ \\ 
            Update the critic network parameters $\boldsymbol{\theta_c}$; \\
            Update the feature network parameters $\boldsymbol{\theta_f}$; \\
            Update the new actor network parameters $\boldsymbol{\theta_{Q_{new}}}$; \\
            \If {$step$ \text{mod} $d$ }{
                Update the old actor network parameters $\boldsymbol{\theta_{Q_{old}}}$ according to Eq.~\eqref{Lclip};\\
                % Update policy parameters $\theta$ \\
                % $step \gets 0$
            }
            \textbf{end} \\
            $step \gets step + 1$
        }
        \textbf{end} \\
    }
    \textbf{end} \\
    % \KwOut{Policy $\pi_{\theta}$}
% \end{algorithmic}
\end{algorithm}

\subsection{Practical Implementation of SoftPPO-LSTM Algorithm}
% \sethlcolor{pink}\hl{[R1-1, R2-1, R3-3]}

\par In this subsection, we provide an edge computing paradigm for applying our method in practical scenarios, particularly regarding the distinction between the training and deployment phases of our algorithm.

\par \textit{1) Training Phrase:} Similar to many advanced machine learning techniques, the SoftPPO-LSTM model requires extensive training to reach convergence before being deployed in practical settings. Therefore, we recommend edge computing paradigms, where training occurs on a central controller with sufficient computing power. In this case, the proposed method interacts with a simulation environment constructed by using pre-collected real-world data such as location and energy.

\par
\begin{figure*}[ht]
    \centering
    \includegraphics[width=0.98\textwidth]{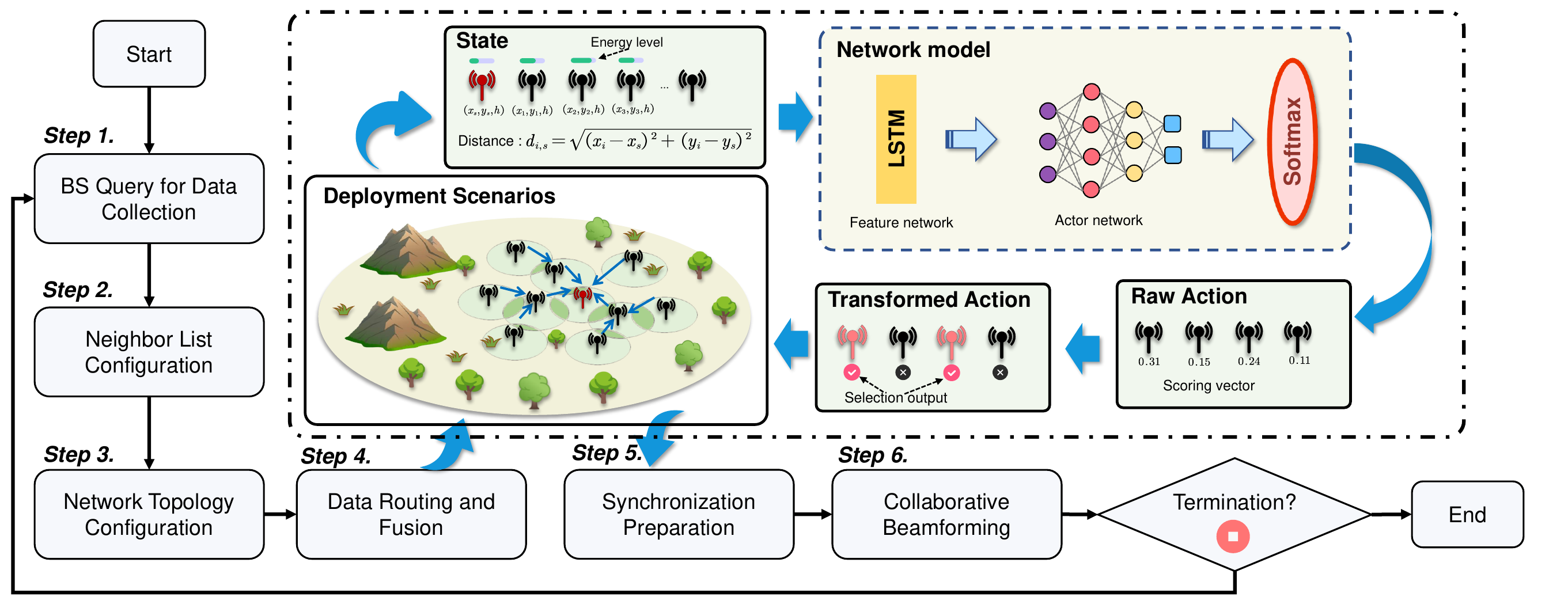}
    \caption{The flow chart of our system in the deployment phase. The OMRP method is performed during Steps 3 and 4. The SoftPPO-LSTM method is performed at the start of Step 5, and the generated action is then executed during Steps 5 and 6.}
    \label{fig:flow_chartdeployment}
\end{figure*}

 \textit{2) Deployment Phrase:} During the deployment phase, the well-trained model receives the state information from the real environment to make decisions. Fig.~\ref{fig:flow_chartdeployment} shows the flow chart of our system in the deployment phase. In this case, the output decisions guide the synchronization
preparation and CB in Steps 5 and 6. Note that these DRL models are executed by the IoT nodes and the deployment experiment is provided in Section~\ref{sec:discussion}.

\par \textit{3) Environment Changes:} In real-world deployment scenarios, environmental changes such as electromagnetic interference and hardware failures can potentially cause beamforming nodes to fail to perform CB accurately~\cite{he2024deep, shinohara2013beam}. As such, the deployed model may need to be frequently updated to address these events. In this case, the BS can act as an intermediate, forwarding real-time network data to a centralized cloud server where the SoftPPO-LSTM model can be periodically re-trained with the latest information. Then, the updated model can be efficiently distributed to the IoT nodes through the network.

\section{Simulation and Analysis}
\label{sec:simulation_analysis}
% 总起段：simulation setting and benchmarks
% TODO: 数据融合 based on r
% \sethlcolor{pink}\hl{[TODO: data fusion based on r, map realization]}
\par In this section, we present the simulation results and analyses for our IoT communication system, focusing on both routing methods and CB methods. We first introduce the simulation setting and benchmarks, and then provide the simulation results and analyses of the OMRP-based routing method and SoftPPO-LSTM-based CB method.

\subsection{Simulation Setups}

% \par This section provides an extensive description of the simulation setup, including the simulation platform, environmental details, simulation details, and benchmarks utilized to evaluate the performance of the proposed method.

\par \textit{1) Simulation Platform:} Our experiments are conducted using a computing setup that includes an NVIDIA GeForce RTX 4090 GPU with 24 GB of memory and a 13th Gen Intel(R) Core(TM) i9-13900K 32-core processor with 128 GB of RAM. The operating system on the workstation is Ubuntu 22.04.3 LTS. For our deep learning computations, we use PyTorch 2.0.1, along with CUDA 11.8.

\par \textit{2) Environmental Details:} This study considers an IoT network consisting of 400 homogeneous IoT nodes, each equipped with a transmit power of 0.1 W. The IoT nodes are randomly deployed in a 200 m $\times$ 200 m square region, with the remote BS located 1000 m outside the region at coordinates (100 m, 1200 m). Each IoT node operates at a bandwidth of 100 kHz, and the background noise level is set at -174 dBm/Hz. 
To ensure a balance between energy consumption for routing and CB processes while maintaining the quality of uplink communication, the minimum received power at the remote BS is set to $-52$ dBm~\cite{heinzelman2000application}.
% \color{black}

\par In our scenario, since considerations of radiation interference and sidelobe levels are not necessary for the scenario, the excitation current weight for the nodes participating in CB is set to the maximum value of 1. 
Based on the above configurations, Eq.~\eqref{beam_two} shows that the involvement of 10 nodes in CB are sufficient to maintain normal communication quality, achieving a signal-to-noise ratio of approximately 24~dB. 
% \color{black}
In addition, we consider the energy dissipation of the CB process as zero and the initial energy of each node $E_0$ as 4.0~J for the comparative analysis of the routing policies. Furthermore, the routing policy of the environment in the simulation of CB policies is configured to the proposed OMRP. Since synchronization and transmission energy dissipation in the CB process are considered, we set the initial energy of each node $E_0$ to 6.0~J. Table~\ref{table:parameter} provides other details about the radio model and the network.

\begin{table}[!t]
\centering
\caption{Simulation Parameters}
\begin{tabular}{l l}
\toprule
\textbf{Parameter} & \textbf{Value} \\ \hline
Network size       & 200 m × 200 m\\ 
Position of the BS        & (100 m, 1200 m)\\ 
Number of nodes: $N$       & 400\\ 
The monitor radius of IoT nodes & 6.0 m \\
Initial energy of each node: $E_0$      & 4.0 J / 6.0 J\\ 
Altitudes of the IoT nodes: $h_t$  & 1.5 m \\ 
Altitudes of the BS: $h_r$  & 20 m \\ 
Energy dissipation-electronic circuit: $E_{elec}$        & 50 nJ/bit        \\ 
Energy dissipation-free space amplifier: $\varepsilon_{fs}$        & $10\ \text{pJ/bit/m}^2$   \\ 
Energy dissipation-multipath amplifier: $\varepsilon_{mp}$        & $0.0013\ \text{pJ/bit/m}^4$ \\ 
Energy dissipation-data fusion: $q_0$       & 20 nJ/bit        \\ 
Length of data packet           & 10000 bits        \\ 
Length of control packet           & 200 bits        \\ 
% Data spatial correlation coefficient : $R_c$ & 6.0 m    \\ 
\bottomrule
\end{tabular}
\label{table:parameter}
\end{table}

\subsection{Simulation and Analysis of OMRP}

\begin{figure*}[!t]
    \centering
    % 第一行的两幅图
    \begin{minipage}[b]{0.24\textwidth}
        \centering
        \includegraphics[width=\linewidth]{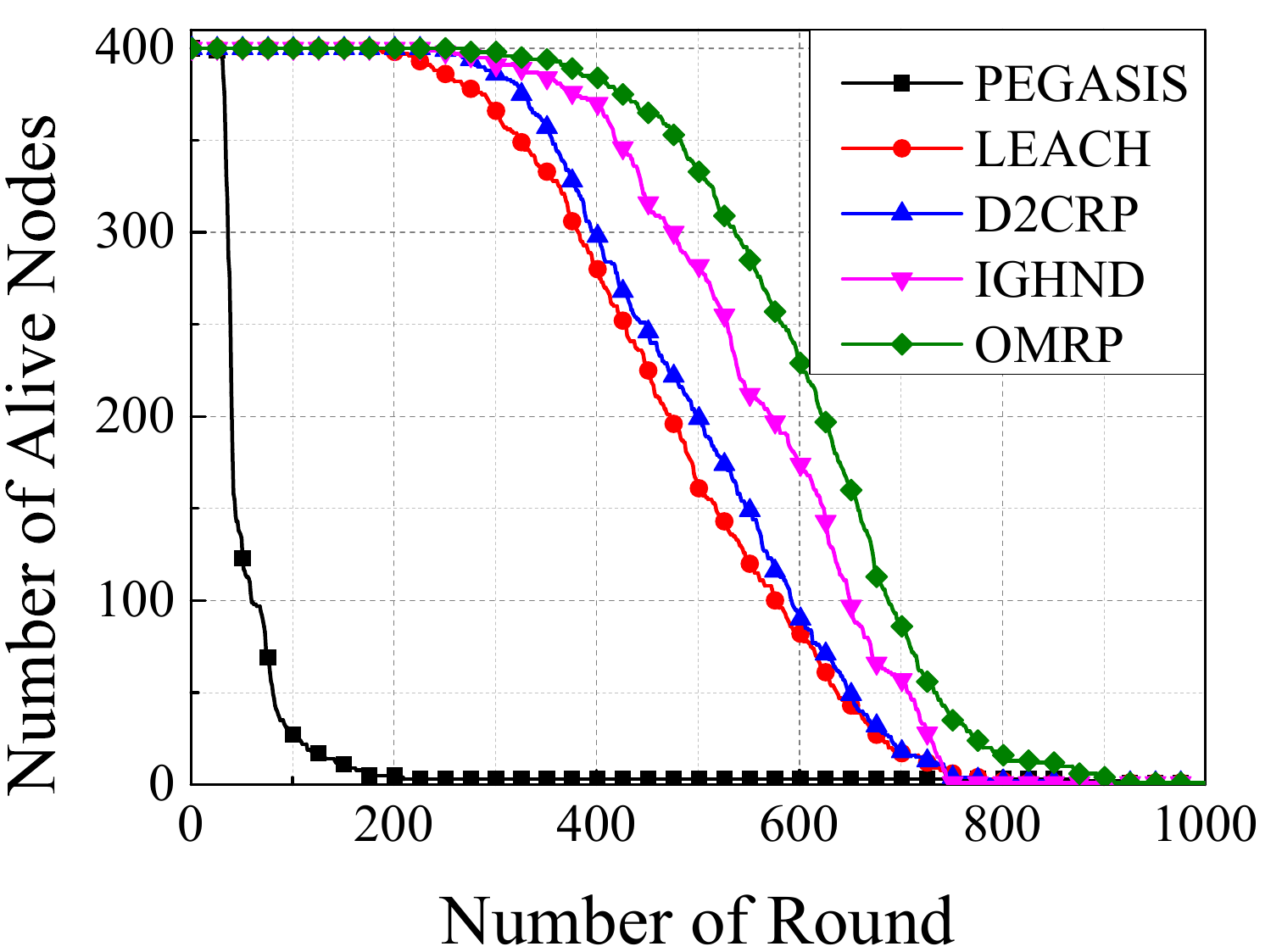}
        \subfloat(a)
        \label{rout.sub.live_node_round}
    \end{minipage}%
    \hfill
    \begin{minipage}[b]{0.24\textwidth}
        \centering
        \includegraphics[width=\linewidth]{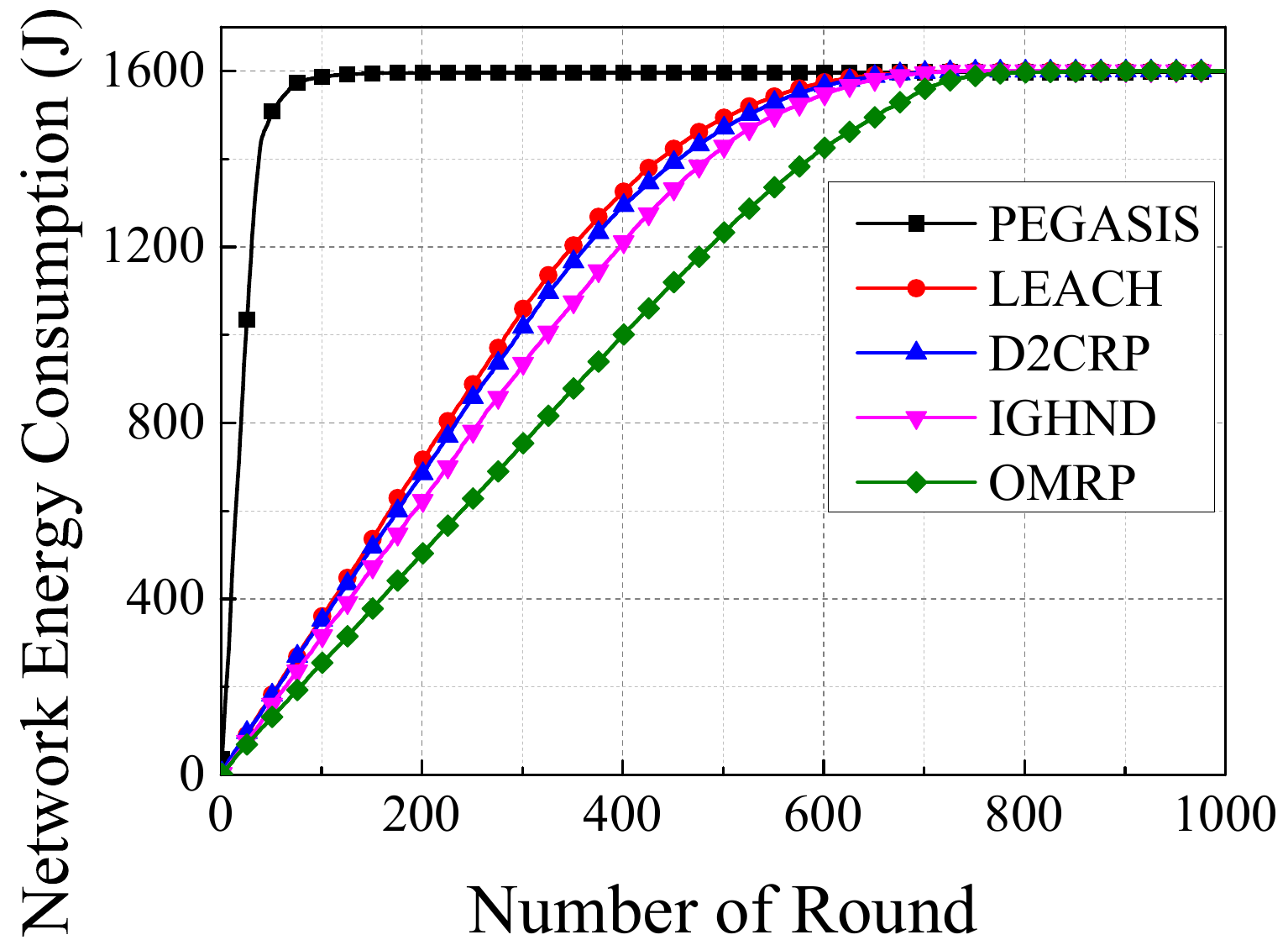}
        \label{rout.sub.energy_consumption_round}
        \subfloat(b)
    \end{minipage}
    \hfill
    \begin{minipage}[b]{0.24\textwidth}
        \centering
        \includegraphics[width=\linewidth]{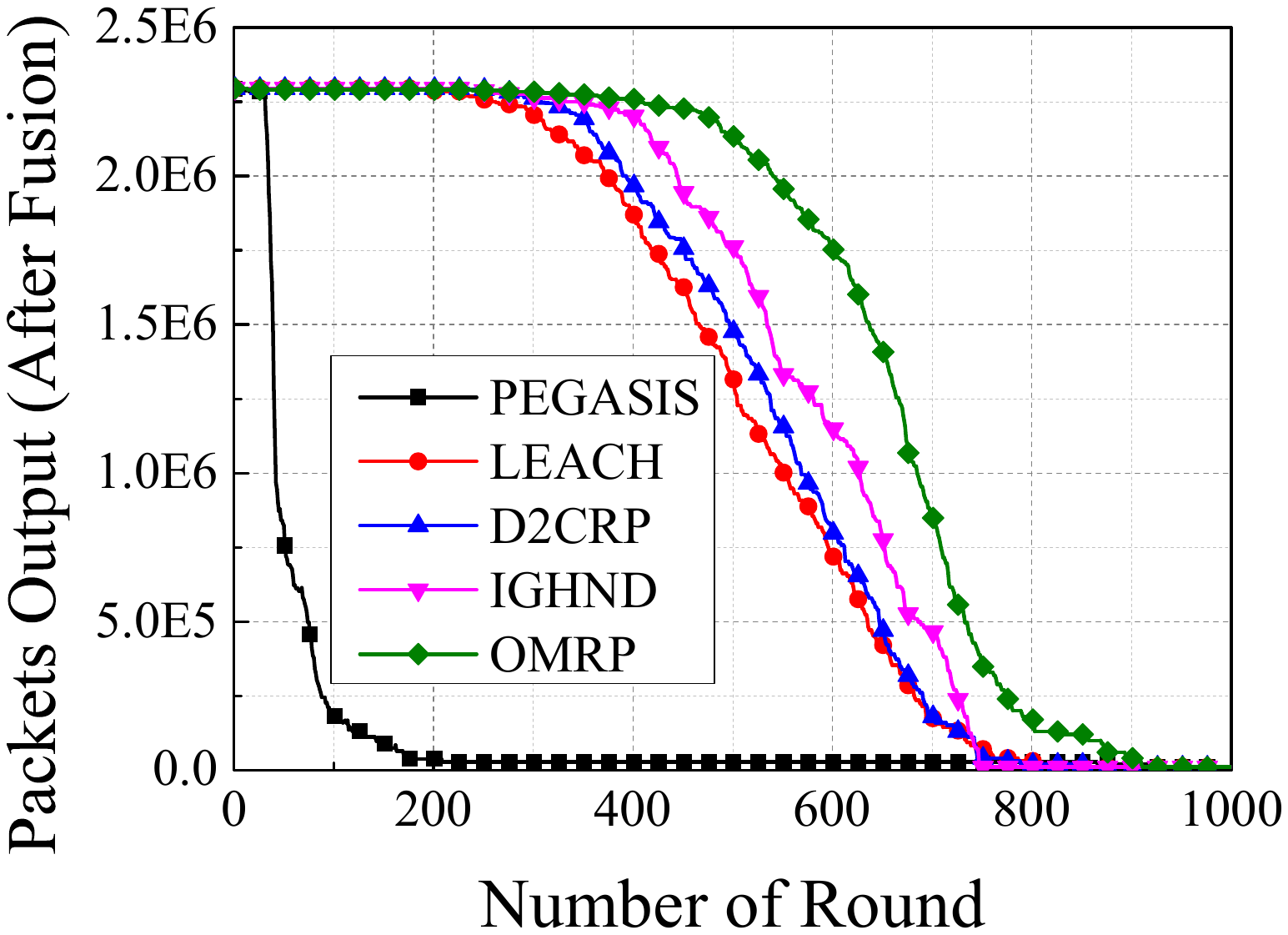}
        \label{rout.sub.packet_size_af_round}
        \subfloat(c)
    \end{minipage}%
    \hfill
    \begin{minipage}[b]{0.24\textwidth}
        \centering
        \includegraphics[width=\linewidth]{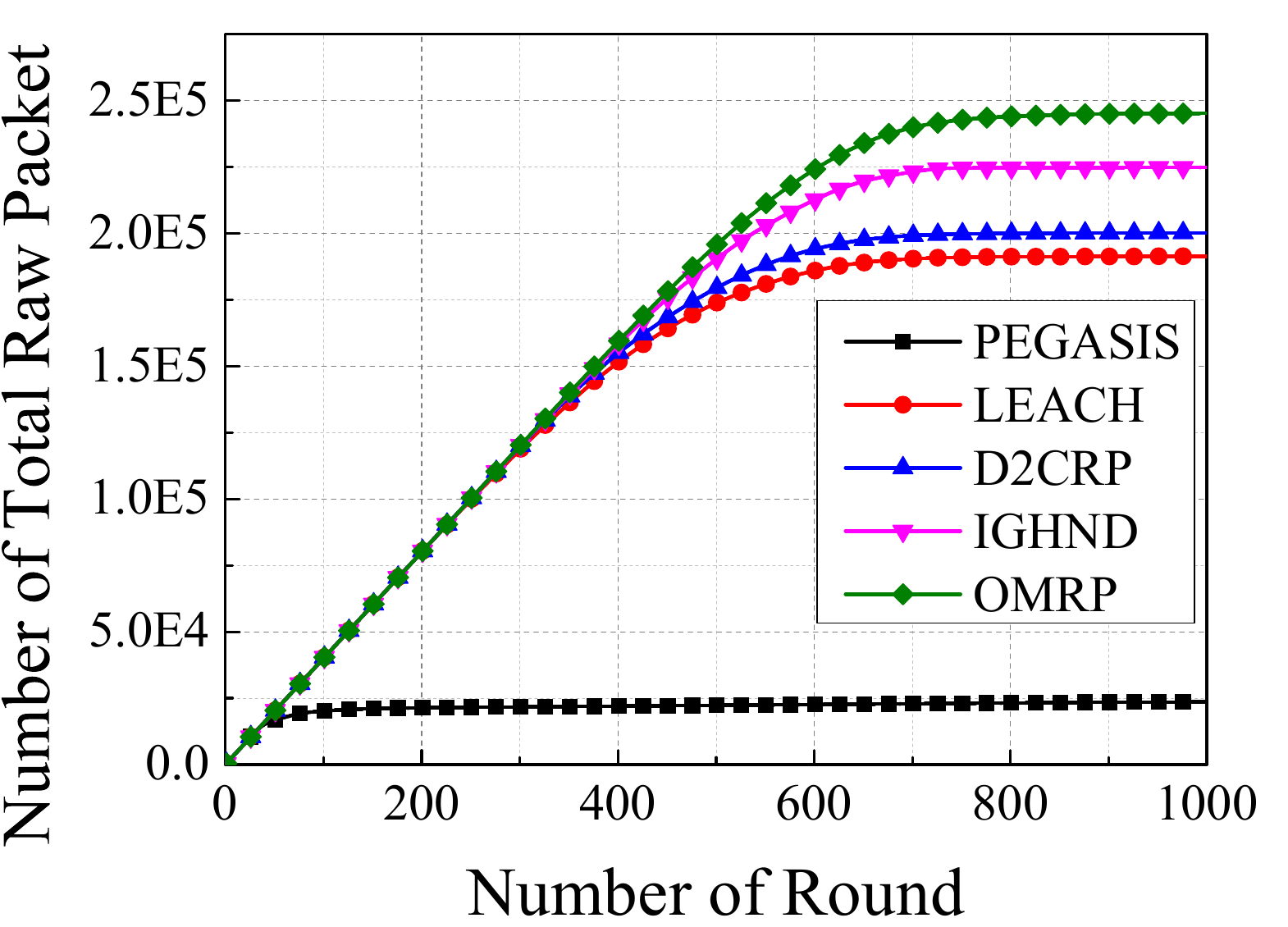}
        \label{rout.sub.raw_packet_size_round}
        \subfloat(d)
    \end{minipage}%
    \caption{Performance of different routing protocols. (a) Comparison of network lifetime. (b) Comparison of network energy consumption. (c) Comparison of packet size output to BS. (d) Comparison of the number of raw packet to BS.}
    \label{perform_routing}
\end{figure*}

% \begin{figure*}[!t]
%     \centering
%     % 第一行的两幅图
%     \begin{minipage}[b]{0.49\textwidth}
%         \centering
%         \includegraphics[width=0.8 \linewidth]{images/live_node_round_0827.pdf}
%         \caption{Comparison of network lifetime.}
%         \label{rout.sub.live_node_round}
%     \end{minipage}%
%     \hfill
%     \begin{minipage}[b]{0.49\textwidth}
%         \centering
%         \includegraphics[width=0.8\linewidth]{images/energy_consumption_round.pdf}
%         \caption{Comparison of network energy consumption.}
%         \label{rout.sub.energy_consumption_round}
%     \end{minipage}
%     % 第二行的两幅图
% \end{figure*}

% 路由仿真 整体介绍
\par To demonstrate the superiority of our proposed OMRP, we compare OMRP with four well-known hierarchical routing protocols: PEGASIS~\cite{lindsey2002pegasis}, LEACH~\cite{heinzelman2000application}, D2CRP~\cite{chen2022d2crp} and IGHND~\cite{farman2018multi}. The experimental results are then analyzed with a focus on network lifetime, packet throughput, and the number of raw packets to the BS.

% 介绍网络生存周期，FND, HND, AND 三个指标
\subsubsection{Comparisons of Network Lifetime} To demonstrate that the OMRP protocol can effectively slow down the death rate of nodes and extend the network lifetime, we show the relationship between the number of alive nodes and the number of running rounds of the five protocols in Fig.~\ref{perform_routing}(a). Additionally, the simulation results of the round of first node death (FND), half node death (HND), and all node death (AND) are presented in Table~\ref{table:nodedeath}.

% \sethlcolor{pink}\hl{[TODO: ADD data]}

% 分析Live node曲线以及 XND 表格数据
\par We can see from Fig.~\ref{perform_routing}(a) that PEGASIS is the first protocol to experience node deaths, followed by LEACH, D2CRP, IGHND, and the proposed OMRP. As the rounds increase, the PEGASIS curve shows a steep decline, which can be attributed to the formation of long chains in PEGASIS, leading to higher energy dissipation during data fusion. As is shown in Table~\ref{table:nodedeath}, the overall network lifetime (measured by the round of AND) under OMRP is approximately 19\% longer than LEACH and 17\% longer than both D2CRP and IGHND. It is clear that OMRP achieves the best performance in these three metrics due to its lower node death rate and longer network lifetime.

% 分析为什么OMRP表现最好 - overlapping degree 减少集群成员所需的传输能量
\par We attribute the slower node death rate of OMRP is mainly due to the introduction of the overlapping degree in the CH competition process. Nodes exhibiting higher overlapping degrees tend to be positioned closer to geographical center of the cluster, making their election as CHs conducive to reducing the transmission energy required by cluster members. By linking the likelihood of the election of a node as a CH to its overlapping degree, energy dissipation in the initial phases of the network is made more efficient.

% Noting that, compared with the other two protocols, OMRP greatly extends the emergence time of FND and HND, which is of great significance for maintaining network stability.

\begin{table}[!t]
\centering
\caption{Comparison of FND, HND, and AND}
\begin{tabularx}{\linewidth}{l X X X}
\toprule
\textbf{Protocol} & \textbf{FND} & \textbf{HND} & \textbf{AND} \\ \hline
PEGASIS       & 21 & 41 & 155 \\ 
LEACH       & 187 & 473 & 733 \\ 
D2CRP       & 236 & 499 & 743 \\ 
IGHND       & 229 & 573 & 742 \\ 
OMRP         & 271 & 624 & 870 \\ 
\bottomrule
\end{tabularx}
\label{table:nodedeath}
\end{table}

\begin{table}[ht]
\centering
\caption{Comparisons of FND-HND, HND-AND, and 1-AND average network energy consumption rate.}
\begin{tabularx}{\linewidth}{l X X X}
\toprule
\textbf{Protocol} & \textbf{FND-HND} & \textbf{HND-AND} & \textbf{1-AND} \\ \hline
PEGASIS       & 30.501 J/round & 1.263 J/round & 5.634 J/round \\ 
LEACH       & 2.762 J/round & 0.547 J/round & 1.707 J/round \\ 
D2CRP       & 2.509 J/round & 0.539 J/round & 1.561 J/round \\ 
IGHND       & 2.346 J/round & 0.463 J/round & 1.728 J/round \\ 
OMRP         & 2.215 J/round & 0.571 J/round & 1.540 J/round \\ 
\bottomrule
\end{tabularx}
\label{table:energy_consumption}
\end{table}

% 介绍网络能耗情况，J/round
\subsubsection{Comparisons of Network Energy Consumption}

\par The network energy consumption is used to measure the performance of the routing protocols. With no external energy supply, the protocol performs better if less energy is consumed in the network after the same rounds. The network energy consumption comparison of the five protocols is shown in Fig.~\ref{perform_routing}(b) and Table~\ref{table:energy_consumption}. Specifically, the energy consumption rate of PEGASIS is very high during the period from the first node death to half of the dead node (FND-HND), which is almost six times the average energy consumption of the overall lifetime of the network (1-AND). In contrast, IGHND and OMRP display much smoother energy consumption rates. In particular, during the FND-HND phase, the energy consumption rate of OMRP is approximately 6\%, 12\%, 20\%, and 92\% lower than that of the IGHND, D2CRP, LEACH, and PEGASIS, respectively. Simulation results show that OMRP has the lowest energy consumption rate throughout the network lifetime.

% \begin{figure*}
%     \begin{minipage}[b]{0.49\textwidth}
%         \centering
%         \includegraphics[width=0.8\linewidth]{images/packet_round_af_0827.pdf}
%         \caption{Comparison of packet size output to BS.}
%         \label{rout.sub.packet_size_af_round}
%     \end{minipage}%
%     \hfill
%     \begin{minipage}[b]{0.49\textwidth}
%         \centering
%         \includegraphics[width=0.8\linewidth]{images/raw_packet_size_round.pdf}
%         \caption{Comparison of the number of raw packet to BS.}
%         \label{rout.sub.raw_packet_size_round}
%     \end{minipage}
%     % \caption{Performance of different routing protocols.}
%     \label{perform_routing}
% \end{figure*}
% 分析 FND-HND-AND 期间的平均能耗(J/round), 说明OMRP的优越性
% \par Table~\ref{table:energy_consumption} shows that the main energy consumption of the networks under the five network protocols mainly occurs in the early stage. In particular, the energy consumption rate of PEGASIS is very high during the period from the first node death to half of the dead node (FND-HND), which is almost six times the average energy consumption of the overall lifetime of the network (1-AND). In contrast, IGHND and OMRP are much smoother. To be specific, the energy consumption rate of OMRP is down about 6\%, 12\%, 20\%, and 92\% in the FND-HND phase when compared to IGHND, D2CRP, LEACH, and PEGASIS, respectively. The simulation results indicate that OMRP has the lowest energy consumption rate in the network lifetime.

% 分析OMRP为何节能：1）多跳、3）簇内TDMA
% 2）CH选举 上一个subsection说过了，所以首句先排除
\par Beyond CH election, the lower energy consumption of OMRP is attributed to the use of a multi-hop strategy and a communication sequence. Specifically, CHs adopt the available multi-hop mode to communicate with the sink node, establishing an optimized communication path within the cluster, which reduces transmission distance and energy consumption for inter-cluster communications. Additionally, based on the overlapping degree, the TDMA communication sequence within the cluster allows data with more redundancy to be compressed earlier, thereby reducing the energy consumption of CHs.

% 介绍网络的吞吐量：融合前+融合后
% data size指标: 融合前-传输效率、融合后-感知能力
\subsubsection{Comparisons of Throughput to BS} 

\par We evaluate network throughput by comparing the size of fused data packets and the count of raw packets sent to the BS across different routing protocols. The packet size reflects how well a protocol maintains data perception, while the raw packet count indicates its transmission efficiency. Figs.~\ref{perform_routing}(c) and \ref{perform_routing}(d) show these comparisons, highlighting the performance of each protocol in terms of data perception and transmission efficiency, respectively.

\begin{table}[ht]
\centering
\caption{Comparisons on data perception maintenance ability.}
\begin{tabularx}{\linewidth}{l X X X}
\toprule
\textbf{Protocol} & \textbf{75\%} & \textbf{50\%} & \textbf{25\%} \\ \hline
PEGASIS       & 36 & 40 & 70 \\ 
LEACH       & 427 & 524 & 627 \\ 
D2CRP       & 453 & 554 & 641 \\ 
IGHND       & 508 & 601 & 672 \\ 
OMRP         & 608 & 674 & 723 \\ 
\bottomrule
\end{tabularx}
\label{table:maintaince_ability}
\end{table}

% 分析一下这个表中数值的提升
\par Table~\ref{table:maintaince_ability} shows the maximum number of rounds various routing protocols can sustain at different throughput ratios. Specifically, OMRP shows an improvement in data perception maintenance ability by approximately 20\%, 34\%, and 42\% at the 75\% throughput ratio; 12\%, 22\%, and 29\% at the 50\% throughput ratio; and 8\%, 13\%, and 15\% at the 25\% throughput ratio compared to IGHND, D2CRP, and LEACH, respectively. Furthermore, the total number of raw packets for OMRP increases by 9\%, 22\%, and 28\% compared to IGHND, D2CRP, and LEACH, respectively.

% 【TODO】似乎来一个总结段比较合适？ 

% \begin{figure}[!t]
%     \centering
%     \includegraphics[width=0.82\linewidth]{images/R_Protocol_lifetime.pdf}
%     \caption{Comparison of network lifetime under different monitoring radii. }
%     \label{R_Protocol_lifetime}
% \end{figure}

% \begin{figure}[!t]
%     \centering
%     \includegraphics[width=0.85\linewidth]{images/R_Protocol_PacketOut.pdf}
%     \caption{Comparison of network throughput under different monitoring radii. }
%     \label{R_Protocol_PacketTotal}
% \end{figure}

\begin{figure}
    \centering
        \centering
    \includegraphics[width=0.95\linewidth]{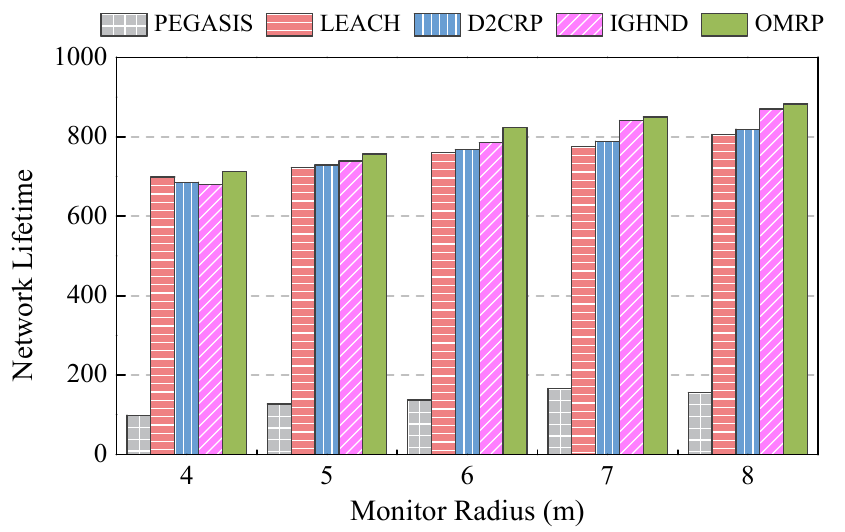}
        \centering
        \subfloat(a)
        \label{R_Protocol_lifetime}
        \centering
    \includegraphics[width=0.96\linewidth]{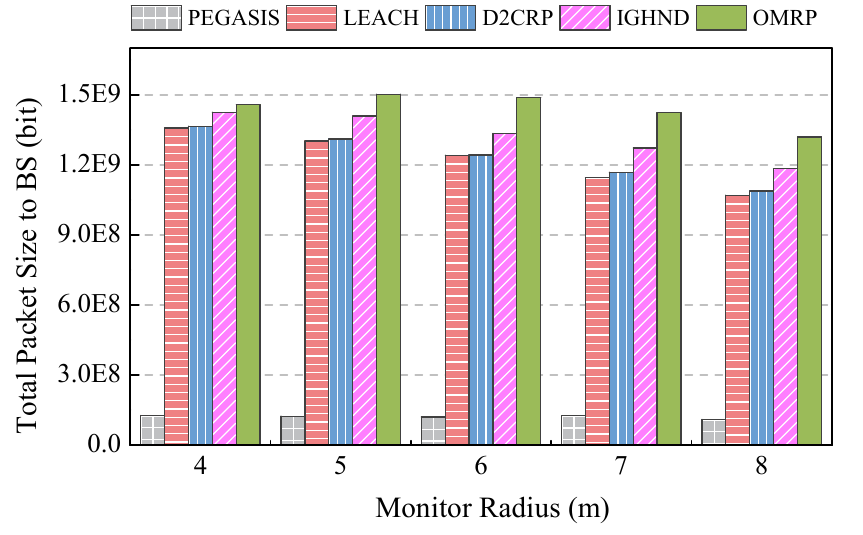}
        \centering
        \subfloat(b)
        \label{R_Protocol_PacketOut}
    \caption{Comparison of routing protocols under different monitoring radii. (a) Network lifetime. (b) Throughput.}
    \label{R_proto}
\end{figure}

% 0826 不提 $R_c$ 改提 $r$
% 不同 r, 也就是不同的sensing area 对应的OMRP鲁棒性分析
\subsubsection{Analysis of the spatial correlation} 

\par With the fixed node distribution and data packet size, the monitor radius $r$ of the IoT nodes affects the degree of data spatial correlation. A larger $r$ leads to more redundant data within the network. Fig.~\ref{R_proto} shows the network lifetime and total throughput of the routing protocol under different values of $r$. It can be seen that OMRP performs best in network lifetime and total throughput in different scenarios.

\subsection{Simulation of CB Method of SoftPPO-LSTM}

\begin{figure*}[h]
    \centering
    % 第一行的两幅图
    \begin{minipage}[b]{0.33\textwidth}
        \centering
        \includegraphics[width=\linewidth]{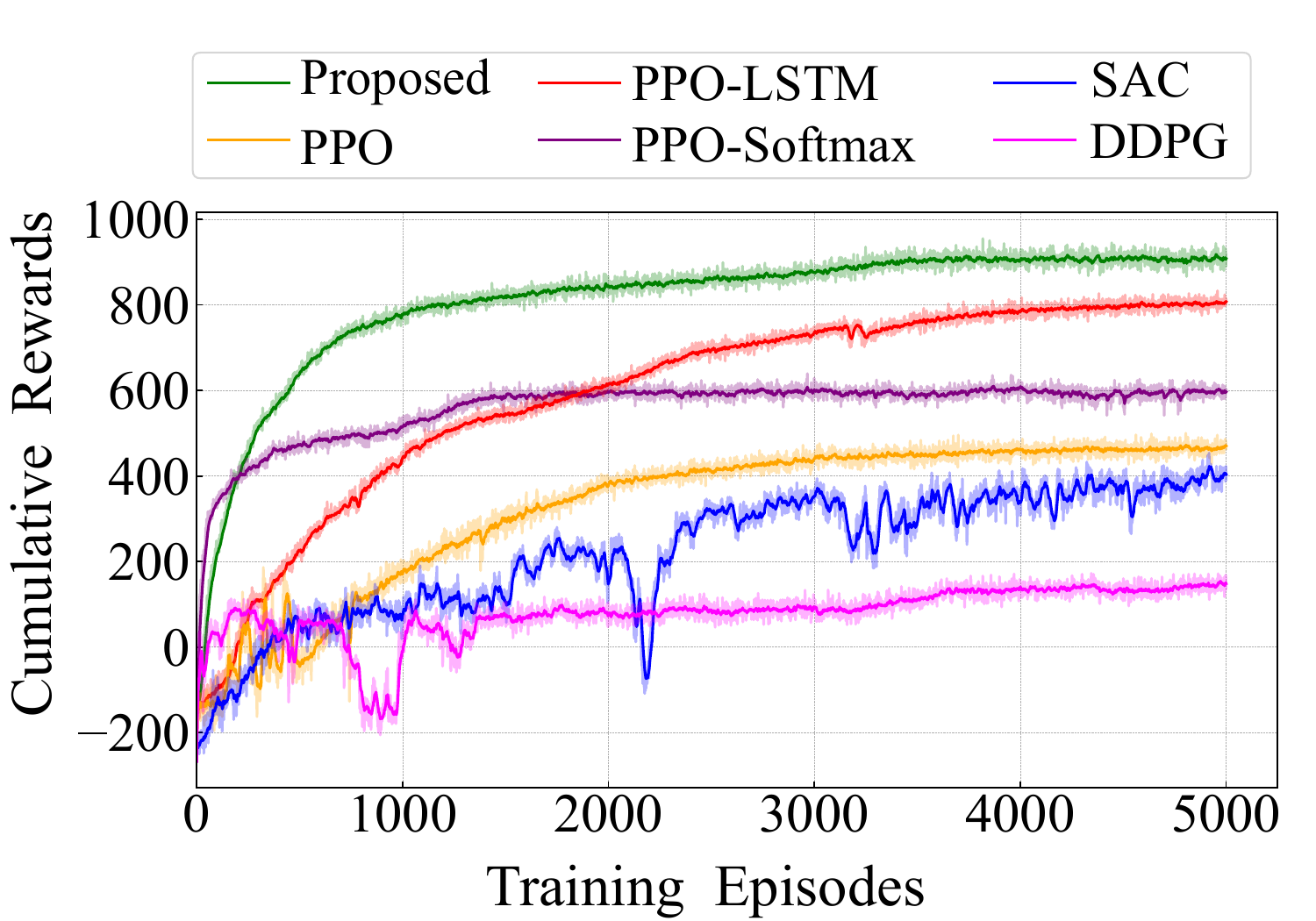}
        % \caption{Comparison of cumulative rewards.}
        \label{cb.sub.episode_reward}
        \subfloat(a)
    \end{minipage}%
    \hfill
    \begin{minipage}[b]{0.33\textwidth}
        \centering
        \includegraphics[width=\linewidth]{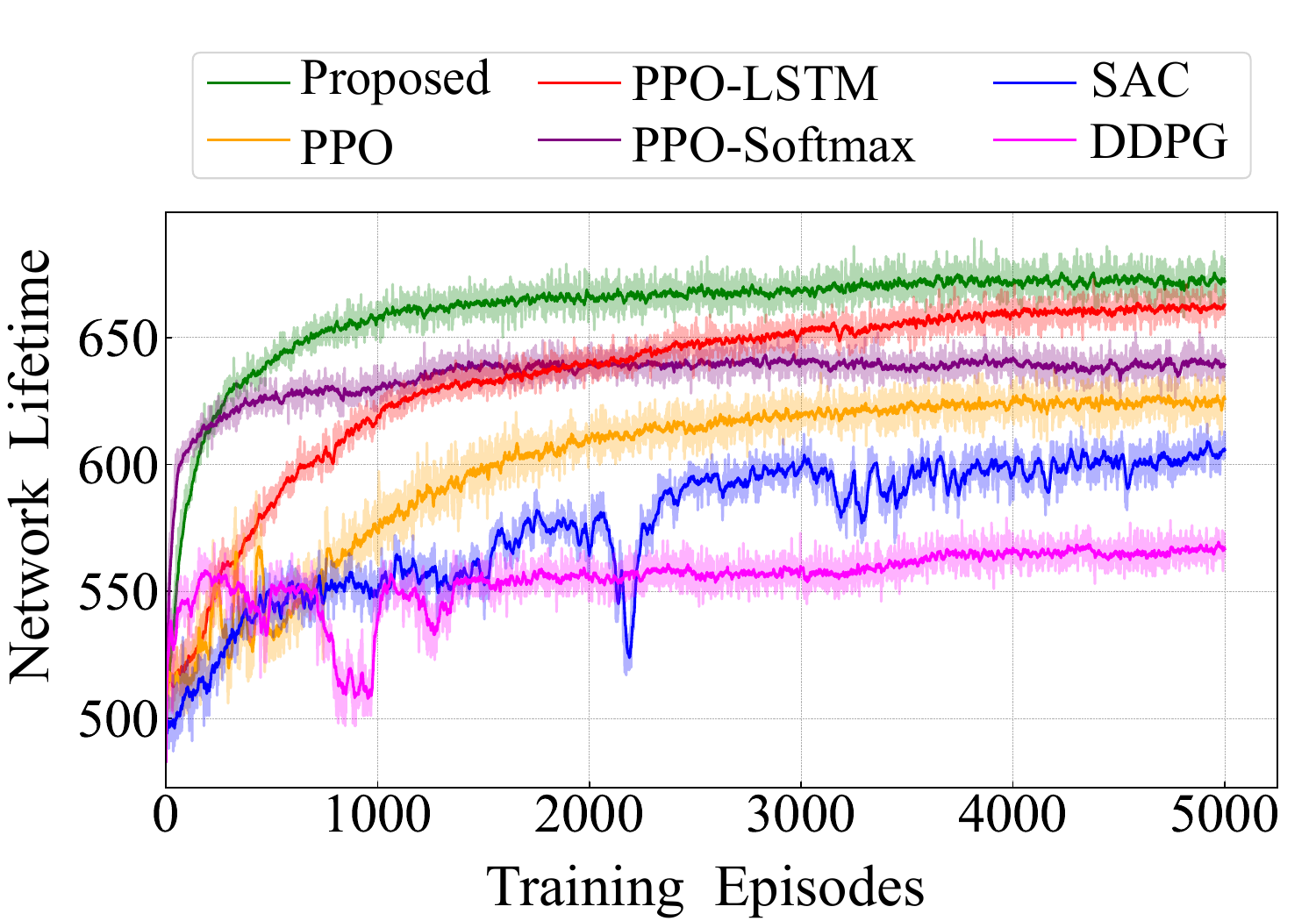}
        % \caption{Comparison of network life time.}
        \label{cb.sub.rounds}
        \subfloat(b)
    \end{minipage}
    \hfill
    \begin{minipage}[b]{0.33\textwidth}
        \centering
        \includegraphics[width=\linewidth]{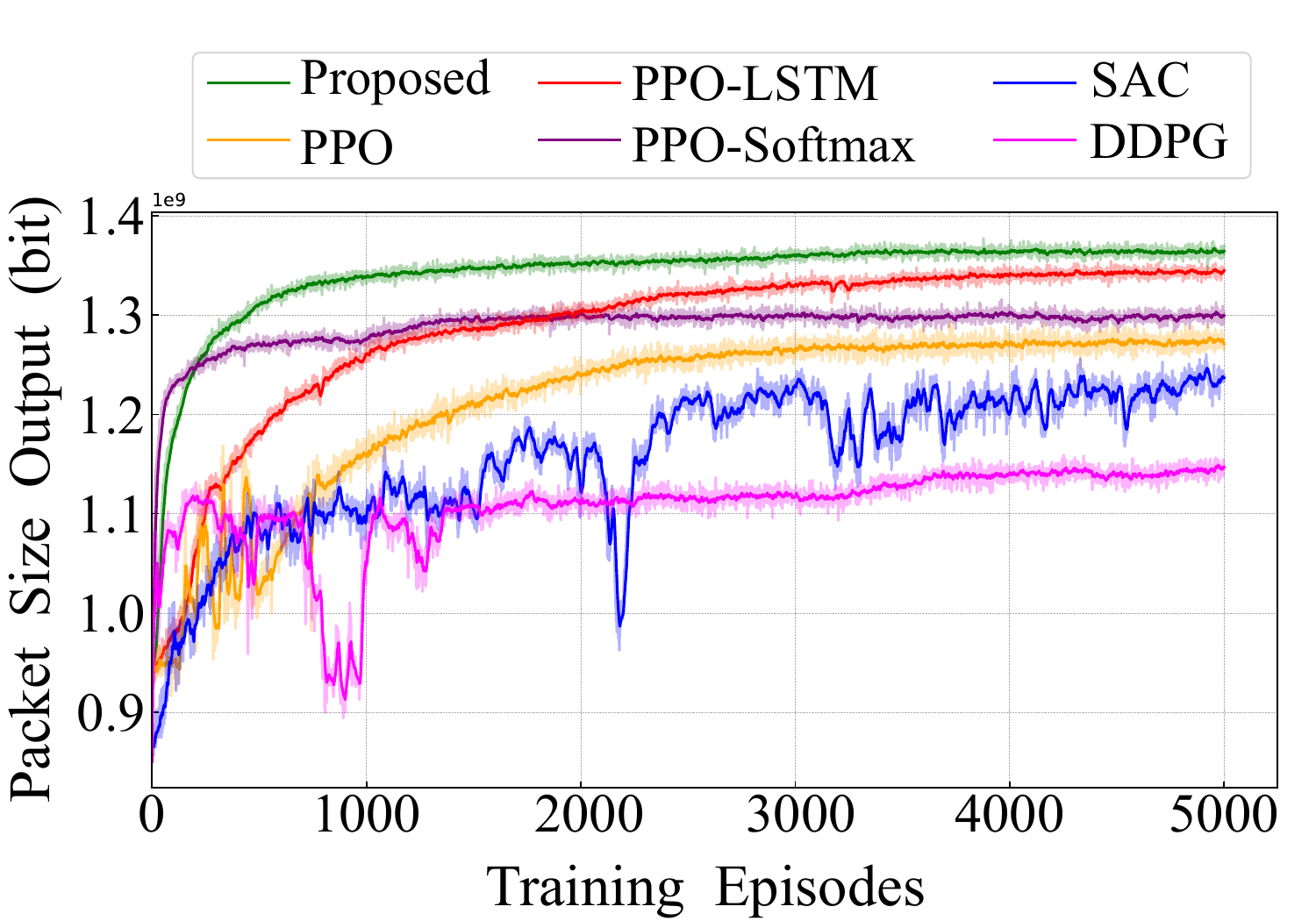}
        % \caption{Comparison of throughput to BS.}
        \label{cb.sub.packets_total}
        \subfloat(c)
    \end{minipage}%
    \caption{Performance of different DRL algorithms. (a) Comparison of cumulative rewards. (b) Comparison of network life time. (c) Comparison of throughput to BS.}
    \label{perform_drl}
\end{figure*}

% 总起段 TODO： CB Policy是在 OMRP的基础上搞得
% DRL对比策略：DDPG、SAC、PPO；传统greedy-based 
\par For DRL comparative analysis, we employ benchmark methods including Deep Deterministic Policy Gradient (DDPG) \cite{lillicrap2015continuous}, Soft Actor-Critic (SAC) \cite{haarnoja2018soft}, and Proximal Policy Optimization (PPO)\cite{schulman2017proximal} as reference models. We also compared the SoftPPO-LSTM-based CB method with the traditional greedy algorithm.

\subsubsection{Comparisons with DRL Policies} 
\par Fig.~\ref{perform_drl}(a) illustrates the cumulative rewards for each episode of SoftPPO-LSTM compared to other benchmark algorithms. Notably, the algorithms do not converge to the same solutions, which is due to the long episode lengths, the complexity of the MDP, and the high dimension of the action space, highlighting the challenge of the problem.

% 分析三个 benchmark 的表现: DDPG、SAC、PPO
\par For the benchmark algorithms, DDPG employs Gaussian noise to diversify strategic actions but struggles to fully explore the high-dimensional action space. It quickly converges to the local optimal strategy and shows less stable learning of superior strategies. Although the SAC algorithm can identify relatively advantageous strategies, its learning trajectory exhibits significant fluctuations in our environment. Similarly, the learning trajectory of PPO exhibits fluctuations in the early training stages. However, PPO ultimately stands out for its robust performance in our environment with intricate action spaces.

% 基于上一段 PPO 的良好表现 => 进一步改进： softmax and LSTM
% \par Based on the strengths of PPO, the SoftPPO-LSTM algorithm further improves stability and performance by introducing softmax and LSTM techniques. Specifically, softmax greatly improved the stability of the agent in the early stages of training, demonstrating a fast and steady learning pace initially. The performance gains from LSTM are more pronounced in the middle and later stages of training. While the initial learning speed improvement from LSTM is not as impressive as that from softmax, its consistent and stable learning enables the agent to achieve better convergence strategies and higher rewards. The SoftPPO-LSTM algorithm integrates both softmax and LSTM enhancements, allowing the agent to develop a nuanced understanding of actions and a deep perception of the environment. As a result, it demonstrates the most stable learning ability and the highest reward score in Fig.~\ref{cb.sub.episode_reward}. Furthermore, as shown in Figs.~\ref{cb.sub.rounds} and~\ref{cb.sub.packets_total}, the trends of network lifetime and throughput are consistent with the cumulative rewards performance.
% 原本一段定位不清晰，现改为一段描述性能，一段分析原因

\par The proposed SoftPPO-LSTM algorithm demonstrates the most stable learning ability and the highest reward scores, as shown in Fig.~\ref{perform_drl}(a). Specifically, SoftPPO-LSTM algorithm outperforms existing DRL methods in cumulative rewards, reflecting its superior training efficiency and learning stability. Additionally, as evidenced in Figs.~\ref{perform_drl}(b) and~\ref{perform_drl}(c), the trends of network lifetime and throughput are consistent with that of cumulative rewards. Specifically, SoftPPO-LSTM achieves a throughput of $1.37 \times 10^9$ bit, representing an 8.3\% increase over PPO, a 10.9\% increase over SAC, and a 19.5\% increase over DDPG. Furthermore, LSTM contributes a 6.5\% improvement, and Softmax provides a 2.6\% improvement compared to the PPO algorithm.

\par The superior performance of SoftPPO-LSTM can be attributed to three aspects. First, in our MDP analysis, we simplify the complex subset selection problem into a continuous scoring problem of dimension $N$, where the top $N_{CB}$ nodes with the highest scores are selected as the final beamforming nodes. This scoring-based action output introduces challenges to the stability of gradient updates, but the clipped mechanism of PPO ensures that the magnitude of gradient changes remains controlled during updates, providing greater stability compared to other algorithms. Second, based on PPO, we integrate softmax control, which further compresses score differences, reduces gradient fluctuations, and highlights higher scores. This enhancement helps the agent better understand the complexity of the problem. Finally, the inclusion of LSTM enables the agent to process time-series data and adapt to long-term training, effectively handling the dynamic nature of our problem. It also allows the agent to better adapt to heuristic routing strategies, further enhancing overall performance.

% Besides, the introduction of a more complex LSTM network into the actor network of the SoftPPO-LSTM algorithm results in a slower initial convergence compared to the PPO algorithm. However, with the combined guidance of LSTM and Softmax, the SoftPPO-LSTM algorithm surpasses other algorithms by achieving higher rewards and enhanced stability.

\begin{figure}[!t]
    \centering
        \centering
    \includegraphics[width=0.95\linewidth]{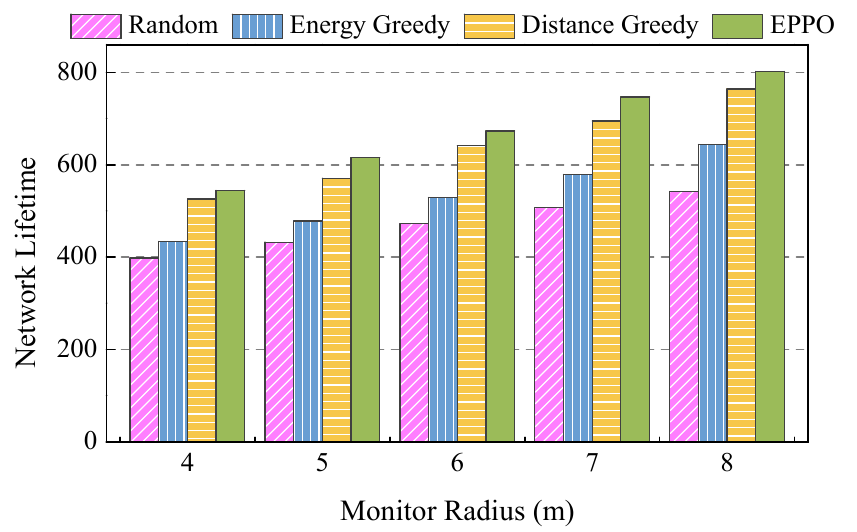}
        \centering
        \subfloat(a)
        \label{traditional_lifetime_out}
        \centering
    \includegraphics[width=0.95\linewidth]{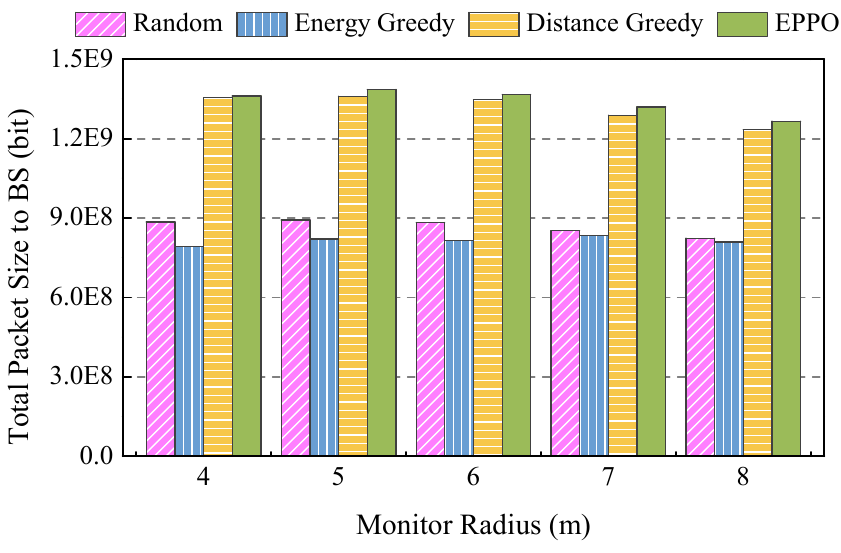}
        \centering
        \subfloat(b)
        \label{traditional_packet_out}
    \caption{Comparison of traditional CB strategies under different monitoring radii. (a) Network lifetime. (b) Throughput.}
    \label{R_traditional}
\end{figure}

\subsubsection{Comparisons with Traditional Policies} We evaluate three traditional strategies including random strategy, energy-based greedy strategy, and distance-based greedy strategy. The energy-based greedy strategy selects the nodes with the highest residual energy for CB, while the distance-based greedy strategy selects the nodes nearest to the sink node. In addition, Fig.~\ref{R_traditional} compares the network lifetime and throughput of these traditional strategies with SoftPPO-LSTM under different values of $r$. The simulation results show that SoftPPO-LSTM consistently achieves the highest throughput and the longest network lifetime.

% 不同r下SoftPPO-LSTM与传统方法的对比
% \par Fig.~\ref{traditional_lifetime_out} and Fig.~\ref{traditional_packet_out} shows the comparison of network lifetime and throughput of different CB policies based on OMRP routing method with different $R_c$. From the simulation results, we can note that SoftPPO-LSTM still has the highest throughput and longest network lifetime.
% \begin{figure}[!t]
%     \centering
%     \includegraphics[width=0.82\linewidth]{images/R_Lifetime_Traditional.pdf}
%     \caption{Comparison of network lifetime under different monitoring radii. }
%     \label{traditional_lifetime_out}
% \end{figure}

% \begin{figure}[!t]
%     \centering
%     \includegraphics[width=0.80\linewidth]{images/R_PacketTotal_Traditional.pdf}
%     \caption{Comparison of network throughput under different monitoring radii. }
%     \label{traditional_packet_out}
% \end{figure}

% 深入到某一轮中分析每轮的 Packet Size
\par
\begin{figure}[!t]
    \centering
    \includegraphics[width=0.8\linewidth]{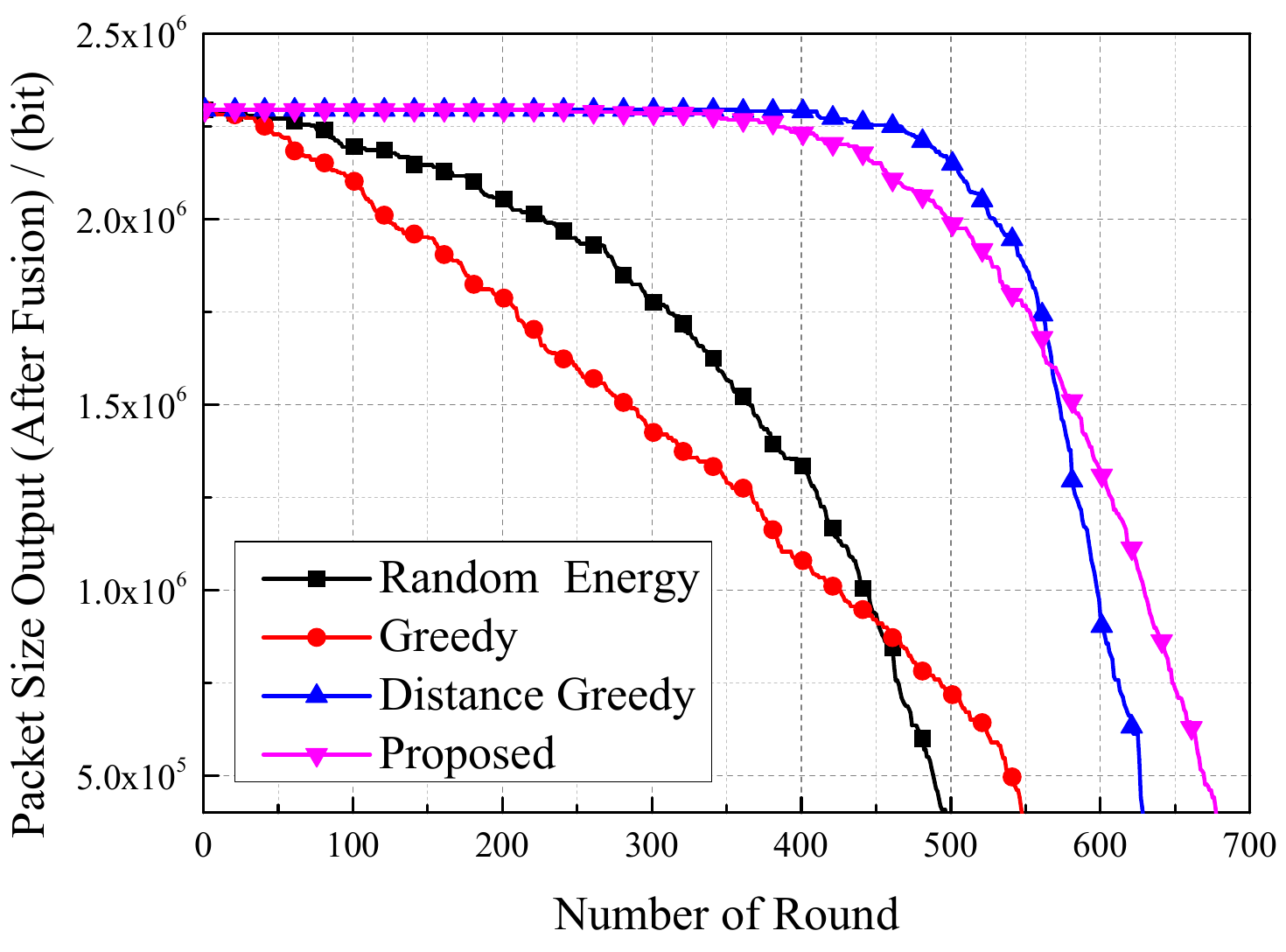}
    \caption{Comparison of packet size output to BS in each round.}
    \label{traditional_pkg_round}
\end{figure}

 Additionally, we evaluate the perception ability of SoftPPO-LSTM alongside these traditional strategies and analyze the performance. As is shown in Fig.~\ref{traditional_pkg_round}, the random and energy-based greedy strategies exhibit a rapid decline in the early stages due to their failure to consider distance in broadcasting, leading to a large broadcast radius and high energy consumption. In contrast, the distance-based greedy strategy slows node death in the early stages but falls quickly in the later stages. We recorded the simulation process of the distance-based greedy strategy and found that the surviving nodes in the later stages are often found at the edge of the network, which increased the routing distance and accelerated node death. Although SoftPPO-LSTM does not outperform the distance-based greedy strategy initially, its comprehensive consideration of both locations and residual energy levels can ease the hot spot issue that the distance-based greedy strategy has. As a result, SoftPPO-LSTM maintains slower node death rates in the later stages, ultimately achieving the longest network lifetime and highest throughput.

% \begin{figure}[!t]
%     \centering
%     \includegraphics[width=1.0\linewidth]{images/Traditional_alive_round.pdf}
%     \caption{Comparison of network alive nodes.\sethlcolor{pink}\hl{[can delete]}}
%     \label{traditional_live}
% \end{figure}

% \subsection{Algorithm Stability Tests}
% 考虑不同 q0和R的OMRP性能？
% 融合系数 q_0
% 感知半径 R（信息密度）

% \sethlcolor{pink}\hl{[Consider Merge]}

% \subsection{Discussions of Simulation Parameters}
% \sethlcolor{pink}\hl{[DO IT OR NOT ?]}
% % 讨论 半径导致的信息密度改变，冗余信息占比
% % 讨论 BS处的最低接收能耗Pr以及N_cb=10的合理性

\section{Discussion}
\label{sec:discussion}

\par In this section, we discuss the scalability, robustness and validation of the proposed methods as well as the adaptability of our optimization framework, and the details are shown as follows.

\subsection{The Scalability of OMRP and SoftPPO-LSTM Methods}

\par
\begin{figure}[h]
    \centering
    \begin{minipage}[b]{0.48\linewidth}
        \centering
        \includegraphics[width=\linewidth]{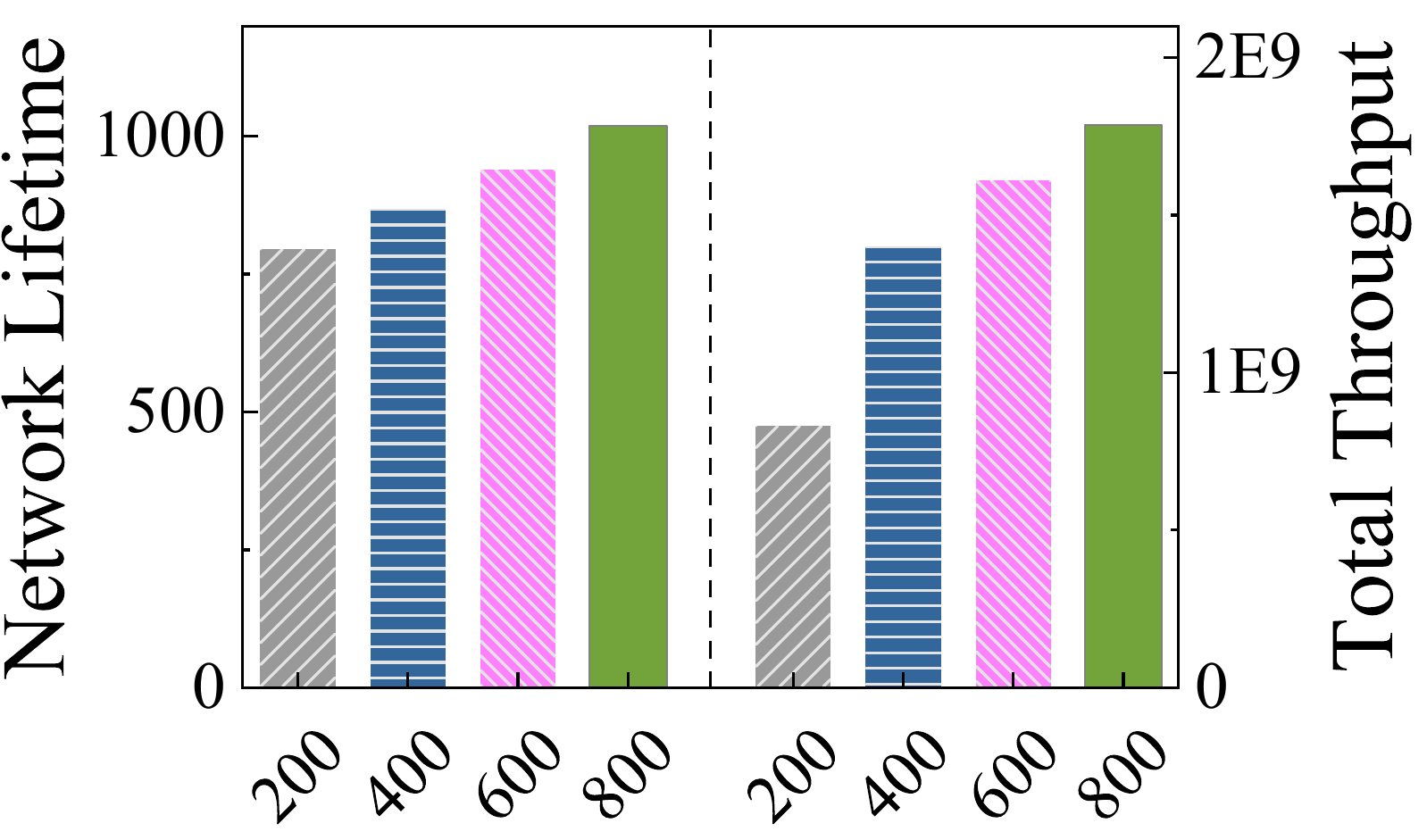}
        \subfloat(a)
        \label{fig:CB_sys_fig:sub1}
    \end{minipage}%
        \centering
        \begin{minipage}[b]{0.48\linewidth}
        \centering
        \includegraphics[width=\linewidth]{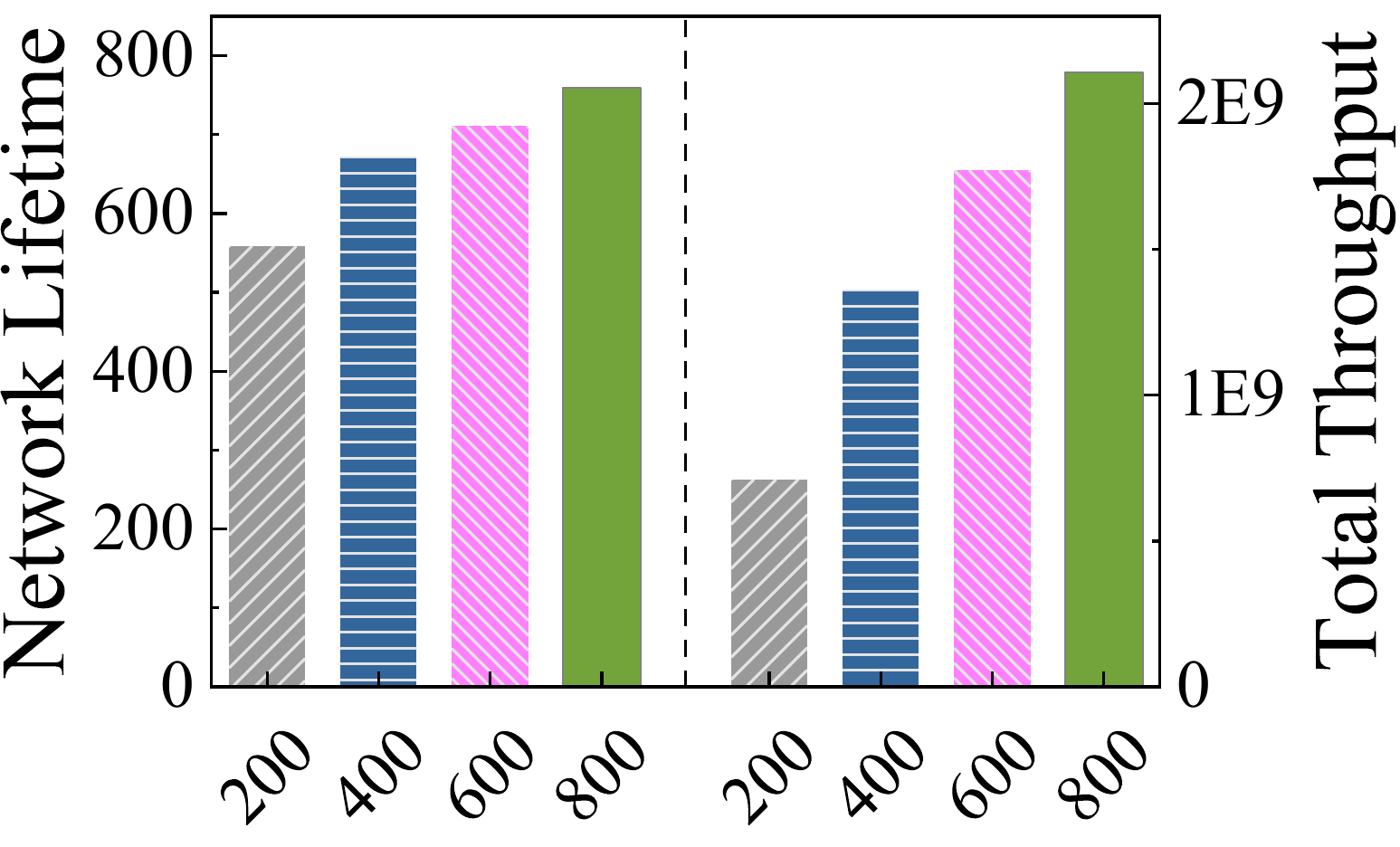}
    \subfloat(b)
        \label{fig:CB_sys_fig:sub2}
    \end{minipage}%
    \caption{Simulation results in different scales of IoT nodes. (a) OMRP method without energy consumption of CB. (b) SoftPPO-LSTM method with OMRP based routing policy.}
    \label{fig:scalability}
\end{figure}

 We evaluate simulations with 200, 400, 600, and 800 IoT nodes to evaluate scalability. Specifically, we simulate both the OMRP method without energy consumption of the CB and the SoftPPO-LSTM method with the OMRP routing policy. Fig.~\ref{fig:scalability} shows that the two methods perform well in terms of lifetime and throughput, indicating that our system can scale effectively to support a larger number of IoT devices.

\subsection{The Imperfect Phase Synchronization for Robustness}

\par In real-world CB implementations, imperfect phase synchronization arises from time and frequency offsets, initial phase errors, and channel estimation inaccuracies~\cite{jung2019secrecy}. To evaluate the impact of phase errors on CB, the AF of the beamforming nodes is redefined as follows~\cite{minturn2013distributed}:

% \par The imperfect phase synchronization errors, which can be caused by factors such as time and frequency synchronization issues, initial phase offsets, and channel estimation errors in real-world CB implementation~\cite{R5-1-1}. To evaluate the impact of phase errors on CB, the AF of the beamforming nodes is redefined as follows~\cite{R5-1-2}:
\begin{equation}
\label{AF_rub}
\mathrm{AF}(\phi, \theta, I)=\sum_{k=1}^{N_{\mathrm{CB}}} I_{k} e^{j \Psi_{k}} e^{j [\frac{2 \pi}{\lambda} d_{k}(\phi, \theta) + \zeta_{i} ]},
\end{equation}
where $\zeta_{i}$ represents the phase error of the $i$th IoT node and it is assumed to obey the Tikhonov distribution with parameter $\kappa$ which determines the error size~\cite{jung2020secure}.

\par
\begin{figure}
    \centering
    \includegraphics[width=\linewidth]{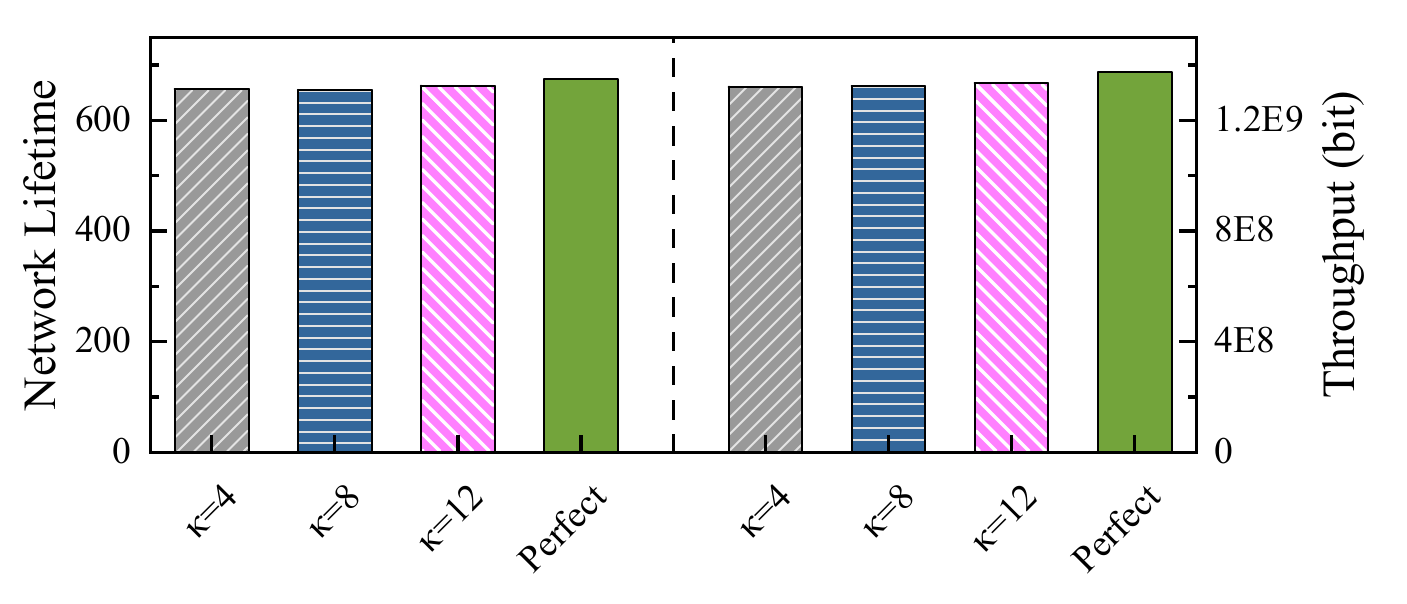}
    \caption{Performance analysis influenced by imperfect phase synchronization.}
    \label{fig:phase_error}
\end{figure}

 Fig.~\ref{fig:phase_error} shows the system performance of network lifetime and throughput considering the phase errors under different values of $\kappa$, the result demonstrates that our solution exhibits negligible performance loss under phase synchronization errors.

\subsection{The Validation of OMRP and SoftPPO-LSTM Methods}

\par In this part, the validation of hardware integration, real-world IoT distribution, and method parameters are discussed as follows.

\subsubsection{Validation of Hardware Integration}

\par Integration tests are critical to ensure the operation of the software on hardware platforms like the IoT~\cite{shetiya2024verification}. To validate the hardware integration of our proposed method, we conduct a deployment experiment using a Raspberry Pi, which is widely recognized as a reliable IoT validation platform in existing works (e.g.~\cite{ponnuru2025baap, chai2025iot, ni2024novel}). Specifically, we deploy the SoftPPO-LSTM model on a Raspberry Pi 4B, featuring an ARM Cortex-A72 1.5 GHz CPU and 8 GB of memory. The SoftPPO-LSTM network parameters occupy 16 MB of disk space, while the entire computing environment requires 835 MB. In this case, the device takes 10.9 seconds to execute the program and generate inference results, with a maximum memory usage of 324 MB. Notably, the inference time can be reduced to 2.2 seconds by preloading the Python program and further minimized to 0.02 seconds by preloading the model into memory. These results show that the SoftPPO-LSTM model can efficiently operate on resource-constrained IoT devices, which highlights the feasibility of deploying our method in practice environments.

\subsubsection{Validation of Real-world IoT Nodes Distribution} 

\par
\begin{figure}[h]
    \centering
    \includegraphics[width=\linewidth]{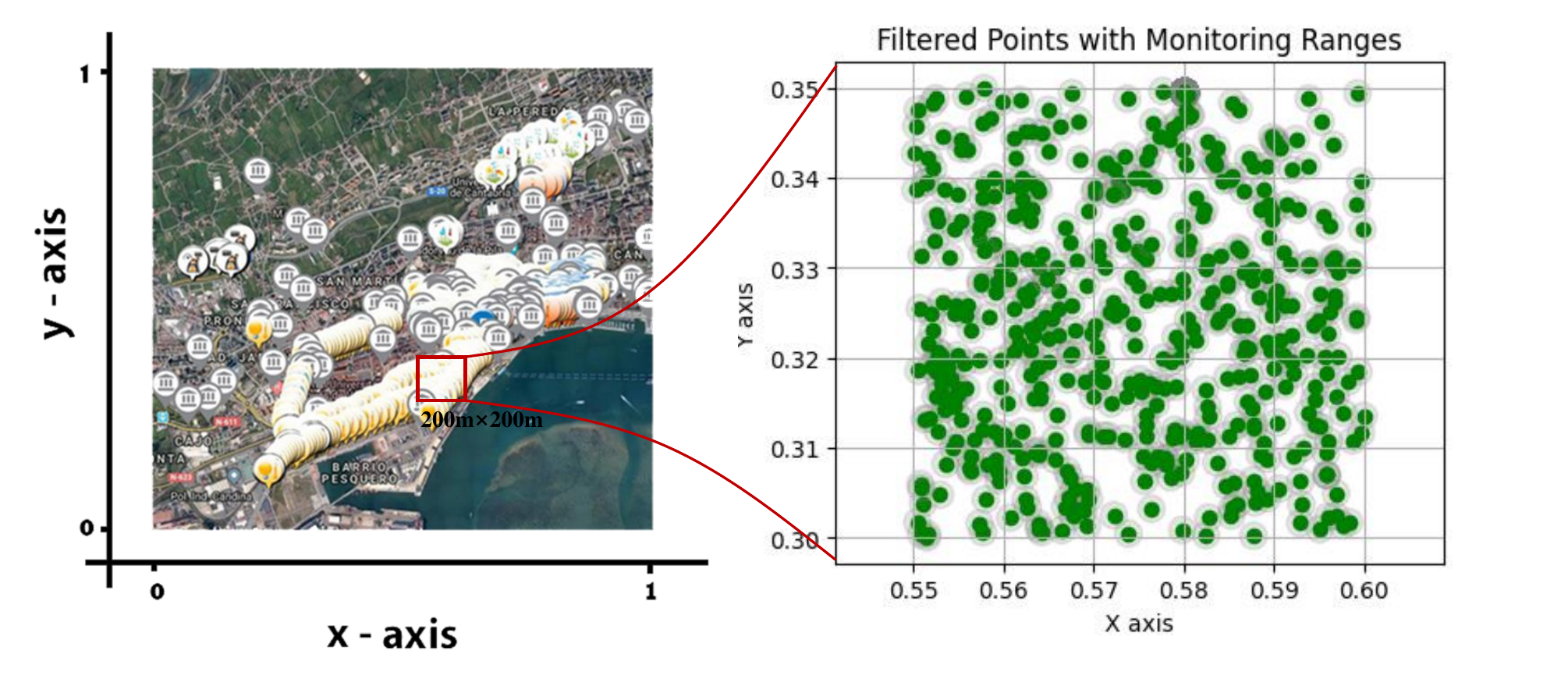}
    \caption{Location map of real-world deployment of IoT devices in the city of Santander. The dataset is provided at https://github.com/Net4uCA/SIoT-IoT-Dataset.}
    \label{fig:IoT_distribution}
\end{figure}

 As shown in Fig.~\ref{fig:IoT_distribution}, we extract a 200 m $\times$ 200 m area containing 588 IoT nodes from a data set that illustrates the distribution of IoT devices in the city of Santander~\cite{marche2020exploit}. As such, we adjust the simulation settings as well as the network sizes to simulate the proposed OMRP routing strategy and the SoftPPO-LSTM algorithm. 

% \begin{figure}[h]
%     \centering
%     % 第一行的两幅图
%     \begin{minipage}[b]{0.48\linewidth}
%         \centering
%         \includegraphics[width=\linewidth]{images/FND_588_nodes_fig.pdf}
%         \subfloat(a)
%     \end{minipage}%
%     \hfill
%     \begin{minipage}[b]{0.48\linewidth}
%         \centering
%         \includegraphics[width=\linewidth]{images/HND_588_nodes_fig.pdf}
%         \subfloat(b)
%     \end{minipage}
    
%     \vspace{2mm} % 控制第一行和第二行的间距

%     % 第二行的两幅图
%     \begin{minipage}[b]{0.48\linewidth}
%         \centering
%         \includegraphics[width=\linewidth]{images/AND_588_nodes_fig.pdf}
%         \subfloat(c)
%     \end{minipage}%
%     \hfill
%     \begin{minipage}[b]{0.48\linewidth}
%         \centering
%         \includegraphics[width=\linewidth]{images/Throughput_588_nodes_fig.pdf}
%         \subfloat(d)
%     \end{minipage}
    
%     \caption{Comparison of FND, HND, AND and throughput performance of different routing strategies based on real-world IoT distribution.}
%     \label{fig:perform_588_routing}
% \end{figure}

\begin{figure}[h]
    \centering
    % 第一行的两幅图
    \begin{minipage}[b]{\linewidth}
        \centering
        \includegraphics[width=\linewidth]{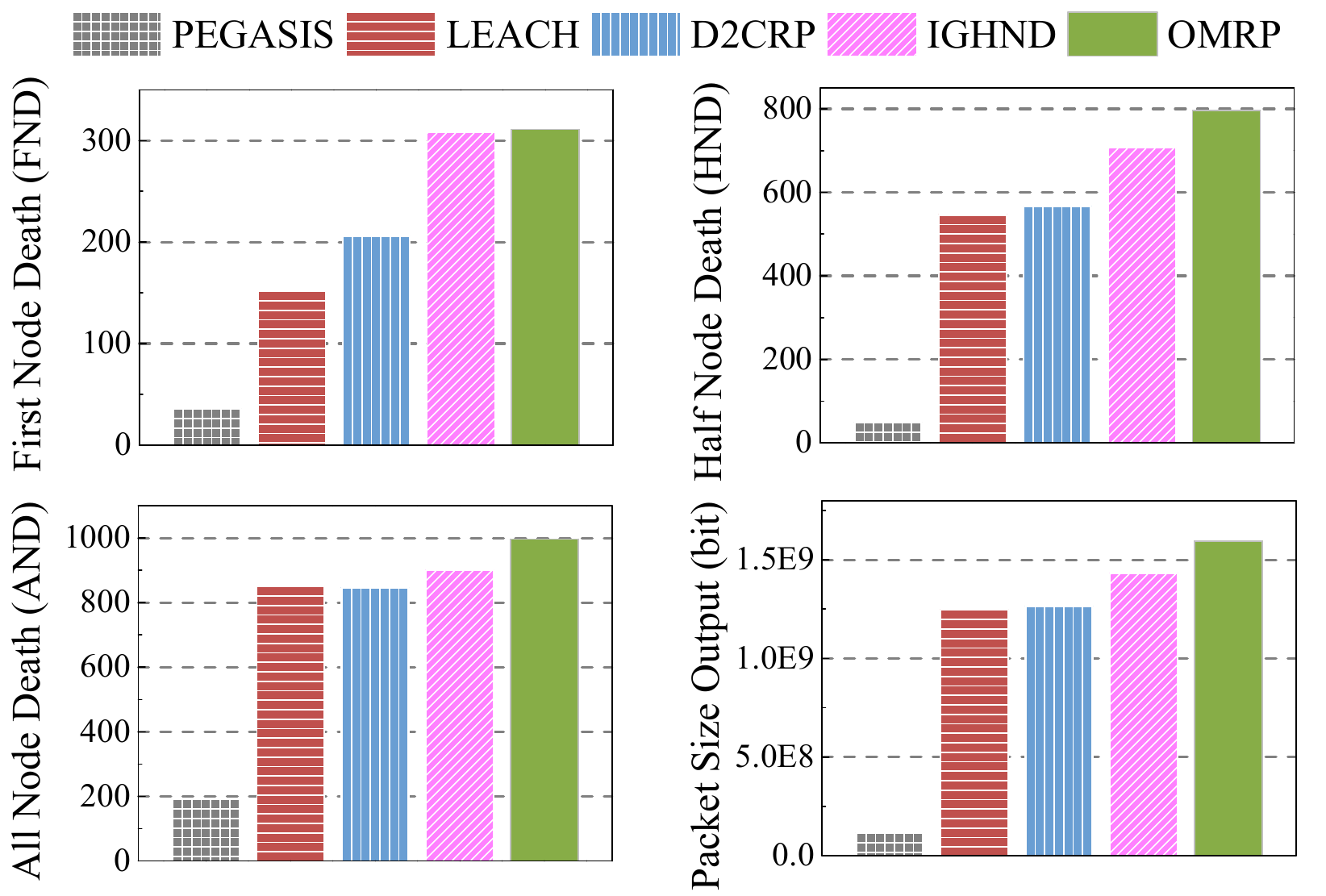}
    \end{minipage}
    \caption{Comparison of FND, HND, AND and throughput performance of different routing strategies based on real-world IoT distribution.}
    \label{fig:perform_588_routing}
\end{figure}

\par
\begin{figure}[h]
    \centering
    % 第一行的两幅图
    \begin{minipage}[b]{\linewidth}
        \centering
        \includegraphics[width=\linewidth]{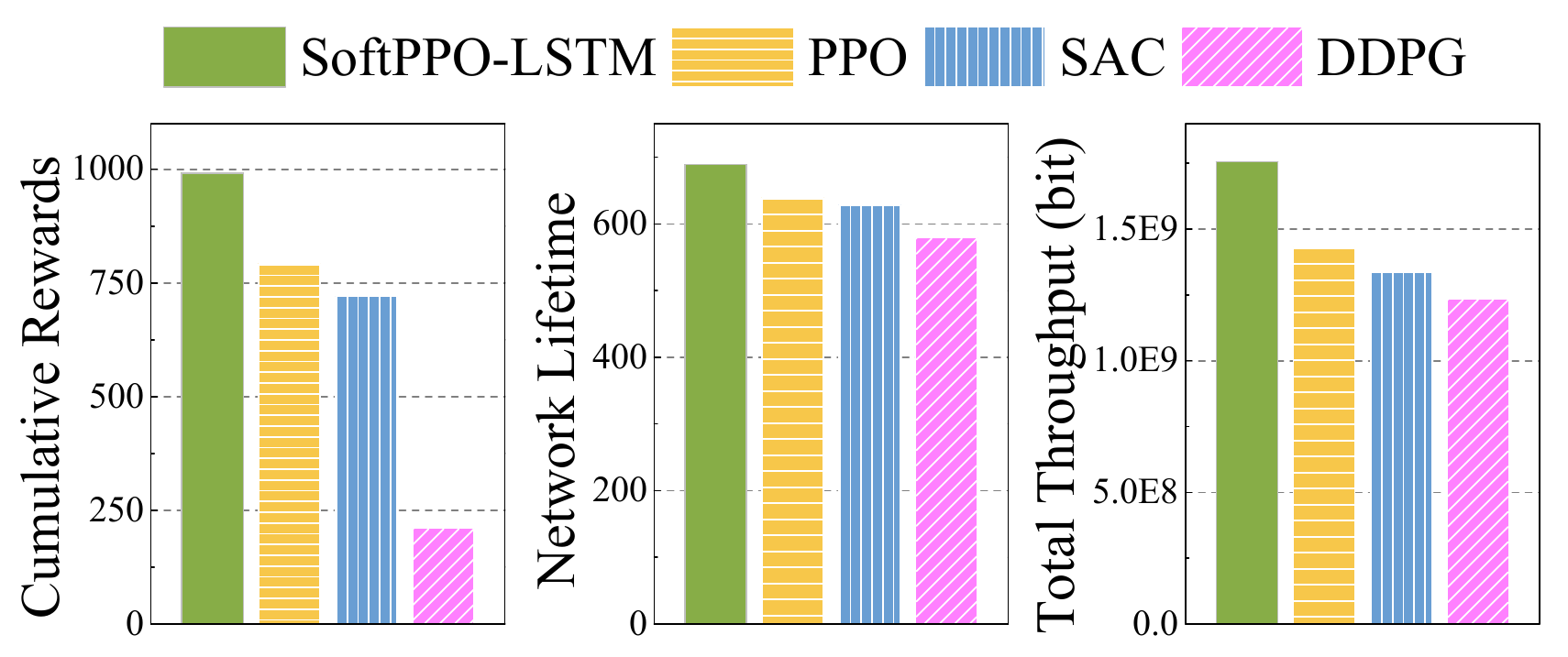}
    \end{minipage}
    \caption{Comparison of cumulative rewards, network lifetime and throughput of different DRL algorithms based on real-world IoT distribution.}
    \label{fig:perform_588_cb}
\end{figure}

 We initialize the IoT distribution based on the dataset, and other settings remain the same as in Section~\ref{sec:simulation_analysis}. In this case, the comparison of FND, HND, AND, and throughput performance in Fig.~\ref{fig:perform_588_routing} shows that our proposed OMRP outperforms other routing policies in all of the above indicators.
Moreover, Fig.~\ref{fig:perform_588_cb} shows the comparison of cumulative rewards, network lifetime, and throughput of different DRL algorithms based on the real-world IoT distribution. The result shows that our proposed SoftPPO-LSTM achieves the highest cumulative rewards, extends the longest network lifetime, and delivers the highest number of data packets.

% \begin{figure}[h]
%     \centering
%     % 第一行的两幅图
%     \begin{minipage}[b]{0.32\linewidth}
%         \centering
%         \includegraphics[width=\linewidth]{images/CB_Reward_588_fig.pdf}
%         \subfloat(a)
%     \end{minipage}%
%     \hfill
%     \begin{minipage}[b]{0.32\linewidth}
%         \centering
%         \includegraphics[width=\linewidth]{images/CB_Lifetime_588_fig.pdf}
%         \subfloat(b)
%     \end{minipage}
%     \hfill
%     \begin{minipage}[b]{0.32\linewidth}
%         \centering
%         \includegraphics[width=\linewidth]{images/CB_Throughput_588_fig.pdf}
%         \subfloat(c)
%     \end{minipage}
%     \caption{Comparison of cumulative rewards, network lifetime and throughput of different DRL algorithms based on real-world IoT distribution. }
%     \label{fig:perform_588_cb}
% \end{figure}

\subsubsection{Validation of Method Parameters} 

\par
\begin{figure}[h]
    \centering
    % 第一行的两幅图
    \begin{minipage}[b]{0.49\linewidth}
        \centering
        \includegraphics[width=\linewidth]{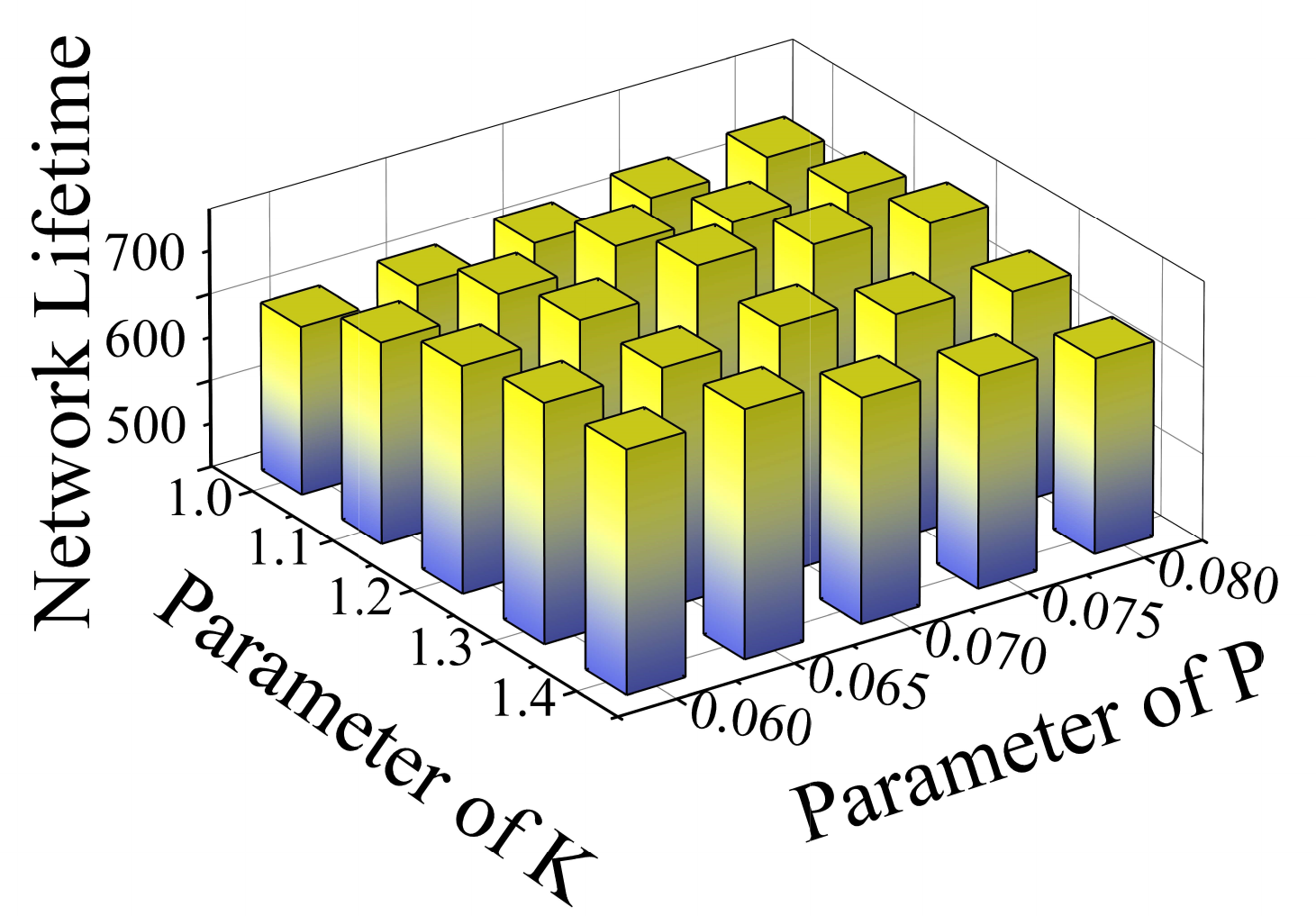}
        \subfloat(a)
    \end{minipage}%
    \hfill
    \begin{minipage}[b]{0.49\linewidth}
        \centering
        \includegraphics[width=\linewidth]{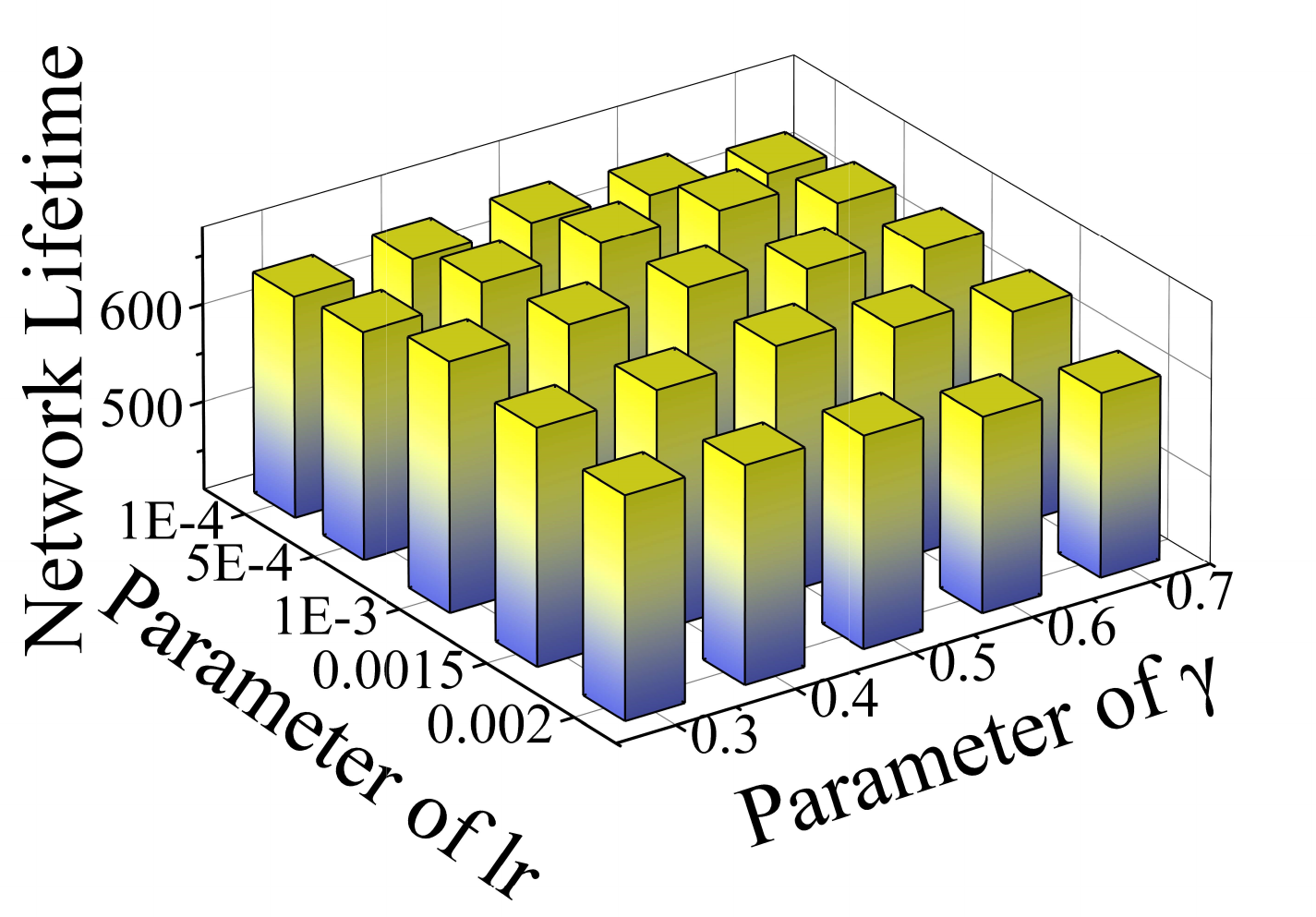}
        \subfloat(b)
    \end{minipage}
    \caption{The optimization performance of network lifetime under different simulation parameters. (a) The validation of amplification factor $K$ and the lower bound ratio $P$ of OMRP. (b) The validation of the discounted factor and the learning rate of SoftPPO-LSTM.}
    \label{fig:param_valida}
\end{figure}

 We validate the amplification factor $K$ and the lower bound ratio $P$ in the OMRP. Specifically, the value of $K$ ranges from 1.0 to 1.4, and the value of $P$ ranges from 0.06 to 0.08. Similarly, we validate the discount factor $\gamma$ and the learning rate $lr$ in the training of the SoftPPO-LSTM algorithm. The value of $\gamma$ ranges from 0.30 to 0.70, and the value of $lr$ ranges from $1e-4$ to $2e-3$. Fig.~\ref{fig:param_valida} shows the validation results, which indicate that $\gamma = 0.50$, $lr = 1e-3$, $K = 1.2$, and $P = 0.07$ achieve the longest network lifetime of system performance.

% \subsection{Adaptability of Optimization Framework}
\subsection{Discussion of Key Assumptions}

\par In this work, we carefully make some assumptions and settings to strike a balance between model tractability and practical relevance. In fact, some of them can be relaxed or replaced with more general ones. For instance, the optimization framework can accommodate dynamic node scenarios by incorporating mobility models, or adjust initial settings to account for differences in node power and energy levels, and even handle heterogeneous data by refining the data correlation modeling for data fusion. These potential adjustments provide flexibility to the optimization framework, allowing it to remain adaptable to varying network conditions and device differences.

\section{Conclusion}
\label{sec:conclusion}

% \par This study introduced a novel correlated data-driven communication scheme for the homogeneous IoT network through the integration of CB and hierarchical routing protocol. The proposed OMRP-based SoftPPO-LSTM strategy leverages the synergy of spatial data correlation and dynamic communication mechanisms to enhance network energy efficiency and data throughput in IoT. 
In this article, we have studied a novel data-driven communication scheme for homogeneous IoT networks by integrating CB with routing protocols. Specifically, we have first utilized the OMRP to aggregate the network data at a specific node. Based on this aggregation, we have formulated a node selection problem for the CB stage aimed at optimizing uplink communication. Given the complexity of this problem, we have introduced the SoftPPO-LSTM algorithm, which intelligently selects CB nodes to enhance transmission efficiency. Simulation results indicate that the proposed OMRP demonstrated superior performance in extending network lifetime, minimizing energy consumption, and optimizing data throughput compared to existing protocols such as PEGASIS, LEACH, D2CRP, and IGHND. The introduction of the overlap degree in the CH election and the dynamic multi-hop strategies significantly contributed to these improvements. Moreover, the SoftPPO-LSTM-based CB policy employs the softmax control and LSTM layers to simplify the solution space and facilitate stable learning over long rounds. These adjustments enable the IoT networks to reduce the energy consumption of the CB process while avoiding possible hot spot problems.

% 局限性
\par While the proposed method demonstrates promising results, certain limitations should be acknowledged. The fault-tolerance mechanisms of the system require further consideration to handle the unpredictable challenges of real-world deployments, such as electromagnetic interference and hardware failures~\cite{he2024deep}. Moreover, the training process of the advanced DRL techniques relies on simulated data, which may not fully capture the complexity and variability of real-world IoT environments~\cite{silvestri2024real}. These factors may impact the stability and performance of the DRL algorithm in practical scenarios.

% 未来工作
\par Future works will focus on addressing these limitations by collecting more realistic deployment data, enhancing system fault tolerance, and improving adaptability through transfer learning and real-time feedback. Extending the model to heterogeneous IoT nodes and dynamic environments will also be explored to further strengthen its applicability. In addition, we plan to conduct more validation experiments to evaluate the scalability and robustness of our system under varying network conditions and hardware configurations.

% \par Future research will explore the adaptability of this mechanism in dynamic and heterogeneous network environments, including considerations of mobility and varying node densities. Further investigation will also focus on the potential of advanced machine learning algorithms to refine the CB strategy and routing protocols in real-time scenarios.

% \newpage

\bibliographystyle{IEEEtran}
% \bibliography{sample.bib}
\bibliography{jab.bib}

% Generated by IEEEtran.bst, version: 1.14 (2015/08/26)
\begin{thebibliography}{10}
\providecommand{\url}[1]{#1}
\csname url@samestyle\endcsname
\providecommand{\newblock}{\relax}
\providecommand{\bibinfo}[2]{#2}
\providecommand{\BIBentrySTDinterwordspacing}{\spaceskip=0pt\relax}
\providecommand{\BIBentryALTinterwordstretchfactor}{4}
\providecommand{\BIBentryALTinterwordspacing}{\spaceskip=\fontdimen2\font plus
\BIBentryALTinterwordstretchfactor\fontdimen3\font minus \fontdimen4\font\relax}
\providecommand{\BIBforeignlanguage}[2]{{%
\expandafter\ifx\csname l@#1\endcsname\relax
\typeout{** WARNING: IEEEtran.bst: No hyphenation pattern has been}%
\typeout{** loaded for the language `#1'. Using the pattern for}%
\typeout{** the default language instead.}%
\else
\language=\csname l@#1\endcsname
\fi
#2}}
\providecommand{\BIBdecl}{\relax}
\BIBdecl

\bibitem{yang2023environment}
F.~Yang, Q.~Sun, Z.~Zhao, X.~Wang, J.~Xu, Y.~Zheng, H.~Zhang, and L.~Wang, ``Environment fusion routing protocol for wireless sensor networks,'' \emph{IEEE Sensors J.}, 2023.

\bibitem{qin2024cdt}
H.~Qin, N.~Li, T.~Wang, G.~Yang, and Y.~Peng, ``{CDT}: {Cross-interface Data Transfer} scheme for bandwidth-efficient {LoRa} communications in energy harvesting multi-hop wireless networks,'' \emph{J. Netw. Comput. Appl.}, p. 103935, 2024.

\bibitem{yang2024effective}
L.~Yang, B.~Ma, L.~Yuan, and B.~Wu, ``Effective application of {IoT} power electronics technology and power system optimization control,'' \emph{Tsinghua Sci. Technol.}, vol.~29, no.~6, pp. 1763--1775, 2024.

\bibitem{verma2017survey}
S.~Verma, Y.~Kawamoto, Z.~M. Fadlullah, H.~Nishiyama, and N.~Kato, ``A survey on network methodologies for real-time analytics of massive {IoT} data and open research issues,'' \emph{J Ambient Intell Humaniz Comput.}, vol.~19, no.~3, pp. 1457--1477, 2017.

\bibitem{jan2021marginal}
S.~R. Jan, R.~Khan, F.~Khan, M.~A. Jan, M.~D. Alshehri, V.~Balasubramaniam, and P.~S. Sehdev, ``Marginal and average weight-enabled data aggregation mechanism for the resource-constrained networks,'' \emph{Comput. Commun.}, vol. 174, pp. 101--108, 2021.

\bibitem{begum2023data}
B.~A. Begum and S.~V. Nandury, ``Data aggregation protocols for {WSN} and {IoT} applications--{A} comprehensive survey,'' \emph{J King Saud Univ Comput Inf Sci.}, 2023.

\bibitem{rawat2023survey}
P.~Rawat and S.~Chauhan, ``A survey on clustering protocols in wireless sensor network: {Taxonomy}, comparison, and future scope,'' \emph{J Ambient Intell Humaniz Comput.}, vol.~14, no.~3, pp. 1543--1589, 2023.

\bibitem{ochiai2005collaborative}
H.~Ochiai, P.~Mitran, H.~V. Poor, and V.~Tarokh, ``Collaborative beamforming for distributed wireless ad hoc sensor networks,'' \emph{IEEE Trans. Signal Process.}, vol.~53, no.~11, pp. 4110--4124, 2005.

\bibitem{sun2020improving}
G.~Sun, X.~Zhao, G.~Shen, Y.~Liu, A.~Wang, S.~Jayaprakasam, Y.~Zhang, and V.~C. Leung, ``Improving performance of distributed collaborative beamforming in mobile wireless sensor networks: {A} multiobjective optimization method,'' \emph{IEEE Internet Things J.}, vol.~7, no.~8, pp. 6787--6801, 2020.

\bibitem{smida2023dual}
O.~B. Smida and S.~Affes, ``Dual-hop robust distributed collaborative beamforming over nominally rectangular wsns in slightly to moderately scattered environments,'' in \emph{Proc. IWCMC}, 2023, pp. 1015--1021.

\bibitem{jung2019secrecy}
H.~Jung and I.-H. Lee, ``Secrecy performance analysis of analog cooperative beamforming in three-dimensional gaussian distributed wireless sensor networks,'' \emph{IEEE Trans. Wireless Commun.}, vol.~18, no.~3, pp. 1860--1873, 2019.

\bibitem{wang2021uplink}
A.~Wang, Y.~Wang, G.~Sun, J.~Li, S.~Liang, and Y.~Liu, ``Uplink data transmission based on collaborative beamforming in {UAV}-assisted {MWSNs},'' in \emph{Proc. IEEE GLOBECOM}, 2021, pp. 1--6.

\bibitem{bao2019stochastic}
X.~Bao, H.~Liang, Y.~Liu, and F.~Zhang, ``A stochastic game approach for collaborative beamforming in {SDN-based} energy harvesting wireless sensor networks,'' \emph{IEEE Internet Things J.}, vol.~6, no.~6, pp. 9583--9595, 2019.

\bibitem{sun2019energy}
G.~Sun, Y.~Liu, Z.~Chen, A.~Wang, Y.~Zhang, D.~Tian, and V.~C. Leung, ``Energy efficient collaborative beamforming for reducing sidelobe in wireless sensor networks,'' \emph{IEEE Trans. Mob. Comput.}, vol.~20, no.~3, pp. 965--982, 2019.

\bibitem{liu2023energy}
T.~Liu, X.~Qu, W.~Tan, R.~Wen, and L.~Yang, ``Energy-efficient joint collaborative and passive beamforming for intelligent-reflecting-surface-assisted wireless sensor networks,'' \emph{IEEE Internet Things J.}, vol.~10, no.~19, pp. 17\,193--17\,205, 2023.

\bibitem{hasan2023power}
M.~Z. Hasan, S.~A. Alabady, and M.~F.~M. Salleh, ``Power control for collaborative sensors in internet of things environments using k-means approach,'' in \emph{Proc. ACN}, 2023, pp. 209--224.

\bibitem{heinzelman2000application}
W.~B. Heinzelman, ``Application-specific protocol architectures for wireless networks,'' Ph.D. dissertation, Massachusetts Institute of Technology, 2000.

\bibitem{behera2019residual}
T.~M. Behera, S.~K. Mohapatra, U.~C. Samal, M.~S. Khan, M.~Daneshmand, and A.~H. Gandomi, ``Residual energy-based cluster-head selection in {WSNs} for {IoT} application,'' \emph{IEEE Internet Things J.}, vol.~6, no.~3, pp. 5132--5139, 2019.

\bibitem{chen2022d2crp}
C.~Chen, L.-C. Wang, and C.-M. Yu, ``{D2CRP}: {A} novel distributed 2-hop cluster routing protocol for wireless sensor networks,'' \emph{IEEE Internet Things J.}, vol.~9, no.~20, pp. 19\,575--19\,588, 2022.

\bibitem{lin2023cmstr}
D.~Lin, Z.~Lin, L.~Kong, and Y.~L. Guan, ``{CMSTR}: {A} constrained minimum spanning tree based routing protocol for wireless sensor networks,'' \emph{Ad Hoc Netw.}, vol. 146, p. 103160, 2023.

\bibitem{wang2012coverage}
B.~Wang, H.~B. Lim, and D.~Ma, ``A coverage-aware clustering protocol for wireless sensor networks,'' \emph{Comput. Netw.}, vol.~56, no.~5, pp. 1599--1611, 2012.

\bibitem{tao2013flow}
Y.~Tao, Y.~Zhang, and Y.~Ji, ``Flow-balanced routing for multi-hop clustered wireless sensor networks,'' \emph{Ad hoc networks}, vol.~11, no.~1, pp. 541--554, 2013.

\bibitem{song2016coverage}
X.~Song, T.~Wen, W.~Sun, D.~Zhang, Q.~Guo, and Q.~Zhang, ``A coverage-aware unequal clustering protocol with load separation for ambient assisted living based on wireless sensor networks,'' \emph{China Commun.}, vol.~13, no.~5, pp. 47--55, 2016.

\bibitem{Kavitha2021AiII}
D.~Kavitha and A.~Chinnasamy, ``Ai integration in data driven decision making for resource management in {Internet of Things(IoT)}: {A} survey,'' in \emph{Proc. IEMECON}, 2021, pp. 1--5.

\bibitem{snigdh2021energy}
I.~Snigdh, S.~S. Surani, and N.~K. Sahu, ``Energy conservation in query driven wireless sensor networks,'' \emph{Microsyst. Technol.}, vol.~27, no.~3, pp. 843--851, 2021.

\bibitem{biswas2020true}
P.~Biswas and T.~Samanta, ``True event-driven and fault-tolerant routing in wireless sensor network,'' \emph{Wireless Pers. Commun.}, vol. 112, no.~1, pp. 439--461, 2020.

\bibitem{saeidi2014fusion}
V.~Saeidi, B.~Pradhan, M.~O. Idrees, and Z.~Abd~Latif, ``Fusion of airborne {LiDAR} with multispectral {SPOT} 5 image for enhancement of feature extraction using {Dempster--Shafer} theory,'' \emph{IEEE Trans. Geosci. Remote Sens.}, vol.~52, no.~10, pp. 6017--6025, 2014.

\bibitem{wang2019new}
J.~Wang, O.~T. Tawose, L.~Jiang, and D.~Zhao, ``A new data fusion algorithm for wireless sensor networks inspired by hesitant fuzzy entropy,'' \emph{Sensors-basel.}, vol.~19, no.~4, p. 784, 2019.

\bibitem{yu2024data}
X.~Yu, W.~Peng, K.~Zhang, Z.~Zhou, and Y.~Liu, ``Data fusion algorithm of wireless sensor network based on clustering and fuzzy logic,'' \emph{Telecommun. Syst.}, pp. 1--10, 2024.

\bibitem{haro2013energy}
B.~B. Haro, S.~Zazo, and D.~P. Palomar, ``Energy efficient collaborative beamforming in wireless sensor networks,'' \emph{IEEE Trans. Signal Process.}, vol.~62, no.~2, pp. 496--510, 2013.

\bibitem{fei2025deep}
J.~Fei, X.~Zhang, C.~Li, F.~Hao, Y.~Guo, and Y.~Fu, ``A deep data fusion-based reconstruction of water index time series for intermittent rivers and ephemeral streams monitoring,'' \emph{ISPRS J. Photogramm. Remote Sens.}, vol. 220, pp. 339--353, 2025.

\bibitem{john2023evaluation}
A.~John, A.~Padinjarathala, E.~Doheny, B.~Cardiff, and D.~John, ``An evaluation of {ECG} data fusion algorithms for wearable iot sensors,'' \emph{INFORM FUSION}, vol.~96, pp. 237--251, 2023.

\bibitem{adade2024advanced}
S.~Y.-S.~S. Adade, H.~Lin, N.~A.~N. Johnson, X.~Nunekpeku, J.~H. Aheto, J.-N. Ekumah, B.~A. Kwadzokpui, E.~Teye, W.~Ahmad, and Q.~Chen, ``Advanced food contaminant detection through multi-source data fusion: Strategies, applications, and future perspectives,'' \emph{Trends Food Sci. Technol.}, p. 104851, 2024.

\bibitem{hua2008optimal}
C.~Hua and T.-S.~P. Yum, ``Optimal routing and data aggregation for maximizing lifetime of wireless sensor networks,'' \emph{IEEE ACM Trans. Netw.}, vol.~16, no.~4, pp. 892--903, 2008.

\bibitem{jayaprakasam2015psogsa}
S.~Jayaprakasam, S.~K.~A. Rahim, and C.~Y. Leow, ``{PSOGSA-Explore}: {A} new hybrid metaheuristic approach for beampattern optimization in collaborative beamforming,'' \emph{Appl. Soft Comput.}, vol.~30, pp. 229--237, 2015.

\bibitem{heinzelman2002application}
W.~B. Heinzelman, A.~P. Chandrakasan, and H.~Balakrishnan, ``An application-specific protocol architecture for wireless microsensor networks,'' \emph{IEEE Trans Wirel Commun.}, vol.~1, no.~4, pp. 660--670, 2002.

\bibitem{luo2006adaptive}
H.~Luo, J.~Luo, Y.~Liu, and S.~K. Das, ``Adaptive data fusion for energy efficient routing in wireless sensor networks,'' \emph{IEEE Trans. Comput.}, vol.~55, no.~10, pp. 1286--1299, 2006.

\bibitem{chen2023balancing}
Q.~Chen, Q.~Zhang, and Y.~Liu, ``Balancing exploration and exploitation in episodic reinforcement learning,'' \emph{Expert Syst. Appl}, vol. 231, p. 120801, 2023.

\bibitem{he2024deep}
H.~He, X.~Meng, Y.~Wang, A.~Khajepour, X.~An, R.~Wang, and F.~Sun, ``Deep reinforcement learning based energy management strategies for electrified vehicles: Recent advances and perspectives,'' \emph{Renew Sustain Energy Rev.}, vol. 192, p. 114248, 2024.

\bibitem{shinohara2013beam}
N.~Shinohara, ``Beam control technologies with a high-efficiency phased array for microwave power transmission in japan,'' \emph{Proc. IEEE}, vol. 101, no.~6, pp. 1448--1463, 2013.

\bibitem{lindsey2002pegasis}
S.~Lindsey and C.~S. Raghavendra, ``{PEGASIS}: {Power-efficient} gathering in sensor information systems,'' in \emph{Proc. IEEE AeroConf}, vol.~3, 2002, pp. 3--3.

\bibitem{farman2018multi}
H.~Farman, B.~Jan, H.~Javed, N.~Ahmad, J.~Iqbal, M.~Arshad, and S.~Ali, ``Multi-criteria based zone head selection in internet of things based wireless sensor networks,'' \emph{Future Gener. Comp. Sy.}, vol.~87, pp. 364--371, 2018.

\bibitem{lillicrap2015continuous}
T.~P. Lillicrap, J.~J. Hunt, A.~Pritzel, N.~Heess, T.~Erez, Y.~Tassa, D.~Silver, and D.~Wierstra, ``Continuous control with deep reinforcement learning,'' \emph{arXiv preprint arXiv:1509.02971}, 2015.

\bibitem{haarnoja2018soft}
T.~Haarnoja, A.~Zhou, K.~Hartikainen, G.~Tucker, S.~Ha, J.~Tan, V.~Kumar, H.~Zhu, A.~Gupta, P.~Abbeel \emph{et~al.}, ``Soft actor-critic algorithms and applications,'' \emph{arXiv preprint arXiv:1812.05905}, 2018.

\bibitem{schulman2017proximal}
J.~Schulman, F.~Wolski, P.~Dhariwal, A.~Radford, and O.~Klimov, ``Proximal policy optimization algorithms,'' \emph{arXiv preprint arXiv:1707.06347}, 2017.

\bibitem{minturn2013distributed}
A.~Minturn, D.~Vernekar, Y.~L. Yang, and H.~Sharif, ``Distributed beamforming with imperfect phase synchronization for cognitive radio networks,'' in \emph{Proc. IEEE Int. Conf. Commun. (ICC)}, 2013, pp. 4936--4940.

\bibitem{jung2020secure}
H.~Jung, S.-W. Ko, and I.-H. Lee, ``Secure transmission using linearly distributed virtual antenna array with element position perturbations,'' \emph{IEEE Trans. Veh. Technol.}, vol.~70, no.~1, pp. 474--489, 2020.

\bibitem{shetiya2024verification}
S.~S. Shetiya, V.~Vyas, and S.~Renukuntla, ``Verification and validation of autonomous systems,'' \emph{arXiv preprint arXiv:2411.13614}, 2024.

\bibitem{ponnuru2025baap}
R.~B. Ponnuru, S.~A. Kumar, M.~Azab, and G.~R. Alavalapati, ``{BAAP-FIoT}: Blockchain assisted authentication protocol for fog-enabled internet of things environment,'' \emph{IEEE Internet Things J.}, 2025.

\bibitem{chai2025iot}
X.~Chai, B.~G. Lee, C.~Hu, M.~Pike, D.~Chieng, R.~Wu, and W.-Y. Chung, ``{IoT-FAR}: A multi-sensor fusion approach for iot-based firefighting activity recognition,'' \emph{INFORM FUSION}, vol. 113, p. 102650, 2025.

\bibitem{ni2024novel}
P.~Ni, R.~Zhou, Q.~Han, X.~Du, K.~Xu, Z.~Jia, and Y.~Bai, ``A novel wireless iot sensing system for cable force identification and monitoring,'' \emph{Eng. Struct.}, vol. 314, p. 118318, 2024.

\bibitem{marche2020exploit}
C.~Marche, L.~Atzori, V.~Pilloni, and M.~Nitti, ``How to exploit the social internet of things: Query generation model and device profiles’ dataset,'' \emph{Comput. Netw.}, vol. 174, p. 107248, 2020.

\bibitem{silvestri2024real}
A.~Silvestri, D.~Coraci, S.~Brandi, A.~Capozzoli, E.~Borkowski, J.~K{\"o}hler, D.~Wu, M.~N. Zeilinger, and A.~Schlueter, ``Real building implementation of a deep reinforcement learning controller to enhance energy efficiency and indoor temperature control,'' \emph{Appl. Energy}, vol. 368, p. 123447, 2024.

\end{thebibliography}
% \bibliography{arxiv.bbl}

\vspace{-10 mm}
\begin{IEEEbiography}[{\includegraphics[width=1in,height=1.25in,clip,keepaspectratio]{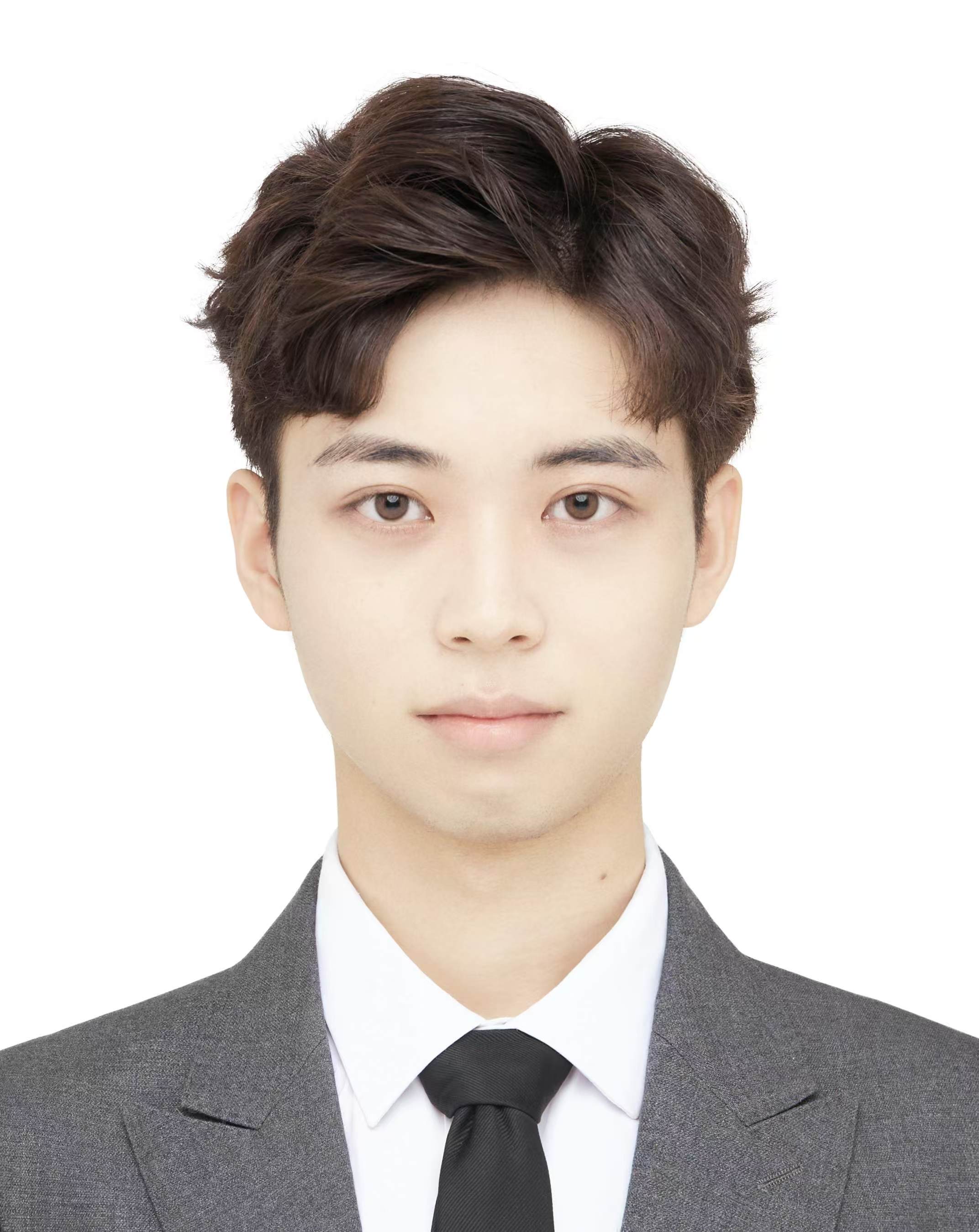}}]{Yangning Li} received the BS degree in software engineering from Jilin University, Changchun, China, in 2023. He is currently working toward the MS degree in computer science and technology with Jilin University, Changchun, China. His research interests include wireless networks and collaborative beamforming.
\end{IEEEbiography}

\vspace{-10 mm}
\begin{IEEEbiography}[{\includegraphics[width=1in,height=1.25in,clip,keepaspectratio]{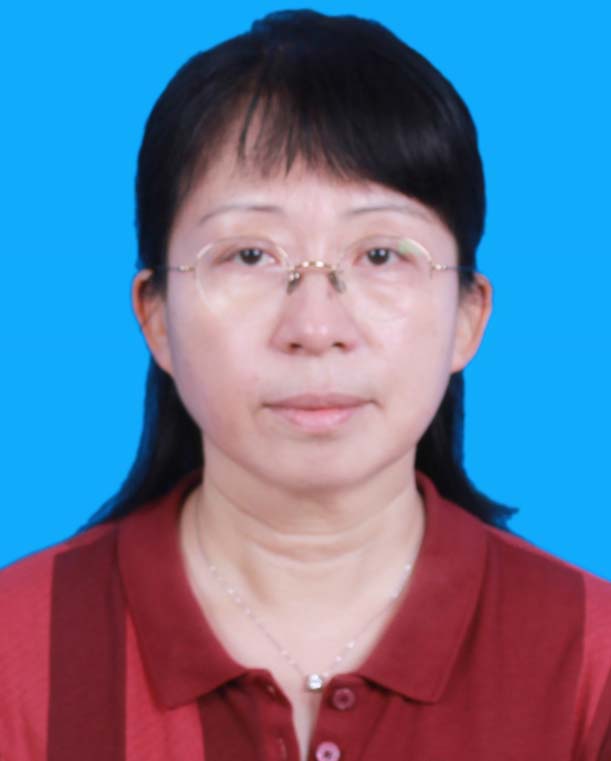}}]{Hui Kang} received the M.E. and Ph.D. degrees from Jilin University in 1996 and 2007, respectively. She is currently a Professor with the College of Computer Science and Technology, Jilin University. Her research interests include information integration and distributed computing.
\end{IEEEbiography}

\vspace{-10 mm}
\begin{IEEEbiography}
[{\includegraphics[width=1in,height=1.25in,clip,keepaspectratio]{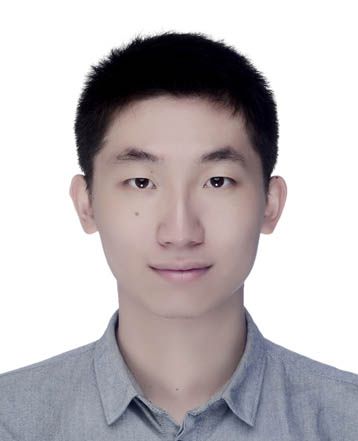}}]
{Jiahui Li} received his B.S. in Software Engineering, and M.S. and Ph.D. in Computer Science and Technology from Jilin University, Changchun, China, in 2018, 2021, and 2024, respectively. He was a visiting Ph.D. student at the Singapore University of Technology and Design (SUTD). He currently serves as an assistant researcher in the College of Computer Science and Technology at Jilin University. His current research focuses on integrated air-ground networks, UAV networks, wireless energy transfer, and optimization.
\end{IEEEbiography}

\vspace{-10 mm}
\begin{IEEEbiography}[{\includegraphics[width=1in,height=1.25in,clip,keepaspectratio]{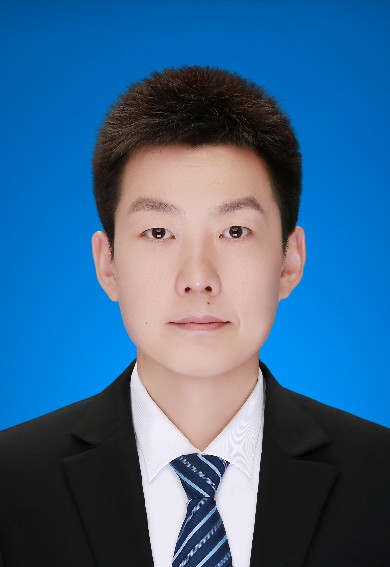}}]{Geng Sun} (Senior Member, IEEE) received the B.S. degree in communication engineering from Dalian Polytechnic University, and the Ph.D. degree in computer science and technology from Jilin University, in 2011 and 2018, respectively. He was a Visiting Researcher with the School of Electrical and Computer Engineering, Georgia Institute of Technology, USA. He is a Professor in College of Computer Science and Technology at Jilin University, and His research interests include wireless networks, UAV communications, collaborative beamforming and optimizations.
\end{IEEEbiography}

\vspace{-10 mm}
\begin{IEEEbiography}[{\includegraphics[width=1in,height=1.25in,clip,keepaspectratio]{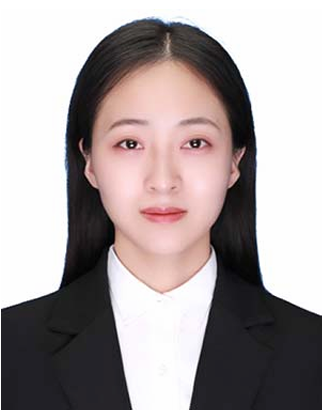}}]{Zemin Sun} received a BS degree in Software Engineering, an MS degree and a Ph.D degree in Computer Science and Technology from Jilin University, Changchun, China, in 2015, 2018, and 2022, respectively. Her research interests include vehicular networks, edge computing, and game theory.
\end{IEEEbiography}

\vspace{-10 mm}
\begin{IEEEbiography}[{\includegraphics[width=1in,height=1.25in,clip,keepaspectratio]{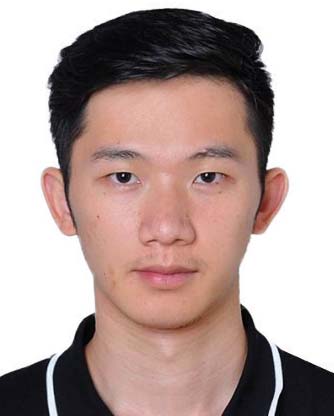}}]{Jiacheng Wang} is a Research Associate in Computer Science and Engineering at Nanyang Technological University, Singapore. Prior to that, he received his Ph.D. at the School of Communication and Information Engineering, Chongqing University of Posts and Telecommunications, Chongqing, China. His research interests include wireless sensing and semantic communications.
\end{IEEEbiography}

\vspace{-10 mm}
\begin{IEEEbiography}[{\includegraphics[width=1in,height=1.25in,clip,keepaspectratio]{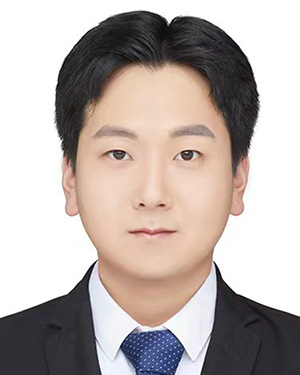}}]{Changyuan Zhao} received the B.Sc. degree in computing and information science from the University of Science and Technology of China, Hefei, China, in 2020, and the MA.Eng. degree in computer science from the Institute of Software, CAS, Beijing, China, in 2023. He is currently pursuing the Ph.D. degree with the College of Computing and Data Science, Nanyang Technological University, Singapore. His research interests include generative AI, communication security, and resource allocation.
\end{IEEEbiography}

\vspace{-10 mm}
\begin{IEEEbiography}[{\includegraphics[width=1in,height=1.25in,clip,keepaspectratio]{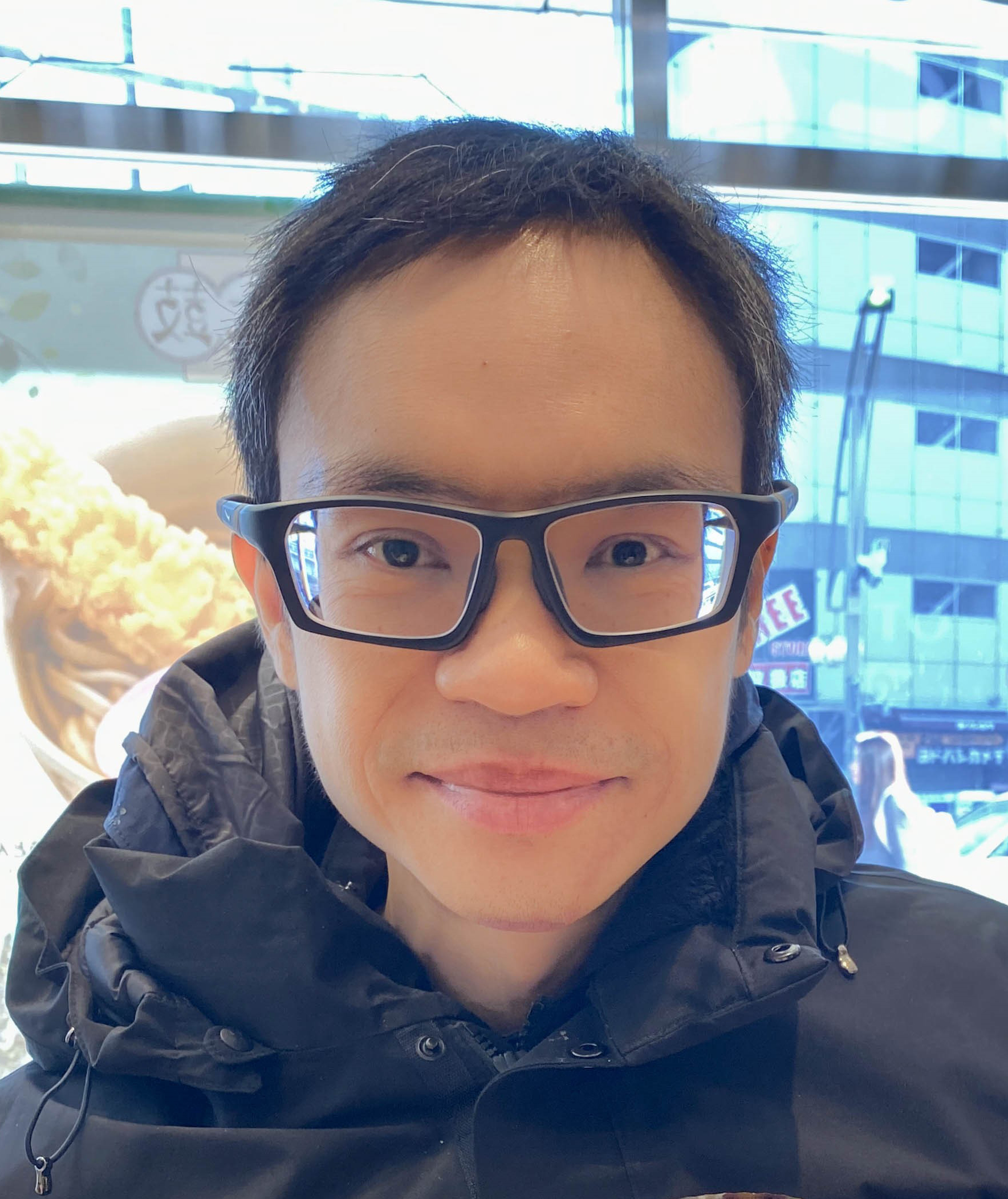}}]{Dusit Niyato} (Fellow, IEEE) received his Bachelor degree in Computer Engineering from King Mongkut's Institute of Technology Ladkrabang, Thailand, Master and Ph.D. degrees from the University of Manitoba in 2005 and 2008, respectively. He is currently a professor in the College of Computing and Data Science, at Nanyang Technological University, Singapore. His research interests are in the areas of sustainability, edge intelligence, decentralized machine learning, and incentive mechanism design.
\end{IEEEbiography}

\end{document}